\documentstyle[epsfig]{osudissert96}

\renewcommand\typesetChapterTitle[1]{\uppercase{#1}}
\newcommand {\al} {\alpha}
\newcommand {\xmes} {\frac{1}{2}\int d^2x^\perp dx^-}
\newcommand {\der} {\partial}
\newcommand {\hr} {H_{\lambda}}

\newcommand {\bea}{\begin{eqnarray}}   
\newcommand {\eea}{\end{eqnarray}}
\newcommand {\beaa}{\begin{eqnarray*}}   
\newcommand {\eeaa}{\end{eqnarray*}}
\newcommand {\be}{\begin{equation}}   
\newcommand {\ee}{\end{equation}}
\newcommand {\eps} {\epsilon}

\newcommand {\del} {\delta}
\newcommand {\sig} {\sigma}
\newcommand  {\lam} {\lambda}
\newcommand  {\pr} {\prime}
\newcommand  {\kap} {\kappa}
\newcommand  {\Lam} {\Lambda}
\newcommand  {\gam} {\gamma}

\newcommand  {\veps} {\varepsilon}
\newcommand   {\pp}   {({\bf p}-{{\bf p}^\pr})^2}
\newcommand   {\easy}  {\del M_{_{1}}^2}
\newcommand    {\hard} {\del M_{_{2}}^2}
\newcommand    {\p} {\Omega_p}
\newcommand    {\ppr} {\Omega_{p^\prime}}

\newcommand    {\pig} {\left(\int_0^\pi \frac{d\omega \sin^2\omega}
{1+\cos\omega} f_{2,1}(\omega) f_{n,1}(\omega)\right)^2}
 \newcommand    {\tp} {{\tilde{\phi}}}

\newcommand {\Del} {\Delta}
\newcommand {\nn} {\nonumber}
\newcommand  {\thh} {P_H}
\newcommand  {\tll} {P_L}
\newcommand {\da} {\dagger}
\newcommand  {\bp}  {|{\bf p}|}
\newcommand   {\hp}  {\hat{{\bf p}}}
\newcommand   {\lr}  {\leftrightarrow}

\begin{document}

\graduationyear{1997}
\author{Billy Darwin Jones}
\title{LIGHT-FRONT HAMILTONIAN APPROACH TO 
THE BOUND-STATE PROBLEM IN QUANTUM
ELECTRODYNAMICS}
\authordegrees{A.S., B.S.}  
\unit{Department of Physics}

\advisorname{Dr. Robert J. Perry}
\member{Dr. Thomas J. Humanic, Dr. Furrukh S. Khan,
Dr. Stephen S. Pinsky, Dr. Junko Shigemitsu}

\maketitle

%
%

\disscopyright


\begin{abstract}

This dissertation presents the  first theoretical investigation
of the Lamb shift in a {\em light-front 
hamiltonian approach}: the dominant part of the 
splitting between the $2 S_{\frac{1}{2}}$ and 
$2 P_{\frac{1}{2}}$ energy
levels in hydrogen is calculated. 
Also presented for the  first time 
is an {\it analytic} calculation in a {\em 
light-front hamiltonian
approach} of 
the singlet-triplet spin splitting in the ground
state of positronium through order $\al^4$. 

We study the QED bound-state problem  in
 a  light-front
  hamiltonian approach. 
 We start from a canonical QED Hamiltonian, and  
  set up a general formalism for
 deriving the effective  Hamiltonian $\hr$
 to some prescribed order in $\alpha$ (with $\alpha \ll 1$).
 $\hr$ is renormalized 
  by requiring it to satisfy coupling coherence.
  Then we use bound-state perturbation theory (BSPT)
  to compute the low-lying spectrum of interest 
  in a consistent set of approximations
  to some
 prescribed order
 in $\al$ and $\al \,\ln(1/\al)$.
 The general formulas are  applied explicitly to the 
 positronium and hydrogen systems.
 Renormalization is  carried out  through order $e^2$, and a 
 nonrelativistic limit of the theory is taken:
 $
 \bp\ll m
 \;,
 $
 where $\bp$ is a typical electron momentum and $m$ is the 
 electron mass.
  Also, in order 
  to derive the results in the few-body sector of interest---$|e{\overline e}\rangle$
   for positronium
  and $|ep\rangle$ for hydrogen---we require our
  final cutoff to satisfy
 $ m \alpha^2\ll
  \tilde{\lam} \ll m \alpha 
 $.
 This upper bound is the dominant energy of emitted photons, and
   this lower bound ensures that we do not  remove
  the non-perturbative energy scale of interest with our renormalization group
  transformation
  that we run perturbatively.

\end{abstract}

%
%


\dedication{To my family}

%
%

\begin{acknowledgements}

 A special thanks to my adviser Robert  Perry for
his guidance,
  wisdom (especially regarding renormalization) and friendship.

I would like to thank Brent H. Allen for helping me to
learn what quantum field theory is all about.

 I would like to thank Stan  G{\l}azek for useful
  discussions and for collaboration on our positronium paper.

A special thanks
to  Koji Harada for introducing the
idea behind the Lamb shift calculation, and also for stimulating 
 conversations over the past few years.

Thanks to my fellow nucs Martina Brisudova, Werner Koepf, Lisa Kurth, 
David Makovoz, John Rusnak, Sergio Szpigel and Trey White
for useful discussions, and Bunny Clark and Dick Furnstahl for their guidance.

There are many coneheads with whom I have had
profitable  discussions.
I would like to mention Edsel Ammons,
Stan Brodsky, Matthias Burkardt, John Hiller,
Gerry Miller, Steve Pinsky,  Wayne Polyzou,
Dave Robertson (who really helped clarify renormalization to me over and over and ...) and Ken Wilson.

I would like to thank Tom Humanic, Furrukh Khan,
Steve Pinsky and Junko Shigemitsu for 
serving on my committee.

And last but certainly not least,
thanks to my wife Saeko for all her patience and love, and
  my parents and siblings
 for their continued
love and support over the years. 

\end{acknowledgements}

\begin{vita}

\dateitem{June 28, 1968}{Born -- Powell, Wyoming}

\dateitem{1987--1989}{A.S., Western
Nebraska Community College, Scottsbluff, Nebraska}

\dateitem{1989--1992}{B.S., Department of Physics,
Colorado School of Mines, Golden, Colorado}

\dateitem{1992--1993}{Graduate Fellow,
The Ohio State University, Columbus, Ohio}

\dateitem{1993--present}{Research Assistant,
The Ohio State University, Columbus, Ohio}

\begin{publist}
\researchpubs

\pubitem{Billy D. Jones, 
\newblock ``Singlet-triplet splitting of positronium in light-front QED,''
\newblock in {\em 25th Coral Gables Conference on High Energy Physics and
Cosmology}, Proceedings, Miami Beach, Florida,
edited by Behram N. Kursunoglu, Stephen Mintz and
Arnold Perlmutter, (Plenum Press, New York, 1997).}

\pubitem{Billy D. Jones,    
Robert J. Perry and Stanis{\l}aw D. G{\l}azek,
\newblock ``Analytic treatment of positronium spin splittings in 
light-front QED,'' 
\newblock {\em Physical Review D} {\bf 55}, 6561 (1997).}

\pubitem{Billy D. Jones and Robert J. Perry,  
\newblock ``Lamb shift in a light-front hamiltonian approach,'' 
\newblock {\em Physical Review D} {\bf 55}, 7715 (1997).}

\end{publist}
\newpage
\begin{fieldsstudy}

\majorfield{Physics}

Studies in light-front field theory: Professor Robert J. Perry

\end{fieldsstudy}

\end{vita}

%
%

\tableofcontents
\listoffigures

%
%
\chapter{Introduction} 

There is much effort being put into solving for the hadronic spectrum from first
principles of Quantum Chromodynamics (QCD)
 in 3+1-dimensions 
using a light-front hamiltonian approach \cite{osu}. However, 
low-energy QCD is challenging,
and a realistic analytic calculation may be impossible.
   There is a need for  exact analytic calculations that test and illustrate
   the
approach. 
The core of this dissertation provides two examples of this in Quantum Electrodynamics
(QED): 
analytic
calculations of (i)
 positronium's ground state spin splitting
through order $\al^4$ and (ii) the dominant part of the  Lamb shift in hydrogen through order
$\al^5\, \ln(1/\al)$. 
These calculations are based on previously published work,
\cite{me1} and \cite{me2} respectively. The specific framework
of calculation, a hamiltonian light-front renormalization group approach,
was set up in 
the invited lectures of
  Perry \cite{perrybrazil}, where
the
leading order calculation 
(deriving ${\cal H}_0$)
was completed, and the two calculations of this
dissertation were mentioned as future prospective calculations---therein lies
the historical motivation.

Why is a calculation of the Lamb shift in hydrogen---which at the level of
detail found in this dissertation was largely
completed by Bethe in 1947 \cite{bethe}---or the ground state spin
splitting in positronium---which through order $\al^4$ was  calculated
by Ferrell over 40 years ago \cite{oldpos}---of any real
interest today?  While completing such calculations using new techniques may be
very interesting for formal and academic reasons, our primary motivation is to
lay groundwork for precision bound-state calculations in QCD.  These
calculations
provide an excellent pedagogical tool for illustrating light-front hamiltonian
techniques, which are not widely known; but more importantly, it presents three
of the central dynamical and computational problems that we must face to make
these techniques useful for solving QCD: How does a constituent picture emerge
in
a gauge field theory?  How do bound-state energy scales emerge
nonperturbatively?  How does rotational symmetry emerge in a nonperturbative
light-front calculation? These questions can be answered directly in QED, as this dissertation
shows. In QCD, the answers clearly change, but the overall computational framework does not,
and thus an analytic understanding of the framework is essential. And, as already mentioned,
QED allows this analytic understanding.

   An outline of this dissertation follows.
   There are six chapters of which this is the first. The final chapter
  contains a general discussion and summary.
  In Chapter~2 we introduce light-front field theory: A pedagogical introduction
  to light-front coordinates is given, and  we present a simple tree-level
   example
  illustrating the vanishing of vacuum mixings in light-front perturbation theory;
  scalar theory is used to illustrate the division between kinematic
  and dynamic  Poincar\'{e} generators in light-front field theory;
  a derivation of a canonical QED Hamiltonian in the light-cone gauge is given.
  In Chapter~3 we give an overview of the light-front hamiltonian bound-state
  problem and then discuss three renormalization group transformations
  in hamiltonian theory that are of interest; in the final section we
  discuss renormalization in light-front field theory which leads us
  to  introduce  ``coupling coherence" which is elucidated with
  a simple one-loop example in coupled scalar $O(2)$ theory.
   Chapters~4 and 5 contain the heart of this dissertation where the 
   aforementioned calculations in positronium and hydrogen respectively
   are presented.
   In Chapter~4 we also present a simpler method of calculating the spin
   splitting,
   which may turn out to be useful in carrying out
     future higher-order calculations.

\chapter{Light-front 
field theory and canonical QED Hamiltonian}\label{ch:2}

In this chapter a pedagogical introduction to light-front field theory, including
a simple example to illustrate the vanishing of vacuum mixings
in amplitudes and a discussion of the ten Poincar\'{e} generators, is
presented. Then a derivation of a canonical QED Hamiltonian is given.
\section{Light-front field theory}\label{sec2.1}

{\it Light-front} coordinates (also called {\it front form},
{\it null plane}, or 
{\it light-cone} coordinates---the usual coordinates are called 
{\it instant form} or
{\it equal-time}) were first presented by Dirac in 1949
 in his pursuit of
  alternative forms of relativistic dynamics that 
combine ``the restricted principle of relativity with the Hamiltonian
formulation of dynamics \cite{dirac, dirac2}." Recent interest in light-front coordinates continues
this pursuit in mainly two arenas: the low-energy nonperturbative
bound-state problem of QCD, where light-front coordinates may allow a simpler vacuum
structure (replacing the vacuum structure with the appropriate effective interactions through
renormalization)
than with equal-time coordinates, and high-energy scattering observables
where light-front
coordinates are the natural coordinates of the system. For an extensive list of light-front
references through the early 90s see \cite{lfref}. 
We must apologize for the inadequate recent references given to this diverse and active area
of research, but for a fairly complete
 list see the following recent reviews \cite{lfreview} and 
references within.

An introduction to light-front field theory will now be 
given.\footnote{For some other works with nice introductions see
 for example
\cite{lfped} and references
within.}
An ``ET" label implies an equal-time vector; a ``LF" 
label implies a light-front vector.
Light-front coordinates are defined  by
\bea
&\bullet&x^{0}_{LF} = x^0_{ET} + x^3_{ET}\\
&\bullet&x^{3}_{LF} = x^0_{ET} - x^3_{ET}\\
&\bullet&x^{i}_{LF} = x^{i}_{ET} ,\; i=1,2\\
&\bullet&x^\mu_{LF}=(x^0_{LF},x^3_{LF},x^i_{LF})\\
&\bullet&  x^{\mu}_{ET}=(x^0_{ET},x^3_{ET},x^i_{ET}).
\eea
Carrying around these ``LF" and ``ET"
labels is tedious, so following convention (an 
obvious convention given the signs on the
right-hand-side of the top two equations above), we define
\bea
x^+ &\equiv& x^{0}_{LF}\;,\\
x^- &\equiv& x^{3}_{LF}
\;,
\eea
and then drop the labels. There is no notational ambiguity
because $x^{i}_{LF} = x^{i}_{ET}$. In this introduction we will
keep the labels if possible confusion may arise otherwise.
More formally the above is written 
\be
x^\mu_{LF}=A^\mu_{~\nu} ~x^\nu_{ET}
\;.\label{eq:Amunu}
\ee
Actually $A^\mu_{~\nu}$ is defined to be
the same for all 4-vectors, but here
we just show the transformation on the
coordinate $x^\mu$. Note that $x^\mu_{LF}$ and $x^\mu_{ET}$ are not
related by a Lorentz transformation since $\det(A)=-2$, while a Lorentz
transformation $L$ must satisfy $\det(L)=\pm 1$.
The next step is the requirement that all Lorentz
scalars are equivalent---done by adjusting the metric tensor.
For the $x^\mu x_\mu$ scalar we have
\be
g_{\mu\nu}^{ET} x^\mu_{ET}
x^\nu_{ET}\equiv g_{\mu\nu}^{LF} x^\mu_{LF}
x^\nu_{LF}\;.
\ee
Since $A=A^{tr}$ (but note $A\neq A^{-1}$) this simply implies
\be
g_{\al\beta}^{LF}=(A^{-1})^\mu_{~\al} ~g _{\mu\nu}^{ET}~ (A^{-1})^\nu_{~\beta}
\longrightarrow g^{LF}_{lower}=A^{-1}\cdot g^{ET}_{lower}\cdot A^{-1}
\ee
and
\be
g^{\al\beta}_{LF}=A^\al_{~\mu} ~g ^{\mu\nu}_{ET}~ A^\beta_{~\nu}
\longrightarrow g_{LF}^{upper}=A \cdot g_{ET}^{upper}\cdot A
\;,
\ee
where the terms on the far right are written
in convenient matrix notation.
Given the standard equal-time metric tensor, $g_{ET}^{\mu \nu}=g_{\mu \nu}^{ET}=
(1,-1,-1,-1)$, we follow convention and drop the LF labels,
take $0\longrightarrow +$ and $3\longrightarrow -$ and end up with\footnote{Note that this
replacement 
    `$0\longrightarrow+$' and `$3\longrightarrow-$' applies whether it be an upper or
   lower index---simple example: $x_{LF}^0\equiv
   x^+$ and $x^{LF}_0\equiv x_+=g_{+-}x^-=x^-/2$.\label{footnotehey}} 
\be
1=\frac{g^{+-}}{2}=\frac{g^{-+}}{2}=2 g_{+-}=2 g_{-+}=-g^{11}=-g^{22}=-g_{11}=-g_{22}
\;.
\ee
The components of the light-front metric tensor
not mentioned are zero.
Note that these factors of two in the metric tensor
lead to factors of two in other places like
\beaa
d^4x=\frac{1}{2}dx^+ dx^- d^2x^{\perp}\;,\; x_{-}=g_{-+}x^{+}=\frac{1}{2}
(x^0+x^3)=\frac{1}{2}(x_0-x_3)\;,\;etc.
\eeaa

Conventionally, $x^+$ is chosen to be the light-front time coordinate. 
Thus $x^-$ is the light-front longitudinal space coordinate. Also, from the $p\cdot x$ scalar,
\beaa
p_\mu x^\mu&=& g_{+-} p^- x^{+} + g_{-+} p^+ x^{-}-p^i x^i=
\frac{1}{2} p^- x^{+} + \frac{1}{2} p^+ x^{-}-p^i x^i
~,\eeaa
we see that $p^-$ is 
the light-front energy coordinate, and $p^+$ is  the light-front
longitudinal momentum coordinate. The light-front dispersion relation
for an on mass-shell particle is interesting:
\be
p^\mu p_\mu=m^2~\Longrightarrow~p^-=\frac{{p^\perp}^2+m^2}{p^+}\;.
\ee
There is no square-root (compare $p^0=\pm \sqrt{{\bf p}^2+m^2}$ in equal-time)
and {\it small} longitudinal momentum $p^+$ implies {\it high} energy $p^-$ 
(except for a set of measure zero for a massless particle, i.e. $m=p^\perp=0$).

 All trajectories
in the forward light-cone in light-front coordinates have
$p^+ \geq 0$. The Lorentz-invariant 3-momentum integral---which is 
the method of summing over
 particle momenta whenever required---shows this (it also reminds us that
particle lines in hamiltonian diagrams are on mass-shell):
\beaa
&&\int\frac{d^4p}{(2 \pi)^4}2 \pi \del (p^2-m^2) \theta(p^0)
f(p)=\int\frac{d^3p}{(2\pi)^32p_0}f(p){\Biggr{|}}_{p_0=\sqrt{{\vec{p}}^2+m^2}}\\
&&\hskip1in=\;\int\frac{d^2p^\perp dp^+ \theta(p^+)}{16\pi^3p^+}f(p)
{\Biggr{|}}_{p^-=\frac{{p^\perp}^2+m^2}{p^+}}\equiv \int_{p}f(p)
\;~~.
\eeaa
 Especially note this last definition of $\int_p$~, which is a shorthand used
 in the dissertation often. 
\begin{figure}[t]\hskip0in
\centerline{\epsfig{figure=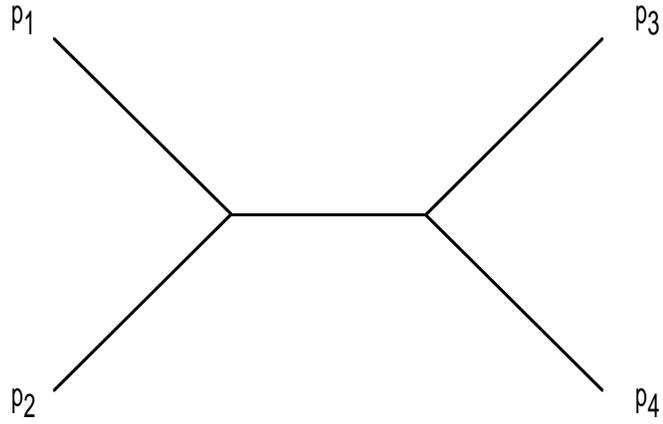,height=2.5in,width=3.5in}}
\caption{Feynman diagram 
 for the tree-level annihilation amplitude in $\phi^3$
 theory.} 
\label{fig:kogut1}
\end{figure}

The consequences of $p^+ > 0$ are illustrated by a simple example \cite{kands}.
 Consider the Feynman diagram of Figure~\ref{fig:kogut1}
 for the tree-level annihilation amplitude in $\phi^3$
 theory. This amplitude is a Lorentz scalar (and thus light-front and 
 equal-time field theory should
 give the same result for this amplitude) given by
 \bea
 \frac{g^2}{s-m^2}\;,
 \eea
 where
 \bea
 s=(p_1+p_2)^\mu\,(p_1+p_2)_\mu
 \;.\label{eq:s}
 \eea
 \begin{figure}[t!]\hskip0in
\centerline{\epsfig{figure=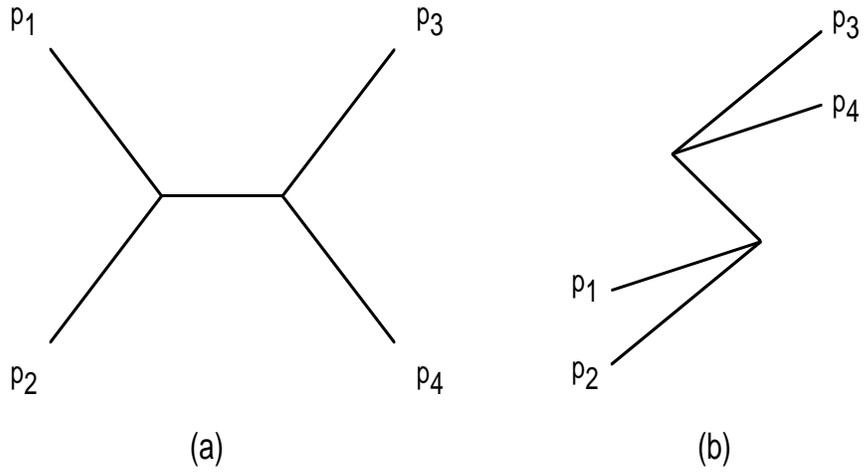,height=3in,width=4.5in}}
\caption{(a) Time-ordered diagram
 for the tree-level annihilation amplitude in $\phi^3$
 theory. (b) Time-ordered diagram containing vacuum mixings. It vanishes
 in light-front field theory.} 
\label{fig:kogut2}
\end{figure}

 In a hamiltonian approach all time orderings must be included which leads
 to the two diagrams of Figure~\ref{fig:kogut2}. 
 In
 equal-time field theory these diagrams contribute
 \bea
 \frac{g^2}{2 E({\bf p}_1+{\bf p}_2)\left[
 E({\bf p}_1)+E({\bf p}_2)-E({\bf p}_1+{\bf p}_2)\right]}
 \label{eq:kogut1}
 \eea
 and
 \bea
 \frac{g^2}{2 E({\bf p}_1+{\bf p}_2)\left[
 -E({\bf p}_1)-E({\bf p}_2)-E({\bf p}_1+{\bf p}_2)\right]}
 \label{eq:kogut2}
 \eea
 respectively, where $E({\bf p})=\sqrt{{\bf p}^2+m^2}$.
 Summing Eqs.~(\ref{eq:kogut1}) and (\ref{eq:kogut2}) gives $\frac{g^2}{s-m^2}$.
 In light-front field theory Eq.~(\ref{eq:kogut2}) vanishes and Eq.~(\ref{eq:kogut1})
 becomes
 \bea
 \frac{g^2}{(p_1+p_2)^+\left[P^-(p_1)+P^-(p_2)-P^-(p_1+p_2)\right]}=
 \frac{g^2}{{\cal P}^+\left[\frac{{{\cal P}^\perp}^2+s}{{\cal P}^+}-
 \frac{{{\cal P}^\perp}^2+m^2}{{\cal P}^+}
 \right]}=\frac{g^2}{s-m^2}
 \:,~~~
 \eea
 where $P^-(p)=\frac{{p^\perp}^2+m^2}{p^+}$, the 3-momentum ${\cal P}=p_1+p_2$,
 and $s$ is the same variable as defined in Eq.~(\ref{eq:s}).
 Thus the positivity of $p^+$ makes the diagram with vacuum mixings---as in Figure~\ref{fig:kogut2}b---vanish, and all 
 the amplitude is in the one time-ordering alone.
 This observation alone sparked much of the initial
 interest in light-front field theory \cite{lfref},
 and continues to be a topic of focus today.
 Today this topic deals with
  a subtlety that will not be discussed much in this dissertation but
 is intensely
 studied by the practitioners of light-front field theory; this subtlety
 deals with $p^+=0$ \cite{zm1} (the so called zero-modes) or 
 $p^+ \longrightarrow 0$ \cite{longpaper} (renormalization group approaches), and whether or not
  there is
  any nonperturbative physics hidden in this sector of the theory. From the example
  just shown it is clear that at least perturbatively the zero-modes do not contribute
  to the amplitudes. We will drop zero-modes initially and replace their effects
  through renormalization counterterms fixed by coupling coherence.
  Coupling coherence is explained in Section~\ref{coherence}, and for a
  discussion on ``$p^+$ renormalization group" see Subsection~\ref{relevant}.
  
  To complete this light-front
   introduction we will discuss the ten
   generators of the Poincar\'{e} group and how the division between the
   kinematic and dynamic operators is different in equal-time and light-front
   field theory. 
   The discussion is not intended to be complete. Transverse and longitudinal surface terms
   are just dropped, and a flavor of the subject is given.
   For further reading and references consult \cite{poincare}.
   
   The Poincar\'{e} generators are constructed from the stress tensor
    $T^{\mu \nu}$---for a concrete example take scalar
   theory
   \be
   T^{\mu \nu}=\frac{\partial {\cal L}}{\partial(\partial_\mu \phi)}~
   \partial^\nu \phi-g^{\mu \nu} {\cal L}=\partial^\mu\phi
   \partial^\nu\phi-g^{\mu \nu} {\cal L}
   \;.\label{eq:stress}
   \ee
   
   On an equal-time surface the generators are expressed as
   \be
   P^\mu=\int d^3xT^{0\mu}~,~J^{\mu\nu}=\int d^3x(x^\mu T^{0\nu}-
   x^\nu T^{0\mu})
   \;,\label{eq:2.20}
   \ee
   with the usual definition of boosts and rotations as 
   $J^{0i}=K^i$ and $J^{ij}=\epsilon_{ijk} J^k$  
   respectively.\footnote{$\epsilon_{123}=1$.
   Roman indices i, j, ... take on the values 1, 2 and 3 in equal-time discussions
   but only the values 1 and 2 for light-front discussions.}
   Note that $(P^i,J^i)$ are kinematic and $(P^0,K^i)$ 
   are dynamic generators as can be seen
   by the fact that the kinematic ones do not contain the hamiltonian 
   density $T^{00}$
   and therefore do not shift states off the initial surface $x^0=0$.
   
   On a light-front surface the generators are expressed as
   \be
   P^\mu=\frac{1}{2}\int d^2x^\perp dx^-T^{+\mu}~,
   ~J^{\mu\nu}= \frac{1}{2}\int d^2x^\perp dx^-(x^\mu T^{+\nu}-
   x^\nu T^{+\mu})
   \;,
   \label{eq:pmu}
   \ee
   where the boosts are $K^3=-\frac{1}{2} J^{+-}$ 
   and $E^i=J^{+i}$, and the rotations are
    $J^3=J^{12}$ and $F^i=J^{-i}$---see below for the
   relation between the equal-time and light-front generators. 
   Note that $(P^+,P^i,E^i,J^3)$ are kinematic as can seen by the
   fact that they do not contain the hamiltonian density $\frac{1}{2}T^{+-}$ 
   and 
   therefore do not shift states off the initial surface $x^+=0$.
   The transverse rotation generators $F^i$ are dynamic as can be simply
   seen by the fact that they contain the hamiltonian density 
    and thus
   shift states off the initial surface. The longitudinal boost generator
   $K^3=-\frac{1}{2} J^{+-}$ appears 
   to be dynamic since it contains the hamiltonian
   density; to see that it is kinematic it is important to note
   that the initial surface is $x^+=0$ (if it were $x^+={\rm constant}$, then
   $K^3$ would move states off the initial surface).
   A standard Lorentz transformation shows that the time coordinates
   in two frames boosted by $K^3$ are related by
   \be
   {x^+}^\pr=x^+ \exp(-\omega)
   \;,
   \ee
   where $\omega$ is the rapidity of the transformation [recall $v=\tanh(\omega)$
   where $v$ is the relative speed of the two frames]. So $x^+=0$ in the
   one frame is seen as ${x^+}^\pr=0$ in the other frame: 
   $K^3$ is kinematic.\footnote{This simple example shows that even
   in free field theory in order for $K^3$ to be kinematic, $x^+=0$ must
   be the choice for the initial surface.}
   This point will be discussed with interactions below.
   Note that $K^3$ should perhaps be called a longitudinal
   scaling generator instead
   of a boost generator since the action of $K^3$ in momentum space on all
   particle momenta is
   \be
   p^+\longrightarrow p^+ \:\exp(-\omega)~,~{\bf p}^\perp\longrightarrow{\bf p}^\perp
   \;,
   \ee
   and the exact Hamiltonian of the theory is transformed as
   \be
   P^-\longrightarrow P^-\:\exp(\omega)
   \;.
   \ee
   
   The relation between the equal-time and light-front generators is easily
   found by using $A^\mu_{~\nu}$ of Eq.~(\ref{eq:Amunu}). 
   $P^\mu$ is a four-vector, so we have simply 
   \be
   P_{LF}^\mu=A^\mu_{~\nu}P_{ET}^\nu\;,
   \ee
   which gives
   \bea
   &&P^+\equiv P^0_{LF}=P^0_{ET}+P^3_{ET}~,~P^-\equiv P^3_{LF}=P^0_{ET}-P^3_{ET}\;,\nn\\
   &&P^1\equiv P^1_{LF}=P^1_{ET}~,~P^2\equiv P^2_{LF}=P^2_{ET}\;.
   \eea
   For the boosts and rotations, which form a tensor, we have
   \be
   J^{\mu\nu}_{LF}=A^\mu_{~\alpha}~J^{\al\beta}_{ET}~A^\nu_{~\beta}\longrightarrow
   J_{LF}^{upper}=A\cdot J_{ET}^{upper}\cdot A
   \;,
   \ee
   where the relation on the right is written in convenient matrix notation.
   If the lower indices are desired we can just use the light-front
   metric tensor (it keeps track of all the factors of $2$) which in
   convenient matrix notation gives
   \be
   J^{LF}_{lower}=g^{LF}_{lower}\cdot J_{LF}^{upper}\cdot g_{lower}^{LF}
   \;.
   \ee
   Recalling footnote~\ref{footnotehey}
   and our conventions below Eqs.~(\ref{eq:2.20}) and (\ref{eq:pmu}),
   this gives
   \bea
  && E^1\equiv J^{01}_{LF}=J^2_{ET}+K^1_{ET}~,~E^2\equiv J^{02}_{LF}=-J^1_{ET}+K^2_{ET}~,~
   -2 K^3\equiv J^{03}_{LF}=-2K^3_{ET}~,~\nn\\
   &&F^1\equiv J^{31}_{LF}=-J^2_{ET}+K^1_{ET}~,~
   F^2\equiv J^{32}_{LF}=J^1_{ET}+K^2_{ET}~,~J^3\equiv J^{12}_{LF}=J^3_{ET}
   \eea
   for the upper light-front indices, and
   \bea
  && J_{01}^{LF}=\frac{J^2_{ET}-K^1_{ET}}{2}~,
   ~J_{02}^{LF}=\frac{-J^1_{ET}-K^2_{ET}}{2}~,~
   J_{03}^{LF}=\frac{K^3_{ET}}{2}~,~\nn\\
   &&J_{31}^{LF}=\frac{-J^2_{ET}-K^1_{ET}}{2}~,~
   J_{32}^{LF}=\frac{J^1_{ET}-K^2_{ET}}{2}~,~J_{12}^{LF}=J^3_{ET}
   \eea
   for the lower light-front
   indices.  To insure that all the
    factors of two are right, a nice check is
    \be
    Tr\left(J^{upper} \cdot J_{lower}\right)=
    J^{\mu\nu}_{LF}\:J^{LF}_{\nu\mu}=J^{\mu\nu}_{ET}\:J^{ET}_{\nu\mu}=
    2\:({\bf K}^2-{\bf J}^2)
    \;.
    \ee
    This holds true for the above relations.
   
   To close this section we work out the scalar example more
   explicitly as an illustrative example of how the
   interactions enter the generators. 
   For fermion and vector examples see \cite{rootyan}.
   Recall the form of $P^\mu$ and $J^{\mu\nu}$ 
   from Eq.~(\ref{eq:pmu}), where the stress tensor $T^{\mu\nu}$ was written in
   Eq.~(\ref{eq:stress}). For concreteness, say the lagrangian density is
   \be
   {\cal L}=\frac{1}{2}\:\partial_\mu\phi\:\partial^\mu\phi-\frac{m^2}{2}\:
   \phi^2-{\cal H}_{int}(\phi)
   \;.
   \ee
   Thus we have
   \bea
   &&T^{++}=(\partial^+\phi)^2\;,\\
   &&T^{+i}=\partial^+\phi~\partial^i\phi
   \eea
   and
   \bea
   T^{+-}&=&\partial^+\phi~\partial^-\phi-2{\cal L}\nn\\
   &=&(\partial_i\phi)^2+m^2\phi^2+2{\cal H}_{int}(\phi)\;.
   \eea
   For scalar theory the kinematic generators are (note there are seven which is one more than in
   equal-time field theory)
   \bea
   &&P^+=\xmes (\partial^+\phi)^2\;,\\
   &&P^i=\xmes \partial^+\phi\:\partial^i\phi\;,\\
   &&J^3=\xmes \left(x^1\der^+\phi\:\der^2\phi-x^2\der^+\phi\:\der^1\phi\right)\;,\\
   &&E^i=x^+ P^i
   -\xmes x^i(\partial^+\phi)^2
   \;,\\
   &&K^3=-\frac{x^+ P^-}{2}+\frac{1}{4}\int d^2 x^\perp dx^-
   x^-(\partial^+\phi)^2\;.\label{eq:K3}
   \eea
   This last equation makes it clear why $K^3$ is kinematic only for
   fields initialized at $x^+=0$; $P^-$ is the exact Hamiltonian of the
   theory and its job is to move states off the initial surface, but at $x^+=0$
   this term in $K^3$ vanishes.\footnote{Interestingly, note that
   $K^3$ is independent of time for all time as detailed
   below, but only if the initial surface is dialed precisely to
    $x^+=0$, does it become kinematic.}
   The dynamic generators are
   \bea
   &&P^-=\xmes\left[(\partial_i\phi)^2+m^2\phi^2\right]+\int d^2 x^\perp dx^-\:{\cal
   H}_{int}(\phi)\;,\\
   &&F^i=\xmes \left(x^-\partial^+\phi~\partial^i\phi-x^i\left[
   (\partial_i\phi)^2+m^2\phi^2\right]\right)\nn\\
   &&~~~~~~~~~~~~~~~~~~~~~~~~~~~-\int d^2 x^\perp dx^-
   x^i\:{\cal H}_{int}(\phi)
   \;.
   \eea
   
   Note that the stress tensor is symmetric and conserved:
   $T^{\mu\nu}=T^{\nu\mu}$ and $\der_\mu T^{\mu\nu}=0$.\footnote{
   The stress tensor is conserved only if the equations of motion
   $\der_\mu\frac{\partial {\cal L}}{\partial(\partial_\mu \phi)}=
   \frac{\der {\cal L}}{\der \phi}$ are satisfied, as is easily verified.}
   Thus the ten Poincar\'{e} generators are independent of time $x^+$.
   For example, we will verify this for
   $K^3$. Explicitly taking a derivative of Eq.~(\ref{eq:K3}) gives
   \be
   \frac{d K^3}{dx^+}=-\frac{P^-}{2}-\frac{x^+}{2}\frac{dP^-}{dx^+}+\frac{1}{4}
   \int d^2x^\perp dx^- x^- \frac{dT^{++}}{dx^+}
   \;.\label{eq:3340}
   \ee
   The second term on the right is zero by
   \be
  0= \xmes \der_\mu T^{\mu-}=\frac{dP^-}{dx^+}+{\rm surface~ terms}
   \;,\label{eq:2.44}
   \ee
   and the third term can be rewritten using
  \bea
&& 0= \frac{1}{4}\int d^2x^\perp dx^-
x^- \der_\mu T^{\mu +}=\frac{1}{4}\int d^2x^\perp dx^-
 x^-(\der_+T^{++}+\der_-T^{-+})\nn\\
&&~~~~~~~~~~~~~~~~~~~~~~~~~~~~~~~~~~~~~~~~~~~~~~~ +~
 {\rm transverse~ surface~ terms}
 \;;
 \eea
 performing an integration by parts over $x^-$, 
 this becomes
 \be
 \frac{1}{4}
   \int d^2x^\perp dx^- x^- \frac{dT^{++}}{dx^+}=
\frac{1}{4}
   \int d^2x^\perp dx^-T^{-+}+{\rm surface~ terms} \;.
   \label{eq:2.46}
 \ee
 Inserting Eqs.~(\ref{eq:2.44}) and (\ref{eq:2.46}) into Eq.~(\ref{eq:3340}) gives
 \be
  \frac{d K^3}{dx^+}=-\frac{1}{4}
   \int d^2x^\perp dx^-\left(T^{+-}-T^{-+}\right)+{\rm surface~ terms}=0
 \;,\label{eq:2.47}
 \ee
 where in this last step surface terms\footnote{The point here is not to discuss
 surface terms, but to note the already present non-trivial cancelation in
 Eq.~(\ref{eq:2.47}).} are  dropped and
 the fact that the stress tensor is symmetric is used.
 Similar algebra  can be used to show that all ten
 Poincar\'{e} generators are independent of time $x^+$.
 As a final remark note that the $-\frac{x^+P^-}{2}$ term in $K^3$ was essential 
 to show that $K^3$ is independent of time
  for all times. 
 
 To review, $P^-$ and $F^i$ are the ``Hamiltonians" of the system, and
 $K^3$, $E^i$, $J^3$, $P^i$ and $P^+$ are the kinematic generators 
 of Lorentz transformations
 of the system.
 The fact that boosts are kinematic leads to a simple change of coordinates\footnote{These
 coordinates are called Jacobi coordinates. More on this later.}
 making this boost invariance manifest and the total momentum of the system is seen to
 factor completely out of the Schr{\"o}dinger equation, lowering the dimensionality
 of the equation and making its analysis simpler (analogous to the nonrelativistic
 hydrogen problem in equal-time coordinates
 where the center-of-mass and relative coordinates factor in the Schr{\"o}dinger equation).
 Note that the eigenvalue of $J^3$ (with $P^i=0$) is the helicity
 of the respective state---in this dissertation, the helicity of the electron
 will be written as $s/2$, where $s=\pm 1$. The fact that the
 rotation generators $F^1$ and $F^2$ are dynamic makes classifying their states complicated.
 However, they are symmetry operators: $[P^-,\,F^i]=0$, thus in principle
  they should lead to  simplification in 
   the diagonalization of $P^-$. In practice, no such simplification has
   been found to date,
   thus we do not discuss them further
 in this dissertation.
\section{Canonical QED Hamiltonian}\label{canonical}

A derivation of the canonical Hamiltonian of hydrogen treating the proton as a 
point particle is presented. For 
positronium neglect all terms that contain a proton field.
A summary of the derivation and results will be given, and
then the details of the derivation will follow.

Starting with the QED lagrangian density
for the electron, proton, and photon system ($e > 0$)
\bea
{\cal L}&=&-\frac{1}{4}F_{\mu \nu} F^{\mu \nu} +
\overline{\psi}_e(i \not\! {\partial } +e \not\!\! {A }- 
 m )\psi_e+
 \overline{\psi}_p(i \not\! {\partial } -e \not\!\! {A }- 
 m_p )\psi_p~,\label{eq:L}
\eea
in a fixed gauge, $A^{+}=0$,\footnote{This derivation will not include a discussion of the gauge
field zero-modes. Initially we drop zero-modes. Their effects are conjectured
to be replaceable by effective interactions that are fixed by coupling coherence.
Coupling coherence is explained in Chapter~\ref{ch3}. For a treatment that incorporates
these gauge field zero-modes from the start in QED see \cite{daveqed} and
references within.}  the
constrained degrees of freedom are removed explicitly, producing 
a canonical Hamiltonian $H$.
$m$ and $m_p$ are the renormalized masses of the electron and proton respectively.
$\not\!\!\! {A }=A^\mu \gamma_\mu$.
Details of the derivation follow below. For our $\gamma$-matrices
we use
the two-component  representation chosen by Zhang 
and Harindranath~\cite{secondharipaper}.
The 
field operator expansions and light-front conventions 
are summarized in Appendix~\ref{appendixlf}. The resulting
canonical Hamiltonian is 
 divided into a free and interacting part
\bea
H=h+v\;.\label{eq:can1}
\eea
  $h$ is the free part given by
\bea
h=
 \int_p \sum_s \left(b_s^\dagger(p) b_s(p)~\frac{{p^\perp}^2+m^2}{p^+}+
 B_s^\dagger(p) B_s(p)~
 \frac{{p^\perp}^2+m_p^2}{p^+}+
  a_s^\dagger(p) a_s(p)~
 \frac{{p^\perp}^2}{p^+}\right)\;,
\label{eq:freebaby}
\eea
plus the anti-fermions.
The notation for our free spectrum is $h |i\rangle = \veps_i |i\rangle$
with $\sum_i |i\rangle\langle i| = 1$, where the sum over $i$ implies a sum over all Fock
sectors, momenta, and spin.
 Recall, we use the shorthand $\int_p = \int \frac{d^2p^\perp dp^+ \theta(p^+)}
 {16 \pi^3 p^+}$. 
$v$ is the interacting part given by
\bea
v&=&
 \int d^2 x^{\perp}dx^- {\cal H}_{int}\;,
\label{eq:46} 
\eea
 where
 \bea
 &&{\cal H}_{int}={\cal H}_{ee\gamma}+{\cal H}_{pp\gamma}+
 {\cal H}_{ee\gamma\gamma}+{\cal H}_{pp\gamma\gamma}+
{\cal H}_{\gamma-{\rm inst}}\;,
\eea
and
\bea
&&\!\!\!\!\!{\cal H}_{ee\gamma}=e\xi_e^{\dag}\left\{-2(\frac{\partial^{\perp}}
{\partial^+}\cdot A^{\perp})+\sig \cdot A^{\perp} 
\frac{\sig \cdot \partial^{\perp}+m}{\partial^+}+
\frac{\sig \cdot \stackrel{\!\!\leftarrow}
{\partial^{\perp}}+m}{\stackrel{\!\!\leftarrow}{\partial^+}} 
\sig \cdot A^{\perp}
\right\}\xi_e\;,
\label{eq:48}\\
&&\!\!\!\!\!{\cal H}_{pp\gamma}=e\xi_p^{\dag}\left\{2(\frac{\partial^{\perp}}
{\partial^+}\cdot A^{\perp})-\sig \cdot A^{\perp} 
\frac{\sig \cdot \partial^{\perp}+m_p}{\partial^+}-
\frac{\sig \cdot \stackrel{\!\!\leftarrow}
{\partial^{\perp}}+m_p}{\stackrel{\!\!\leftarrow}{\partial^+}} 
\sig \cdot A^{\perp}
\right\}\xi_p\;,
\label{eq:48b}\\
&&\!\!\!\!\!{\cal H}_{ee\gamma\gamma}=
-i e^2\left\{\xi_e^{\dag} \sig \cdot A^{\perp}\frac{1}{\partial^+}
 (\sig \cdot A^{\perp} \xi_e)\right\}
\;, \\
 &&\!\!\!\!\!{\cal H}_{pp\gamma\gamma}=
-i e^2\left\{\xi_p^{\dag} \sig \cdot A^{\perp}\frac{1}{\partial^+}
 (\sig \cdot A^{\perp} \xi_p)\right\}
\;, \\
&&\!\!\!\!\!{\cal H}_{\gamma-{\rm inst}}= 
-\frac{1}{2}J^+\frac{1}{\left(\partial^+\right)^2} J^+
\label{eq:instant}\;.
 \eea
$J^+=2 e \left(
\xi_p^\dagger \xi_p-\xi_e^\dagger \xi_e
\right)$ and $\sig^i$ are  the standard $SU(2)$ Pauli matrices.
 $i=1,2$ only; e.g.,  $\sig \cdot \partial^{\perp}=
\sig^i \partial^i = \sig^1 (-\frac{\der}{\der x^1})+\sig^2(-\frac{\der}{\der x^2})$.
The dynamical fields are $A^i$, $\xi_e$ and $\xi_p$, the transverse photon
and two-component electron and proton fields respectively. 
 For the relation between $\psi$ and $\xi$ and a 
  summary of our light-front conventions
see Appendix~\ref{appendixlf}. 

In this dissertation the formal expression for the canonical Hamiltonian 
given by Eqs.~(\ref{eq:can1})--(\ref{eq:instant}) is of little practical use.
Of more practical use are matrix elements of the 
canonical Hamiltonian $H$ in
the free basis of $h$. The matrix elements used in this dissertation, 
as derived by Allen \cite{brent},
are given in Appendix~\ref{brentrules}.

Now the
details of the derivation of the canonical Hamiltonian are presented.
Given ${\cal L}$ of Eq.~(\ref{eq:L}), the equations of motion are
\bea
\partial_\mu F^{\mu \nu} &=&  J^\nu\;,\\
(i \not\! {\partial } +e \not\!\! {A }- 
 m )\psi_e&=&0\;,\\
(i \not\! {\partial } -e \not\!\! {A }- 
 m_p )\psi_p&=&0\;,
\eea
where 
$J^\mu=e\left({\overline \psi}_p \gamma^\mu \psi_p-
{\overline \psi}_e \gamma^\mu \psi_e\right)$. The physical
gauge $A^+=0$ is chosen
and the projection operators $\Lam_+$ and $\Lam_-$ are inserted into the equations
of motion. Note $\psi_-=\Lam_- \psi$ and $\psi_+=\Lam_+ \psi$.
Three of the equations are seen to be constraint equations:
\be
-\frac{1}{2} \partial^+ \partial^+ A^-+\partial^i\partial^+A^i=J^+\;,
\ee
\be
i\partial^+\psi_{e-}=\left(
i\alpha^i \partial^i+e\alpha^iA^i+\frac{m \gamma^+}{2}
\right)\psi_{e+}
\ee
and
\be
i\partial^+\psi_{p-}=\left(
i\alpha^i \partial^i-e\alpha^iA^i+\frac{m_p \gamma^+}{2}
\right)\psi_{p+}\;.
\ee
The fact that these are constraints can be seen from the fact that
 no time derivatives $\partial^-$ 
appear. Note $\alpha^i=\gamma^0\gamma^i$.
Inverting the space
derivative $\partial^+$ gives
\be
A^-= \frac{-2}{( \partial^+)^2} J^+ + 2 \frac{ \partial^i}{ \partial^+} A^i
\;,\label{eq:const1}
\ee
\be
\psi_{e-}=\frac{1}{i\partial^+}\left[
\left(
i\alpha^i \partial^i+e\alpha^iA^i+\frac{m \gamma^+}{2}
\right)\psi_{e+}
\right]
\ee
and
\be
\psi_{p-}=\frac{1}{i\partial^+}\left[
\left(
i\alpha^i \partial^i-e\alpha^iA^i+\frac{m_p \gamma^+}{2}
\right)\psi_{p+}
\right]\;.\label{eq:const2}
\ee

To get a flavor of a position space representation of the
 inverse longitudinal derivative, note that
 it could in principle be defined as follows (see below for what we actually use), where
 $f(x)$ is an arbitrary field:
\beaa
&&\Biggl{(} \frac{1}{\partial^+} \Biggr{)}f(x^-)=\frac{1}{4}
\int_{- \infty}^{+ \infty}dy^{-}\eps(x^{-} - y^{-})\,f(y^-)+g_1\;,\\
&&\Biggl{(} \frac{1}{\partial^+} \Biggr{)}^2f(x^-)=\frac{1}{8}
\int_{- \infty}^{+ \infty}dy^{-}|x^{-} - y^{-}|\,f(y^-)+g_2+x^-g_3\;,\\
 &&\partial^+=2\partial_-=2\frac{\partial}{\partial x^-}\;,\\
&&\partial_- \eps(x^- - y^-)=2 \del(x^- - y^-)\;,\\
&&\eps(x)=\theta(x)-\theta(-x)\;.
\eeaa
$x^\perp$ and $x^+$ are implicitly in the arguments of the $f(x)$s.
$g_1$, $g_2$ and $g_3$ are arbitrary fields independent of $x^-$. For a discussion on these
boundary terms see \cite{hari1}.
Notice that this inverse longitudinal derivative
is non-local.

In practice,
we define the
 inverse longitudinal derivative in momentum space. We 
 explicitly put
the momentum representation of the field operators
into the respective terms of the Hamiltonian, multiply
the fields out explicitly, and then replace the inverse derivatives by
appropriate factors of longitudinal momentum  with
the restriction $|p^+|/{\cal P}^+\geq \eps=0_+$ 
[${\cal P}^+$ is the total longitudinal momentum of the physical state
of interest].
The absolute value sign on $\left|p^+\right|$ is required for the instantaneous interactions.
For example, a product of two fields gives
\be
\frac{1}{i\der^+} \exp[-i(p-k)\cdot x]\longrightarrow
\frac{1}{p^+-k^+}\theta\left(|p^+-k^+|-\eps {\cal P}^+\right)
\exp[-i(p-k)\cdot x]
\;.\label{eq:2.67}
\ee
As a final note,
 we work in continuum field theory and drop all
surface terms initially. If these terms are required for the Hamiltonian to
run {\it coherently}, they are conjectured to arise through the process of renormalization.

The dynamical degrees of freedom are $A^i$, $\psi_{e+}$ and $\psi_{p+}$. 
The canonical
hamiltonian density is defined in terms of these dynamical degrees of freedom
\be
{\cal H}=\frac{\partial {\cal L}}{\partial(\partial^-A^i)}\partial^-A^i
+\frac{\partial {\cal L}}{\partial(\partial^-\psi_{e+})}\partial^-\psi_{e+}
+\frac{\partial {\cal L}}{\partial(\partial^-\psi_{p+})}\partial^-\psi_{p+}
-{\cal L}
\;.
\ee
Taking these derivatives of the lagrangian density and combining terms,
the hamiltonian density takes the following simple form
\be
{\cal H}=\frac{1}{2}(\partial^i A^j)^2+
\psi_{e-}^\dagger i \partial^+ \psi_{e-}+
\psi_{p-}^\dagger i \partial^+ \psi_{p-}
-2 \left(
\frac{1}{\partial^+}\frac{J^+}{2}
\right)^2
+\frac{J^+}{2}A^-
\;,
\ee
where the constraints of Eqs.~(\ref{eq:const1})--(\ref{eq:const2}) 
are assumed to be satisfied. In our $\gam$-matrix representation,
only two of the components of the 4-spinors $\psi_{e+}$ and $\psi_{p+}$ 
are nonzero. Writing
these as the 2-spinors $\xi_e$ and $\xi_p$ respectively,\footnote{
In other words, we are defining 
$\psi_{e+}=\Lam_+ \psi_e=(\psi_{e1},\psi_{e2},0,0)\equiv (\xi_e,0)$,
 with an analogous definition
 for the dynamical proton field.} and inserting the constraints
of Eqs.~(\ref{eq:const1})--(\ref{eq:const2}),
$H$ takes on the form written earlier 
in Eqs.~(\ref{eq:can1})--(\ref{eq:instant}), where surface terms such as in
\be
\int d^2 x^\perp dx^-\left(\frac{1}{\der^+} J^+\right)\left(\frac{1}{\der^+} J^+\right)=
-\int d^2 x^\perp dx^-J^+\left(\frac{1}{\der^+} \right)^2J^++{\rm surface ~terms}
\;,
\ee 
are dropped.

\chapter{Light-front hamiltonian bound-state problem}\label{ch3}

In this chapter, an
outline of the general computational strategy we employ is given. This
includes a derivation of the effective Hamiltonian and a 
discussion of our diagonalization procedure.
 We discuss three hamiltonian transformations which are of interest,\footnote{A fourth
 transformation is mentioned in a footnote.}
 give a general discussion of our renormalization program, and close with two
 simple examples: one introducing the standard renormalization
 group terminology through transverse and longitudinal renormalization group
 discussions, and the other elucidating
  coupling coherence.
\section{General computational strategy}

First, we give a brief overview. See the later sections of this chapter for the appropriate
references and/or if the terminology is new to the reader.
We  use
the renormalization group
to produce a regulated effective Hamiltonian $H_\lambda$, where
$\lambda$ is the final cutoff and renormalization is required to remove cutoff
dependence from all physical quantities. 
This renormalization group transformation is run from the bare (initial)
 cutoff $\Lam$ down
to the effective (final) cutoff $\lam$, with $\lam \ll \Lam$.
At this point we have a regulated effective
Hamiltonian that contains 
a finite number of  independent\footnote{A coupling required in the theory when 
the symmetries of interest
are maintained by the regulator will be called `independent.' Additional
couplings required to restore symmetries broken by the regulator
 will be called `dependent.'
The reasoning behind this terminology is made clear below.
Note that by `symmetries' we include `hidden symmetries': fixed relations between couplings in
the broken phase due to an underlying symmetry of the theory which is  however not maintained by
the vacuum.}
 relevant and marginal couplings---in the transverse renormalization group sense
 which is implied throughout this dissertation except for a brief example 
 in Subsection~\ref{relevant}---and an infinite number of
  independent irrelevant couplings\footnote{
  Since $\lam \ll \Lam$, the {\it independent} irrelevant couplings
   become functions of only the {\it independent}
 relevant and marginal couplings, and are exponentially insensitive to their boundary
 conditions set at the bare cutoff $\Lam$ (universality).} 
 as would occur in any cutoff theory;
  also, there are an infinite number of new
   dependent relevant, marginal and
  irrelevant
couplings. Note that we are speaking in terms of dimensionless couplings (with the
appropriate power of the cutoff explicitly factored out).

The reader may be concerned with the fact that there are an infinite number of dependent
{\it relevant} and {\it marginal} couplings in the final Hamiltonian. This occurs because 
(i) longitudinal
locality is absent and (ii) rotational symmetry is not manifest
in 
light-front field theory. Predictability is not lost however because we require
these dependent couplings to run {\it coherently} with the  independent relevant and
marginal couplings,
that is we do not allow them to have {\it explicit} cutoff dependence.
 These extra conditions placed on the theory are called
 `coupling coherence'. Coupling coherence fixes these extra counterterms,
and
 the symmetries broken by the regulator  (most noteworthy for us: gauge, chiral
   and
rotational invariance) are restored in the final physical solution.
 These conjectures have been found to be true in all examples worked to
date; see Section~\ref{perry} for a simple one-loop example and for references
 to further examples.

  This complicated effective Hamiltonian
cannot be directly diagonalized, and since we want to solve bound-state
problems we cannot solve it using perturbation theory.  The second step is to
approximate the full effective Hamiltonian using
\bea
 &&H_\lam={\cal H}_0 + \left(H_\lam-{\cal H}_0\right)~\equiv~{\cal H}_0+{\cal
 V}~,
\label{eq:1cv}
 \eea
 where ${\cal H}_0$ is an approximation that can be solved
nonperturbatively---for QED analytically, which is one of the primary motivations for
studying QED---and ${\cal V}$ is treated in bound-state perturbation theory
(BSPT). 
The test of ${\cal H}_0$ is whether BSPT converges or not.

We now formulate the following questions.  Is there a range of scales $\lambda$ for which
${\cal H}_0$ does not require particle emission and absorption?  What are the
few-body
interactions in ${\cal H}_0$ that generate the correct nonperturbative
bound-state
energy scales?  Is there a few-body realization of rotational invariance; and
if not, how does rotational symmetry emerge in BSPT? 
We should emphasize that for our purposes we are primarily interested in
answering these questions for low-lying bound-states, and refinements may be
essential to discuss highly excited states or bound-state scattering.

It is essential that $\tilde{\lam}$,\footnote{The tilde on $\tilde{\lam}$ is a
useful notation since we want to discuss a regulator that restricts binding energy
scales
of a typical atom.
Since we use different transformations for the positronium and hydrogen calculations
the relation between $\lam$ and $\tilde{\lam}$ changes accordingly:
$\tilde{\lam}_{hydrogen}=\frac{\lam^2-(m+m_p)^2}{2(m+m_p)}$ 
where a Bloch transformation is used and 
$\tilde{\lam}_{positronium}=\frac{\lam^2}{2(m+m)}$ where a G-W transformation is used.
The difference (besides the trivial $m_p\longleftrightarrow m$ difference) 
arises simply because G-W regulates energy {\it differences} whereas
Bloch regulates energies.\label{tilde}}
 which governs the degree to which states are
resolved,
 be adjusted to obtain a constituent approximation.  If $\tilde{\lambda}$ is
kept large with respect to all mass scales in the problem, arbitrarily large
numbers of constituents are required in the states because constituent
substructure is resolved.  A constituent picture can emerge if high free-energy
states couple perturbatively to the low free-energy states that dominate the
low-lying bound-states.  In this case the cutoff can be lowered until it
approaches the nonperturbative bound-state energy scale and perturbative
renormalization may be employed to approximate the effective Hamiltonian.  In
QED 
a simplification occurs because the system is nonrelativistic.
For example in positronium
\bea
 \frac{p^+_{electron}}{{\cal P}^+} &=& \frac{1}{2} + {\cal O}(\alpha)\;,\label{nr1}\\
 \frac{p^\perp_{electron}}{m} &=& {\cal O}(\alpha)\;,\label{nr2}\\
 \alpha &\ll& 1\;,
 \eea
 where $m$ is the renormalized
 electron mass and ${\cal P}^+$ is the total longitudinal momentum of positronium.
 Also, the dominant photon wavelength that {\it couples} to the bound-state is of order
 the size of the bound-state: wavelength $\sim 1/(m\al)$. There is a natural gap
 in the system between the valence and valence plus one photon states and this 
 above-mentioned
 range into which the cutoff must be lowered is

\begin{equation}
m\alpha^2  \ll \tilde{\lambda} \ll m\alpha \;\;,
\label{eq:2cv}
\end{equation}

\noindent where $m$ is the electron mass and the relation
between $\lam$ and $\tilde{\lambda}$ is explained in footnote~\ref{tilde}.
 If the cutoff is lowered to this range, hydrogen and positronium
 bound-states are well
approximated by electron-proton and electron-positron states respectively,
 and photons and pairs can be included
perturbatively.

It is an oversimplification to say the constituent picture emerges because the
QED coupling constant is very small.  Photons are massless, and regardless of how
small $\alpha$ is, one must in principle use nearly degenerate bound-state
perturbation theory that includes extremely low energy photons
nonperturbatively.  This is not required in practice, because the Coulomb interaction which
sets the important energy scales for the problem produces neutral bound-states
from which long wavelength photons effectively decouple.  Because of this, even
though arbitrarily small energy denominators are encountered in BSPT
 due to mixing of valence bound-states and states
including extra photons, BSPT can converge because
emission and absorption matrix elements vanish sufficiently rapidly.
A nice example of this is seen in the Lamb shift calculation of Chapter~\ref{ch:lamb}.

The well-known answer to the second question above is the two-body Coulomb
interaction sets the nonperturbative energy and momentum scales appropriate
for  QED.  We have already used the results of the Bohr scaling analysis that
reveals the bound-state momenta scale as $p \sim m\alpha$ and the energy
scales as $E \sim m\alpha^2 $.  As a result the dominant photon momenta are
also of order $m\alpha $, and the corresponding photon energies are of
order $m\alpha $.  This is what makes it possible to use renormalization to
replace photons with effective interactions.  The dominant photon energy scale
is much greater than the bound-state energy scale, so that $\tilde{\lambda}$ can be
perturbatively lowered into the window in Eq.~(\ref{eq:2cv}) and photons are not
required in the state explicitly, but only implicitly through  effective interactions.  

Finally we discuss rotational invariance in a light-front approach.  In
light-front field theory, boost invariance is kinematic, but rotations about
transverse axes involve interactions.  Thus rotational invariance is not
manifest and all cutoffs violate rotational invariance in light-front field
theories. In QED it is easy to see how counterterms in $H_\lambda$ arise during
renormalization that repair this symmetry perturbatively; however, the issue of
nonperturbative rotational symmetry is potentially much more complicated.  We
first discuss leading order BSPT and then turn
to higher orders.

To leading order in a constituent picture we require a few-body realization of
rotational symmetry.  This is simple in non-relativistic systems, because
Galilean rotations and boosts are both kinematic.  In QED the constitutuent
momenta in all low-lying bound-states are small, so a non-relativistic reduction
can be used to derive ${\cal H}_0$.  Therefore to leading order in QED we can employ a
non-relativistic realization of rotational invariance. 

At higher orders in BSPT rotational invariance will
not be maintained unless corrections are regrouped.    The
guiding principle in this and all higher order calculations is to expand not in
powers of ${\cal V}$, but in powers of $\alpha$ and $\al \,\ln(1/\al)$.  ${\cal H}_0$ should
provide the leading term in this expansion for BSPT
to be well-behaved, and subsequent terms should emerge from finite orders of
BSPT after appropriate regrouping.  Powers of
$\alpha$ appear through explicit dependence of interactions on $\alpha$, and
through the dependence of leading order eigenvalues and eigenstates on $\alpha$
introduced by interactions in ${\cal H}_0$.  This second source of dependence can be
estimated using the fact that momenta scale as $m\alpha$ in the bound-state
wave functions.  Of more interest for the Lamb shift calculation is the appearance of
$\al \,\ln(1/\al)$, which is signaled by a divergence in unregulated bound-state
perturbation theory.  As has long been appreciated, such logarithms appear
when the number of scales contributing to a correction diverges.

Now we change gears a bit.
 There is an open question as to which transformation should be used to derive
 the renormalized Hamiltonian. We discuss three different transformations:
 G-W, Wegner, and 
 Bloch.\footnote{A fourth transformation of promise is currently under study by Harada,
 Okazaki and co-workers
  \cite{koji1}. They combine the simplicity of Feynman perturbation theory with a
 similarity transformation that moves the effects
 of couplings between few and many-body states into effective interactions
 resulting in an effective Hamiltonian acting in the few-body space alone.}
 The G-W transformation is unitary and
 was developed by
G{\l}azek and Wilson
 \cite{glazekwilson1} to deal with the small energy denominator problem. 
 {\it This transformation will be used for the positronium calculation}.
 An early application which used the G-W transformation \cite{longpaper} was a weak-coupling treatment
 of QCD. 
 The Wegner transformation \cite{wegner} is also unitary and allows
 no small energy denominators---the elegance in the definition of the
 formalism
 is one of the appealing features of the Wegner transformation. 
 The Wegner transformation was developed with condensed matter applications in mind,
 however it has recently been applied to QED \cite{elena}.
 Another transformation---{\it used
  for the Lamb shift calculation}---is the Bloch transformation \cite{bloch}. 
  Recent 
  applications in scalar field theory \cite{ed1} and QED \cite{ed2},    
  utilize the Bloch transformation.
  It is
 not a unitary transformation and may have small energy denominator problems, but
 it is the simplest transformation to apply
 in practice and the natural gap in QED between the valence and valence plus 
  photons sectors 
  avoids problems here. A final note is that the small energy denominator
 problem should
  not arise in general when solving for the low-lying spectrum which we discuss next.
 
 The small energy denominator problem is explained as follows. An effective Hamiltonian
 with energy cutoff $\tilde{\lam}$ has effective interactions in the low-energy space $L$
 at second order of the following generic form
 \be
\langle L \left|H_{eff}^{^{(2)}}\right|L\rangle\sim
 \sum_H \frac{\langle L|V|H\rangle\langle H|V|L\rangle}{E_L-E_H}
 \;,
 \ee
 where $|L\rangle$ and $|H\rangle$ 
 are in the {\it low} and {\it high} energy spaces respectively and 
 $E_H$ is above $\tilde{\lam}$ (and below $\tilde{\Lam}$)
 while $E_L$ is below $\tilde{\lam}$. 
 This effective interaction arises to replace the effects of couplings
 removed by lowering the cutoff from $\tilde{\Lam}$ to
 $\tilde{\lam}$.
 The small energy denominator problem arises because in general
 $E_L$ can go all the way {\it up} to the 
 boundary $\tilde{\lam}$ and $E_H$ can go all the way {\it down} to the boundary $\tilde{\lam}$. This can lead
 to a vanishing denominator which is problematic unless the matrix elements vanish fast enough
 and lead to either a vanishing or finite result. Note that if $E_L$ 
 (which is not integrated over in $\sum_H$ but rather is fixed by the wave function)
 is much less than the
 boundary $\tilde{\lam}$ then there is no problem. Thus we want to keep $\tilde{\lam}$ 
 relatively high in  comparison to the nonperturbative bound-state energy scale. But
 also as detailed above we want to keep $\tilde{\lam}$ relatively low to suppress 
 many-body mixings such as
 photon emission.
 In QED there is a range allowed for $\tilde{\lam}$ that is high enough,
 so the small energy denominator problem is not encountered, and at the same time is
 not too high, and thus photon emission can be removed perturbatively. This
 range is written above in Eq.~(\ref{eq:2cv}). 
 In summary, even with the Bloch transformation, the small energy problem should not arise when
 solving for the low-lying spectrum because $\tilde{\lam}$ can always be adjusted to be
 well above 
 this level and
 $\left|E_L-E_H\right|\sim\tilde{\lam}$ or bigger, and can not vanish. In addition, in QED with 
 the natural
 gap $|E_{e{\overline e}\gamma}-E_{e{\overline e}}|/E_{e{\overline e}}\sim 137$, we can still 
 keep $\tilde{\lam}$ low enough and perturbatively replace emission effects with few-body
 interactions in the valence sector alone.
 The small energy denominator problem should
  only
 begin to arise when for example
  calculating a scattering amplitude of energy near the boundary $\tilde{\lam}$.

\section{Step one: Derivation of $\hr$ with G-W transformation}\label{G-W}

A self-contained discussion on the derivation of the
effective Hamiltonian $H_\lam$ with the G-W transformation
will now be given. The discussion will be general, and in particular will hold
for both QED and QCD.

The derivation starts with the      definition of a bare
 Hamiltonian $H_\Lam$\footnote{In the initial setup 
of the G-W transformation, $\lam$ and $\Lam$ will be used as a shorthand
for $\frac{\lam^2}{{\cal P}^+}$ and $\frac{\Lam^2}{{\cal P}^+}$ respectively 
(in other words they are to be thought of as having the dimension of energy),
where ${\cal P}^+$ is the total longitudinal momentum of the physical state of interest,
and $\lam^2$ and $\Lam^2$ have dimension $({\rm mass})^2$.}
\bea
H_\Lam&\equiv&h+v_{_{\Lam}} \;,\label{eq:hap1}\\
v_{_{\Lam}}&\equiv& f_{_{\Lam}} {\overline v}_{_{\Lam}}\;,\\
{\overline v}_{_{\Lam}}&\equiv& v+\del v_{_{\Lam}}\;,\\
H&\equiv&h+v
\;,
\eea
where $h$ is the free Hamiltonian, $H$ is the canonical Hamiltonian,\footnote{See
Section~\ref{canonical} for a canonical QED Hamiltonian. See \cite{longpaper} for a
canonical QCD Hamiltonian.} $f_{_{\Lam}}$ is 
a regulating function, and $\del v_{_{\Lam}}$ are counterterms defined through the process
of renormalization.
The canonical Hamiltonian $H$ is written in terms of renormalized parameters 
which implicitly depend on the renormalization scale $\mu$ as explained below.
The \mbox{counterterms} $\del v_{_{\Lam}}$ are fixed
by coupling coherence. In Section~\ref{coherence} 
coupling coherence will be explained further 
and also a discussion of the scale dependence
of the theory will be given.

The free Hamiltonian $h$ is given by (using positronium as an illustration)
\bea
h &=& \int_p \sum_s \left \{\frac{{p^\perp}^2+m^2}{p^+}\,\left[b_s^\dagger(p) b_s(p)+
 d_s^\dagger(p) d_s(p)\right]
 +\frac{{p^\perp}^2}{p^+}\,
  a_s^\dagger(p) a_s(p)
 \right \}
 \label{eq:hap2}~,~\\
 &&~~~~~~~~~~~~~~~~~~h |i\rangle= \veps_i |i\rangle\;\;,\;\;\sum_i |i\rangle\langle i | = 1 \;,
 \label{eq:11}\eea
 where the sum over $i$ implies a sum over all Fock sectors and spins, and integrations
 over all momenta in the respective free states.
$m$ is the renormalized fermion mass.
 
 The regulating function 
$f_{_{\Lam}}$ is defined to act in 
the following way
\bea
\langle i | f_{_{\Lam}} {\overline v}_{_{\Lam}} |j \rangle &\equiv& f_{_{\Lam ij}} \langle i | 
{\overline v}_{_{\Lam}} |j\rangle \equiv f_{_{\Lam ij}} {\overline v}_{_{\Lam ij}}~,\\
f_{_{\Lam ij}}&\equiv& \theta(\Lam-|\Delta_{ij}|)\;\;,\;\;\Delta_{ij}\equiv \veps_i-\veps_j
\;.
\eea
Note that this choice of a step function is not necessary
and can lead to pathologies, however it is useful for doing analytic
calculations.

Next, a                         similarity transformation is 
defined that acts on $H_\Lam$ and restricts the energy
widths\footnote{The magnitude of the {\it energy difference} between the free states in a matrix element of a Hamiltonian is defined to be
its {\it energy width}.} in the effective Hamiltonian $H_\lam$ to be 
below the final  cutoff, $\lam$ ($\lam < \Lam$). 
This transformation allows  recursion relationships 
to be set up for $H_\lam$, which can be written in the following general form
\bea
H_\lam&=& h+v_\lam \label{eq:hap3}~,\\
v_\lam&\equiv&f_\lam {\overline v}_\lam\;,\label{eq:3.16}\\
{\overline v}_\lam&=&{\overline v}_\lam^{^{(1)}}+{\overline v}_\lam^{^{(2)}}+\cdots
\;,\label{eq:ttnot}
\eea
where the superscripts imply the respective order in the canonical interaction $v$,
which recall is written in terms of renormalized parameters. 
 
 Now, starting with the above bare Hamiltonian, we will describe this procedure more explicitly.
The similarity transformation is 
defined
to act on a bare cutoff continuum Hamiltonian, $H_\Lam$, in the following way:
\bea
H_\lam&\equiv& S(\lam ,\Lam) H_\Lam S^\dagger(\lam, \Lam)\;,\label{eq:star}
\\
S(\lam, \Lam)  S^\dagger(\lam, \Lam)&\equiv&S^\dagger(\lam, \Lam) S(\lam, \Lam)  \equiv 1
\;.\label{eq:unitary}
\eea
 This transformation is unitary, so  $H_\Lam$ and $H_\lam$ have the
 same spectrum:
 \bea
 H_\Lam |\Psi_\Lam\rangle&=& E |\Psi_\Lam\rangle\;,\\
 \underbrace{S(\lam ,\Lam) H_\Lam S^\dagger(\lam ,\Lam)}_{H_\lam}
 \underbrace{ S(\lam, \Lam)|\Psi_\Lam\rangle}_{|\Psi_\lam\rangle}&=&E \underbrace{S(\lam, \Lam) 
 |\Psi_\Lam\rangle}_{|\Psi_\lam\rangle}
 \;.
 \eea
 Therefore, $E$ is independent of
 the final cutoff
  $\lam$ if an exact transformation is made.
 $E$ is also independent of the bare
 cutoff $\Lam$ after the Hamiltonian has been renormalized. 
 
 To put the equations in a differential framework, note that
  Eqs.~(\ref{eq:star}) and  (\ref{eq:unitary})
  are equivalent to the following equation (proven below)
 \bea
 \frac{d H_\lam}{d \lam}&=& \left [ H_\lam,T_\lam\right] \label{eq:star2}\;,
\label{eq:wnbv} \eea
 with 
 \bea
 S(\lam ,\Lam)&\equiv&{\cal T} \exp \left(\int_\lam^\Lam T_{\lam^\pr} d \lam^\pr\right)\;,
 \label{eq:tdef}
 \eea
 where $T_{\lam^\pr}$ is the anti-hermitian ($T_\lam^\dagger=-T_\lam$) generator of
 energy width transformations, and
 ${\cal T}$ orders operators from left to right in order of {\it increasing} running
 cutoff scale 
 $\lam^\pr$.
 Eq.~(\ref{eq:wnbv}) is a first order differential equation, thus one boundary condition
 must be specified to obtain its solution. This boundary condition is the bare Hamiltonian:
  $H_\lam |_{_{\lam \rightarrow \Lam}} \equiv H_\Lam$.  $H_\Lam$ is determined by
  coupling coherence.\footnote{Note that the bare
  Hamiltonian is not just
   the canonical Hamiltonian written in terms of
    bare parameters. The bare Hamiltonian must be adjusted through the process of
     renormalization
    until the conditions of coupling coherence are satisfied.}
  
  Now we prove that Eqs.~(\ref{eq:star2}) and (\ref{eq:tdef}) are equivalent to
  Eqs.~(\ref{eq:star}) and (\ref{eq:unitary}). First note that to 
  prove unitarity, Eq.~(\ref{eq:unitary}), we need to know
  \be
   S^\dagger(\lam ,\Lam)\equiv{\cal T}^\dagger
    \exp \left(-\int_\lam^\Lam T_{\lam^\pr} d \lam^\pr\right)\;,
  \ee
  where we used the fact that $T_{\lam^\pr}$ is anti-hermitian, and
  ${\cal T}^\dagger$ orders 
  operators from left to right in order of {\it decreasing} running cutoff scale 
 $\lam^\pr$. Now for the proof: Take a derivative of $\hr$ and use unitarity giving
 \bea
 \frac{d \hr}{d \lam}&=& \frac{dS(\lam ,\Lam)}{d \lam} S^\dagger(\lam ,\Lam) 
 S(\lam ,\Lam) H_\Lam
 S^\dagger(\lam ,\Lam)\nn\\
 &&~~~~~~~~~~+
 S(\lam ,\Lam)H_\Lam
 S^\dagger(\lam ,\Lam)S(\lam ,\Lam)\frac{dS^\dagger(\lam ,\Lam)}{d \lam}\nn\\
 &&=\frac{dS(\lam ,\Lam)}{d \lam} S^\dagger(\lam ,\Lam)\hr +\hr S(\lam ,\Lam)
 \frac{dS^\dagger(\lam ,\Lam)}{d \lam}
 \;.
 \eea
  Taking a derivative of the unitarity condition gives
  \be S(\lam ,\Lam)
  \frac{dS^\dagger(\lam ,\Lam)}{d \lam}
  =-\frac{dS(\lam ,\Lam)}{d \lam}S^\dagger(\lam ,\Lam)\;,
  \ee
  and thus we have
  \be
 \frac{d \hr}{d \lam}=\left[\hr,- \frac{dS(\lam ,\Lam)}
 {d \lam}S^\dagger(\lam ,\Lam)\right]
 \label{eq:almost}\;.
  \ee
  Also, explicitly taking a derivative of Eq.~(\ref{eq:tdef}) gives
  \be
  \frac{dS(\lam ,\Lam)}{d \lam}={\cal T}\left\{\left[
   \exp \left(\int_\lam^\Lam T_{\lam^\pr} d \lam^\pr\right)
  \right](-T_\lam)\right\}=
  -T_\lam S(\lam ,\Lam)
  \;,
  \ee
  which after using unitarity combined with Eq.~(\ref{eq:almost}) implies
  \be
 \frac{d H_\lam}{d \lam}=\left [ H_\lam,T_\lam\right] \;,\label{eq:3.29}
  \ee
  as was to be shown.
  So, we must specify $T_\lam$ and the running of $\hr$ with the cutoff is completely
  defined, and it runs unitarily.
   
   To define $T_\lam$ note that it is enough to specify how 
 ${\overline v}_\lam$ and $h$ change with the running cutoff scale $\lam$. This is seen
  by writing out Eq.~(\ref{eq:3.29}) more explicitly using Eq.~(\ref{eq:hap3}):
  \bea
\frac{d h}{d\lam}+\frac{d}{d\lam}\left(f_\lam {\overline v}_\lam \right)&=&
[h,T_\lam]+[v_\lam,T_\lam]\;.\label{eq:19}
\eea
We solve this perturbatively in $v$, and note
 that $h$ depends only on the renormalization scale $\mu$ (more on this later)
 but not on the running cutoff scale $\lam$.  Also, we demand that
  $T_\lam$ and $v_\lam$ do not contain any small energy denominators.  Thus we define
\bea
\frac{d h}{d \lam}&\equiv&0\;,\label{eq:star4}\\
\frac{d {\overline v}_\lam}{d \lam}&\equiv& [v_\lam,T_\lam]\;.
\label{eq:star3}
\eea
Eq.~(\ref{eq:star3}) is a choice such that   
$T_\lam$ and consequently $v_\lam$
do not allow any small energy denominators.
These additional constraints determine $T_\lam$ and ${\overline v}_\lam$, which  are 
 given by the following 
equations
\bea
 \left[h,T_\lam\right]
&=&
 \overline{v}_\lam \frac{d f_\lam}{d \lam} -
 {\overline f}_\lam [v_\lam,T_\lam]\;,  \label{eq:22}\\
{\overline v}_\lam &=& v+\del v_{_{\Lam}} -\int_\lam^\Lam [ v_{\lam^\pr},T_{\lam^\pr}]
d \lam^\pr
\;,\label{eq:23}
\eea
 where $f_\lam+{\overline f}_\lam\equiv 1
 \;$\@.
 Eqs.~(\ref{eq:22}) and (\ref{eq:23}) follow from Eqs.~(\ref{eq:19})--(\ref{eq:star3})
 and the boundary condition $H_\lam |_{_{\lam \rightarrow \Lam}} \equiv H_\Lam$.
 Now we solve Eqs.~(\ref{eq:22}) and (\ref{eq:23}) for $T_\lam$ and $\overline{v}_\lam$.
 Given Eqs.~(\ref{eq:hap3}) and (\ref{eq:3.16}), we need to determine ${\overline v}_\lam$,
and $H_\lam$ is then known. The perturbative
solution to Eqs.~(\ref{eq:22}) and (\ref{eq:23}) is
\bea
&&{\overline v}_{\lam}=
{\overline v}_{\lam}^{^{(1)}}
+{\overline v}_{\lam}^{^{(2)}}+\cdots
\label{eq:yes}~,
\\
&&T_{\lam}=
T_{\lam}^{^{(1)}}
+T_{\lam}^{^{(2)}}+\cdots~,\\
&&\del v_{_{\Lam}} = \del v_{_{\Lam}}^{^{(2)}}
+\del v_{_{\Lam}}^{^{(3)}} +\cdots\;,
\eea
where the superscripts imply the respective order in the canonical interaction, $v$, and
these quantities are given by
\bea
 {\overline v}_\lam^{^{(1)}}&=& v
~, \\
 \left [ h , T_\lam^{^{(1)}} \right]&=& v 
 \frac{d f_\lam}{d \lam}
 ~,\\
 \overline{v}_{\lam}^{^{(2)}}&=& 
 -
\int_{\lam}^{\Lam}
d \lam^\prime [ v_{\lam^\prime}^{^{(1)}} , T_{\lam^\prime}^{^{(1)}}]
+ \delta v_{_{\Lam}}^{^{(2)}}
 ~,\label{eq:vtwo}\\
 \left[h,T_\lam^{^{(2)}}\right]
&=&
 \overline{v}_\lam^{^{(2)}} \frac{d f_\lam}{d \lam} -
 {\overline f}_\lam [v_\lam^{^{(1)}},T_\lam^{^{(1)}}] 
 ~,\\
 \overline{v}_{\lam}^{^{(3)}}&=& 
 -
\int_{\lam}^{\Lam}
d \lam^\prime \left(\left[ v_{\lam^\prime}^{^{(1)}} 
, T_{\lam^\prime}^{^{(2)}}\right] +
\left[ v_{\lam^\prime}^{^{(2)}} 
, T_{\lam^\prime}^{^{(1)}}\right]\right)
+ \delta v_{_{\Lam}}^{^{(3)}} 
~,\\
\left[h,T_\lam^{^{(3)}}\right]
&=&
 \overline{v}_\lam^{^{(3)}} \frac{d f_\lam}{d \lam} -
 {\overline f}_\lam \left(\left[ v_{\lam}^{^{(1)}} 
, T_{\lam}^{^{(2)}}\right] +
\left[ v_{\lam}^{^{(2)}} 
, T_{\lam}^{^{(1)}}\right]\right) 
~,\\
&\vdots&\;.\nonumber
 \eea
 A general form of these effective interactions is
  \bea
 {\overline v}_\lam^{^{(i)}}&=&-\sum_{j,k=1}^\infty \del_{(j+k,i)}
 \int_\lam^\Lam d\lam^\pr[ v_{\lam^\prime}^{^{(j)}} , T_{\lam^\prime}^{^{(k)}}]
+ \delta v_{_{\Lam}}^{^{(i)}}
 \;,
 \eea
 for $i=2,3, \cdots$, with
  ${\overline v}_\lam^{^{(1)}}=v$.
 
 Recall the general form of the effective Hamiltonian given by 
 Eqs.~(\ref{eq:hap3})--(\ref{eq:ttnot}). The explicit perturbative 
 solution through third order now follows.
 ${\overline v}_\lam^{^{(1)}}=v$, and
now we write the explicit form of the second order effective
interaction
  ${\overline v}_\lam^{^{(2)}}$.  From Eq.~(\ref{eq:vtwo}) we obtain
\bea
  {\overline v}_{\lam ij}^{^{(2)}}&=&
  \sum_{k} v_{ik} 
  {v}_{kj} \left(
  \frac{g^{(\lam \Lam)}_{ikj}}{\Delta_{ik}} +
  \frac{g^{(\lam \Lam)}_{jki}}{\Delta_{jk}}
  \right) +\del v_{\Lam ij}^{^{(2)}}\;,\\
  {\rm where}&&g^{(\lam \Lam)}_{ikj} \equiv
  \int_\lam^\Lam d \lam^\prime f_{\lam^\prime jk} 
  \frac{d f_{\lam^\prime ki}}{d \lam^\prime}\;.
  \label{eq:g}
  \eea
  $\del v_{\Lam ij}^{^{(2)}}$ will be determined by requiring 
  the conditions of coupling coherence to be satisfied.\footnote{For the example
  of positronium's self-energy see Subsection~\ref{4.1.1}. See Subsection~\ref{perry} for
  a coupled scalar theory example.}
  These previous equations are valid for an arbitrary regulating
  function $f_\lam$. In this work we will use
   $f_{\lam i j} = \theta(\lam-|\Delta_{ij}|)$ [a
   convenient choice for doing analytic calculations]. This gives
  \bea
  g^{(\lam \Lam)}_{ikj}&=&
  \left( f_{\Lam ik}-f_{\lam ik}
  \right) \Theta_{ikj}\;,\label{eq:abg}\\
  \Theta_{ikj} &\equiv&
  \theta \left( |\Delta_{ik}|-|\Delta_{kj}|
  \right)\;.
  \label{eq:bigtheta}
  \eea
  For completeness, the explicit form of the third order effective
interaction
  ${\overline v}_\lam^{^{(3)}}$ is
  written in Appendix~\ref{app:bloch}.
\section{Step two: Diagonalization of $\hr$}\label{3.3}

The second step in our approach is to
solve for the spectrum of $H_\lam$. 
The Schr{\"o}dinger equation for eigenstates of $H_\lam$  is
\bea
H_\lam| \Psi_{_{\lam, N}}({\cal P})\rangle&=&
E_{_{N}}| \Psi_{_{\lam ,N}}({\cal P})\rangle\;,
\eea
which written out in all its many-body complexity is
\bea
\sum_j \langle i |H_\lam|j \rangle \langle j | \Psi_{_{\lam, N}}({\cal P})\rangle&=&
E_{_{N}} \langle i | \Psi_{_{\lam ,N}}({\cal P})\rangle\;,\label{eq:sad}
\eea
where the sum over $j$ implies a sum over all Fock sectors and spins, and integrations
 over all momenta in the respective free states $|j\rangle$.
`$N$' labels all the quantum numbers of the state, and is discrete for bound states and
continuous
for scattering states.
 $E_{_{N}}\equiv\frac{{{\cal P}^\perp}^2+M_{_{N}}^2}{{\cal P}^+}$, 
where for positronium for example the binding energy $-B_N$ is defined by
$M_{_{N}}^2\equiv(2 m+B_{_{N}})^2$.  `${\cal P}$' is the total momentum of the
 state of physical interest.
Solving this eigenvalue 
equation exactly is not feasible, because  all
sectors are still coupled. For example, for positronium we have
\bea
&&|\Psi_{_{\lam ,N}}({\cal P})\rangle=\sum_i |i\rangle \langle i |\Psi_{_{\lam ,N}}({\cal
P})\rangle =
\sum_{i^\pr}|e {\overline e} ( i^\pr)\rangle \langle e 
{\overline e} (i^\pr)|\Psi_{_{\lam, N}}({\cal P})\rangle\nonumber\\
&&+\sum_{i^\pr}|e {\overline e} \gam( i^\pr)\rangle \langle e 
{\overline e}\gam (i^\pr)|\Psi_{_{\lam, N}}({\cal P})\rangle+
\sum_{i^\pr}|e {\overline e} e {\overline e}( i^\pr)\rangle \langle e 
{\overline e}e {\overline e} (i^\pr)|\Psi_{_{\lam, N}}({\cal P})\rangle+\cdots\;,~~~~
\eea
where the sum over $i^\prime$ implies a sum over all spins, and integrations
 over all momenta in the respective free states $|i^\prime\rangle$.

As already mentioned, we  divide $H_\lam$ into two pieces
\bea
H_\lam&=& {\cal H}_0+\left(H_\lam-{\cal H}_0\right)\equiv{\cal H}_0+{\cal V}\;,
\eea
diagonalize ${\cal H}_0$ exactly for these QED calculations,
and calculate corrections to the spectrum of ${\cal H}_0$ in BSPT
with ${\cal V}$ to some consistent prescribed order in $\al$ and $\al \ln(1/\al)$.

We close this section by writing the standard BSPT 
Raleigh-Schr\"{o}dinger formulas.  For simplicity,
we write the formulas for the non-degenerate case~\cite{baym}:
\bea
&&\left ( {\cal H}_{0}+{\cal V} \right )
|\Psi_{_{\lam, N}} ({\cal P}) \rangle =
E_{_{ N}} | \Psi_{_{\lam, N}} ({\cal P}) \rangle~,\\
&&{\cal H}_{0}  | \psi_{_{ N}} ({\cal P})\rangle =
{\cal E}_{_{ N}} | \psi_{_{ N}} ({\cal P}) \rangle~,\\
&&| \Psi_{_{\lam, N}} ({\cal P})\rangle =
| \psi_{_{ N}} ({\cal P})\rangle + \sum_{_{M \neq N}}
\frac{ | \psi_{_{ M}} ({\cal P})\rangle \frac{\langle
  \psi_{_{M}} ({\cal P})| {\cal V} 
  | \psi_{_{N}} ({\cal P})\rangle}
  {\langle
  \psi_{_{N}} ({\cal P})|  
   \psi_{_{N}} ({\cal P})\rangle}}
   {{\cal E}_{_{N}}-{\cal E}_{_{M}}} +
  { \cal O}({{\cal V}}^{2})~,\\
  && E_{_{N}}={\cal E}_{_{N}} +
   \frac{\langle
  \psi_{_{N}} ({\cal P})| {\cal V} 
  | \psi_{_{ N}} ({\cal P})\rangle}
  {\langle
  \psi_{_{N}} ({\cal P})|  
   \psi_{_{N}} ({\cal P})\rangle} +
   \;\;\;\sum_{_{M \neq N}} 
   \frac{\left|\frac{\langle
  \psi_{_{N}} ({\cal P})| {\cal V} 
  | \psi_{_{ M}} ({\cal P})\rangle}
  {\langle
  \psi_{_{N}} ({\cal P})|  
   \psi_{_{N}} ({\cal P})\rangle}\right|^{2}}
   {{\cal E}_{_{N}}-{\cal E}_{_{M}}} +
   {\cal O}({{\cal V}}^{3})~,~~
\label{eq:hey1}\eea
where ${\cal P}$ is the total three-momentum of the state and ``N" 
labels all the quantum numbers of the respective state.
These  formulas will be used in Chapter~\ref{ch:pos} to solve
for positronium's leading ground state spin splitting (actually degenerate BSPT is required
here), and in Chapter~\ref{ch:lamb}
to calculate the dominant part of the Lamb shift in hydrogen. Note that for the light-front case
$E_{_{N}}\equiv\frac{{{\cal P}^\perp}^2+M_{_{N}}^2}{{\cal P}^+}$ 
and ${\cal E}_{_{N}}\equiv 
\frac{{{\cal P}^\perp}^2+{\cal M}_{_{N}}^2}{{\cal P}^+}$, where $M_{_{N}}^2$ 
and ${\cal M}_{_{N}}^2$ are the
exact and leading-order mass-squared respectively.
\section{Wegner transformation}\label{Wegner}

The Wegner transformation is unitary and thus follows the initial discussion
on the G-W transformation in Section~\ref{G-W}. The Wegner transformation
is defined by $T_s=[H_s,h]$,\footnote{Actually,
Wegner uses $H_d$ (the full diagonal part of the Hamiltonian at scale `$s$') instead of $h$ (the free
Hamiltonian). We choose the free Hamiltonian for its simplicity.} 
and so the Hamiltonian evolves according to
\be
\frac{d H_s}{ds}=[H_s,[H_s,h]]
\;,
\ee
where $s$ is a cutoff of dimension $1/({\rm energy})^2$, which is obvious from the form of
this defining equation. The free Hamiltonian
$h$ only depends on the renormalization scale $\mu$ but not on the running cutoff $s$:
$\frac{dh}{ds}=0$. See Section~\ref{coherence} for further discussion on this point. Proceeding
we define 
\be
v_{sij}=\exp(-s\Del_{ij}^2) {\overline v}_{sij}
\;,
\ee
where $h |i\rangle=\veps_i|i\rangle $, $\Del_{ij}=\veps_i-\veps_j$, $\langle i|v_s|j\rangle=
v_{sij}$ and $H_s=h+v_s$.
After a little algebra we have
\be
\frac{d{\overline v}_{sij}}{ds}=\sum_k\left(\Del_{ik}+\Del_{jk}\right)
{\overline v}_{sik}{\overline v}_{skj}\exp\left[-2s
\Del_{ik}\Del_{jk}\right]
\;.\label{eq:wegnerbigone}
\ee
 It was useful to note
$\Del_{ij}^2-\Del_{ik}^2-\Del_{jk}^2=-2\Del_{ik}\Del_{jk}$. Note that this is completely
nonperturbative so far, however perturbatively it does
 start out at second order on the right-hand-side.
This transformation has
a simple nonperturbative form, but to date  has only been solved perturbatively
in field theory. We make the perturbative ansatz
\be
{\overline v}_{s}=
{\overline v}_{s}^{^{(1)}}
+{\overline v}_{s}^{^{(2)}}+\cdots\;,
\ee
where the superscript implies the order in the bare interaction ${\overline v}_{s_{_{B}}}$.
At first order this gives
\be
\frac{d{\overline v}_{sij}^{^{(1)}}}{ds}=0\;,
\ee 
which implies
\be
{\overline v}_{s_{_{R}}ij}^{^{(1)}}={\overline v}_{s_{_{B}}ij}
\;,
\ee
where $s_{_{R}}$ is the final cutoff.
Note that $s_{_{R}}>s_{_{B}}$ because of the dimension of the cutoffs, 
and the ``no cutoff limit" is $s_{_{B}}\longrightarrow 0$. At second order
we have
\be
\frac{d{\overline v}_{sij}^{^{(2)}}}{ds}=\sum_k\left(\Del_{ik}+\Del_{jk}\right)
{\overline v}_{s_{_{B}}ik}{\overline v}_{s_{_{B}}kj}\exp\left[-2s
\Del_{ik}\Del_{jk}\right]
\;.
\ee
This exponential is easy to integrate with result
\bea
{\overline v}_{s_{_{R}}ij}^{^{(2)}}&=&\frac{1}{2}\sum_k
{\overline v}_{s_{_{B}}ik}{\overline v}_{s_{_{B}}kj}
\left(\frac{1}{\Del_{ik}}+\frac{1}{\Del_{jk}}\right)\times\nn\\
&&~~~~~~~~~~\times \left[
\exp\left(-2s_{_{B}}
\Del_{ik}\Del_{jk}\right)-\exp\left(-2s_{_{R}}
\Del_{ik}\Del_{jk}\right)
\right]
\;.
\eea
We could continue this exercise to arbitrary order 
without too much apparent difficulty because exponentials are easy to integrate,
but we will stop here since
this transformation is not used in the specific QED examples of this dissertation.
Basically, the Wegner transformation is a G-W transformation with the regulating
function $f_{\lam ij}$ chosen to be a Gaussian $\exp(-\Del_{ij}^2/\lam^2)$. 

\section{Bloch transformation}\label{blochsec}

In this Section we discuss a derivation of an effective Hamiltonian
via a Bloch transformation.
As already mentioned, we use this transformation for our Lamb shift calculation.
We use the Bloch transformation to separate the low and high energy scales 
of the problem
and derive an effective Hamiltonian acting in the low-energy space alone with an identical
  low-energy spectrum
 to the bare Hamiltonian. 
We closely follow Section~IV of \cite{whitebook}, where
a  discussion, including the original references, and
derivation of  a general 
effective Bloch Hamiltonian can be found. 

We start with a bare time-independent Schr{\"o}dinger equation
\bea
&&H_\Lam
 |\Psi_\Lam\rangle=E |\Psi_\Lam\rangle
\;.\label{eq:eeeee}
\eea
Then
projection operators onto the low- and high-energy spaces, $P_L$ and 
$P_H$ respectively,
are defined,

\bea
P_L&=&\theta\left(
\frac{\lam^2+{{\cal P}^\perp}^2}{{\cal P}^+}-h
\right)
,\\
P_H&=&
\theta\left(
\frac{\Lam^2+{{\cal P}^\perp}^2}{{\cal P}^+}-h
\right)~\theta\left(h-
\frac{\lam^2+{{\cal P}^\perp}^2}{{\cal P}^+}
\right)
\;,\\
\tll+\thh&=&\theta\left(
\frac{\Lam^2+{{\cal P}^\perp}^2}{{\cal P}^+}-h
\right)
\;,
\eea
where $\theta(x)$ is a step function.  
$\Lam$ and $\lam$ are the bare and effective cutoffs respectively with
 $\lam < \Lam$.\footnote{The shorthands 
 $\frac{\lam^2+{{\cal P}^\perp}^2}{{\cal P}^+} \longrightarrow \lam$ and 
 $\frac{\Lam^2+{{\cal P}^\perp}^2}{{\cal P}^+}\longrightarrow \Lam$ are often used when
 it does not lead to confusion.}  ${\cal P}=\left({\cal P}^+,{\cal P}^\perp\right)$ is the
 total momentum
 of the physical state of interest.
 $h$ is the free Hamiltonian, which for hydrogen is written in Eq.~(\ref{eq:freebaby}). 

An effective Hamiltonian acting in the low-energy 
space  with an equivalent 
low-energy spectrum to $H_\Lam$ is sought.
To proceed, a new operator ${\cal R}$ is defined
that connects the $\tll$ and $\thh$ spaces
\bea
&&\thh |\Psi_\Lam\rangle={\cal R} \tll |\Psi_\Lam\rangle
\;.
\label{eq:RRR}
\eea
For a discussion and construction of ${\cal R}$,  see Eq.~(4.4) and below
in \cite{whitebook}.

This leads to the following 
 time-independent Schr{\"o}dinger equation for the effective Hamiltonian 
\bea
&&H_\lam |\Phi_\lam\rangle=E |\Phi_\lam\rangle
\;,\label{eq:3.71}
\eea
where 
$E$ is the same eigenvalue as in Eq.~(\ref{eq:eeeee}), the state $|\Phi_\lam\rangle$ is a 
projection
onto the low-energy space [with the same norm as $|\Psi_\Lam\rangle$ of
Eq.~(\ref{eq:eeeee})]
\bea
|\Phi_\lam\rangle&=&\sqrt{1+{\cal R}^\dag {\cal R}}~\tll|\Psi_\Lam\rangle
\label{eq:normmm}
\eea
and
 $H_\lam$ is a hermitian effective Hamiltonian given by
\bea
H_\lam=\frac{1}{\sqrt{1+{\cal R}^\dag \,{\cal R}}}
(\tll+{\cal R}^\dag)\,
H_\Lam\,
(\tll+{\cal R})\,
\frac{1}{\sqrt{1+{\cal R}^\dag {\cal R}}}
\;.
\label{eq:ttrr}
\eea
Note that $H_\lam$ acts in the low-energy space. 
In principle, all bare states $|\Psi_\Lam\rangle$ that have
support in the low-energy space have a corresponding eigenvalue given by
solution to the eigenvalue equation (\ref{eq:3.71}).
 
 Defining $H_\Lam=h+v_\Lam$, where $h$ is the free field theoretic Hamiltonian and
 $v_\Lam$ are the bare interactions,\footnote{$h$ is written in terms of renormalized
 parameters,
 and it is convenient to define $v_\Lam = v + \del v_\Lam$, where $v$ is the canonical
 field
 theoretic interactions written in terms of renormalized parameters
  and $\del v_\Lam$ are  the counterterms
 that must be determined through the process of renormalization. See  
 Eqs.~(\ref{eq:can1})--(\ref{eq:instant})
 for the canonical Hamiltonian of the hydrogen system.}
  through fourth order in $v_\Lam$, the effective Hamiltonian is given by
 \bea
 \langle a | H_\lam |b \rangle&=&\langle a|h+v_\Lam|b\rangle
 +\frac{1}{2}\sum_i\left(
\frac{ \langle a | v_\Lam|i\rangle\langle i|v_\Lam|b\rangle}{\Delta_{ai}}+
 \frac{\langle a | v_\Lam|i\rangle\langle i|v_\Lam|b\rangle}{\Delta_{bi}}
 \right)\nn\\
 &&~+\frac{1}{2}\sum_{i,j}\left(
 \frac{\langle a | v_\Lam|i\rangle\langle i|v_\Lam|j\rangle\langle j|v_\Lam|b\rangle
 }{\Del_{ai}\Del_{aj}}+
 \frac{\langle a | v_\Lam|i\rangle\langle i|v_\Lam|j\rangle\langle
 j|v_\Lam|b\rangle}{\Del_{bi}\Del_{bj}}
 \right)\nonumber\\
 &&~-\frac{1}{2}\sum_{c,i}\left(
 \frac{\langle a | v_\Lam|i\rangle\langle i|v_\Lam|c\rangle\langle c|v_\Lam|b\rangle
 }{\Del_{bi}\Del_{ci}}+
 \frac{\langle a | v_\Lam|c\rangle\langle c|v_\Lam|i\rangle\langle
 i|v_\Lam|b\rangle}{\Del_{ai}\Del_{ci}}
 \right)\label{eq:heff}\nn\\
 &&~+~\langle a|v_\lam^{^{(4)}}|b\rangle
 \;,\label{eq:blochtofourth}
 \eea
 where the fourth order terms are written in Appendix~\ref{app:bloch}.
 $\Delta_{ia}= \veps_i-\veps_a$, with $h |i\rangle = \veps_i |i\rangle$.
 We are using $|a\rangle,~|b\rangle,~\cdots$ 
 to denote low energy states (states in $\tll$) and $|i\rangle,~|j\rangle,~\cdots$
 to denote high energy states (states in $\thh$). 
 Note that the denominators are all $\veps_a-\veps_i$---there are no $\veps_a-\veps_b$ 
 or 
 $\veps_i-\veps_j$ 
 denominators from the same space. However, note that there are potentially
 problematic
 \be
  \sum_{c,i}\frac{\langle v\rangle\langle v\rangle\langle v\rangle}{\Del_{ci}\ldots}
  \nn\ee
   and 
   \be
 \sum_{i,c,j}\frac{\langle v\rangle\langle v\rangle\langle v\rangle\langle v\rangle}
 {\Del_{ci}\Del_{cj}\ldots}\nn
 \ee
  type 
 terms in the effective Hamiltonian beyond second
 order.
 See the already 
 mentioned Reference~\cite{whitebook} for a description of an arbitrary order (in perturbation
 theory)
 effective Bloch Hamiltonian
 and also for a convenient diagrammatic representation of the same.
\section{Renormalization issues in 
light-front field theory}\label{coherence}

The renormalization concepts upon which this dissertation are based were introduced
and elucidated  by several examples in \cite{perrywilson}.
In \cite{whitebook}
this work was continued with ``A Renormalization Group Approach to Hamiltonian
Light-Front Field Theory." These studies were a continuation---generalizations to the
light-front arena---of Wilson's seminal work
on the modern renormalization group reviewed in \cite{w1,w2,w3} with a nice simple introduction
given in \cite{w4}. Here we will give an overview of the concepts and elucidate them with
a few examples.

The effective Hamiltonian
$H_\lam$ is renormalized by requiring it 
 to satisfy  ``coupling coherence."
 Simply put, our regulator breaks gauge and rotational invariance, and
 we need some principle in order to specify our counterterms
 in a unique and meaningful way, such that 
  physical results  are cutoff and
 renormalization scale independent, and manifest these symmetries.
  We demand that these additional counterterms required to
 restore the symmetries broken by the regulator
 run coherently with the ``canonical renormalizable 
 couplings"---that is we do not allow them to
  explicitly depend on the cutoff but only implicitly
 through the ``canonical renormalizable 
 couplings" dependences on the cutoff. This is coupling coherence.\footnote{Actually,
 there is an
 additional requirement as explained below.}
 An example will be worked out in Subsection~\ref{perry}. 
 See the previously mentioned \cite{perrywilson} for further examples, and note
 that ``coupling coherence" was first formulated by Oehme, Zimmermann and
 co-workers in \cite{osz}, where additional  worked examples
 may be found.\footnote{Oehme, Zimmermann and
 co-workers use the term ``reduction of couplings" instead
 of ``coupling coherence."}
 
 The additional requirement of coupling coherence is necessary because without this assumption in a
 light-front effective Hamiltonian, there are an infinite number of relevant and marginal couplings and
 predictability is lost.\footnote{The terminology ``relevant," and ``marginal," as well as
 ``irrelevant," is explained below.}
 This necessity was brought on by the fact that the only regulators we know how to employ
 in the nonperturbative bound-state problem in light-front field theory break Lorentz covariance
 and gauge invariance. Also, longitudinal  locality (and perhaps transverse, but in this dissertation
 we assume transverse locality) is lost. The net result is that an infinite number of relevant,
 marginal and irrelevant couplings arise; that is, above and beyond the usual finite number of
 relevant and marginal couplings and the infinite number of irrelevant couplings\footnote{Due
 to universality, predictability is not lost for this infinite set of couplings.}
 as given by any cutoff effective field theory. These extra
 couplings are required to restore  symmetries broken by the regulator or vacuum, and it is conjectured
 that coupling coherence restores these symmetries. For the discussion
 that follows, following Perry and Wilson \cite{perrywilson}, call these extra couplings 
 required to
 restore the symmetries of interest
 {\it dependent} couplings, and call the couplings required even if  the
 symmetries of interest are maintained by the regulator and vacuum {\it independent} couplings.
 Coupling coherence by construction fixes all the dependent couplings to be functions of only the
 independent relevant and marginal couplings near the lower cutoff $\lam$ with $\lam \ll
 \Lam$.

 In the notation of the G-W transformation, 
 a coupling coherent Hamiltonian written in terms of {\it dimensionless} couplings
 for $\lam \ll \Lam$ satisfies
\bea
 H_\lam&=&S(\lam ,\Lam) H_\Lam(\Lam, e_\Lam, m_\Lam, w_\Lam,
c(e_\Lam, m_\Lam)) S^\dagger(\lam ,\Lam)\nn\\
&&\hskip1in\longrightarrow H_\Lam(\lam, e_\lam, m_\lam, w(e_\lam, m_\lam), c(e_\lam, m_\lam))\;,
\label{eq:alot}
\eea
with the additional requirement that all dependent couplings (represented by
 `$c$' in the
argument of the 
Hamiltonians) vanish when the independent marginal
couplings are set to zero. 
In Eq.~(\ref{eq:alot}), $e_\Lam$ and $m_\Lam$ are independent dimensionless marginal and relevant
couplings respectively, and $w_\Lam$ represents the infinite set of independent dimensionless
irrelevant couplings.\footnote{As already mentioned, the
independent irrelevant couplings flow to a function of the independent relevant
 and marginal couplings for $\lam \ll \Lam$ in the standard way according to the renormalization
 group. That is, the independent irrelevant couplings are automatically coherent and additional
 assumptions need not be made here.} 
$c(e_\lam, m_\lam)$ represents the infinite set of dependent
dimensionless relevant, marginal and irrelevant
couplings. Coupling coherence fixes these
couplings through
\be
\frac{dc(e_\lam, m_\lam)}{d\lam}\equiv\frac{\der c(e_\lam, m_\lam)}{\der e_\lam}
\frac{\der e_\lam}{\der\lam}+
\frac{\der c(e_\lam, m_\lam)}{\der m_\lam}\frac{\der m_\lam}{\der\lam}\;,
\ee
that is
\be
\frac{\der c(e_\lam, m_\lam) }{\der \lam}\equiv 0\;,\label{cLAST}
\ee
[note that this previous equation is implied in the notation of Eq.~(\ref{eq:alot})]
and as already mentioned,
\be
 c(0, m_\lam)\equiv 0
 \;.\label{71797}\ee
 
 The initial bare Hamiltonian $H_\Lam$ does not satisfy Eq.~(\ref{eq:alot}), its
 form changes under the action of $S(\lam ,\Lam)$. $H_\Lam$ must be adjusted until its form does
 not change. This ``adjustment" is the process of renormalization.
 Coupling coherence is a highly non-trivial constraint on the
theory and to date has only been solved perturbatively. In this dissertation, 
for QED,
Eq.~(\ref{eq:alot})
is solved
to order $e^2$, which turns out to be fairly simple because $e$ does not run until order
$e^3$. In Subsection~\ref{perry}, solutions to Eq.~(\ref{eq:alot}) are obtained for the
 running of
the marginal coupling through one loop in massless coupled $O(2)$ scalar theory.

As
far as the scale dependence goes, the canonical Hamiltonian
$H$ depends only on the renormalization scale $\mu$ which
actually does not explicitly
 enter discussions of renormalization until including at least
the running 
coupling $e$; the counterterms depend on $\Lam$ and this renormalization scale $\mu$,
but can not depend
 on the final cutoff $\lam$ or the procedure is ill-defined. 
 The effective Hamiltonian $H_\lam$ depends only
on $\lam$, however in order that it does not depend on the renormalization scale $\mu$,
$e$ and $m$ must implicitly depend on $\mu$. If the renormalization procedure has been
completed to some order in $e$, the limit $\Lam \longrightarrow \infty$
can be taken,\footnote{Strictly speaking, {\it if} QED is a `trivial' theory, the limit
$\Lam \longrightarrow \infty$ should not be taken. However, even in the trivial scenario,
as long as $\Lam \ll m \exp(\frac{3 \pi}{2\al})\sim m \times 10^{280}$,
logic prevails in all practical calculations.}
 and  physical observables, in particular the spectrum, are
 independent of the remaining
scales $\lam$ and $\mu$. The $\lam$-independence arises {\it explicitly} through
the diagonalization process, while the $\mu$-independence arises 
through {\it explicit}
$\mu$-dependent terms canceling against {\it implicit} 
$\mu$-dependence in $e$ and $m$.
In practice the scale independence only arises approximately and
the goal is to have a procedure in which one can systematically
remove scale dependence order by order in $e$.

In the next two subsections we present simple examples to explain
the standard renormalization group terminology from the viewpoint
of light-front field theory, and to elucidate coupling coherence.

\subsection{Relevant, marginal and irrelevant terminology 
(on the light-front)
illustrated through 
simple examples}\label{relevant}

Because of the simple analytic
form of the transformation we will use the Wegner transformation
in this subsection. Recall Section~\ref{Wegner}, especially Eq.~(\ref{eq:wegnerbigone}) 
which is the
exact flow equation for the effective interactions where
$H_s=h+v_s$ and 
\be
v_{sij}=\exp(-s \Del_{ij}^2)~ {\overline v}_{sij}
\;,
\ee
with $\Del_{ij}=
\veps_i-\veps_j$.

From Eq.~(\ref{eq:wegnerbigone}) we see that ${\overline v}_{s}$ 
only changes at non-linear order
and we have
\be
\frac{d{\overline v}_s}{ds}= {\cal O}({\overline v}_{s}^2)
\;.
\ee
The standard renormalization group terminology is defined with
the assumption that 
the non-linear terms are small. At linear order we have
\be
\frac{d{\overline v}_s}{ds}=0
\;.\label{eq:999666}
\ee

Now we write a general interaction in scalar light-front field theory
in $3+1$ dimensions.
Since the physical idea behind renormalization group  works best for 
local interactions, and longitudinal locality is absent in a light-front 
approach, 
  we only assume transverse
locality and only
 keep track of  transverse momenta in this initial example 
 (we discuss a longitudinal renormalization group below). 
 Given this, a general interaction (with no zero mode terms
 such as $a_1 a_2 a_3 a_4$)
 has the form
 \bea
 {\overline v}_s&=&\sum_{n=0}^\infty\sum_{M=2}^\infty\sum_{m=1}^{M-1}~c_{nmM}~
 \int_{1,2,\cdots,M}16\pi^3\del^3\left(
 p_1+p_2+\cdots+p_m-p_{m+1}-p_{m+2}-\cdots-p_M
 \right)\nn\\
 &&\hskip2in\times a_1^\da a_2^\da \cdots a_m^\da~ a_{m+1} a_{m+2} \cdots a_M~
 p_\perp^{2 n}~s^D\label{eq:transv}
 \;,
 \eea
 where $\int_1=\int_{p_1}=
 \int\frac{d^2p_1^\perp dp_1^+ \theta(p_1^+)}{16\pi^3p_1^+}$, 
 $a_1=a(p_1)$, the commutation relations are
 \be
 [a_1,a_2^\da]=16 \pi^3 p_1^+ \del^3(p_1-p_2)
 \;,
 \ee
 and the factor $p_\perp^{2n}$
 is a shorthand\footnote{This is `$2n$' based on the kinematic
 rotational symmetry about the longitudinal axis with corresponding  generator
 $J^3$.} for a product of $M$ transverse
 momenta
 each to a particular power 
 with the sum of all the powers being equal to $2n$---this is all that matters
 in order
 to classify these operators; note $n$ can {\it not} be negative based on the
 assumption of transverse locality.
 $c_{nmM}$ are {\it dimensionless} couplings. We have explicitly made these
 couplings dimensionless by including the appropriate power of 
  the cutoff through the factor $s^D$.\footnote{Recall the dimension of
 $s$ is $1/({\rm energy})^2$, which as far as transverse momentum goes
 implies $s \rightarrow 1/p_\perp^4$. $s_{_{B}}$ and $s_{_{R}}$ 
 are the initial and final
 cutoffs respectively. Recall that $s_{_{R}} > s_{_{B}}$ and the
 ``no cutoff limit" is $s_{_{B}}\longrightarrow 0$.} A simple exercise
 shows that to make $c_{nmM}$ dimensionless we must have
 \be
D = \frac{M+2n-4}{4}\;.\label{eq:D}
 \ee
 Taking a derivative of this general form of the interaction gives\footnote{Note
 that $\frac{da^\da_i}{ds}=0$. Recall that the free Hamiltonian $h$ only depends on the
 renormalization scale $\mu$ and not on the running cutoff
 scale $s$: $\frac{dh}{ds}=0$;
 and $a^\da_i|0\rangle$ is a state of $h$ that thus only depends on $\mu$ but not on the
 running cutoff
 scale $s$.}
 \bea
\frac{1}{s^D}\frac{d{\overline v}_s}{d\ln s}&=& 
\sum_{nmM}\int_{1,2,\cdots,M}16\pi^3\del^3\left(
 p_1+p_2+\cdots+p_m-p_{m+1}-p_{m+2}-\cdots-p_M
 \right)\nn\\
&&~~\times a_1^\da a_2^\da \cdots a_m^\da~ a_{m+1} a_{m+2} \cdots a_M~
 p_\perp^{2 n}~\left(
 \frac{dc_{nmM}}{d\ln s}+D~c_{nmM}
 \right)\;.
 \eea
 Thus at linear order [recall Eq.~(\ref{eq:999666})] the flow equation
 implies
 \be
 \frac{dc_{nmM}}{dt}+D~c_{nmM}=0
 \;,
 \ee
 where we have defined $t=\ln s$. The range of $s$ is $0$ to $\infty$
 which in terms of $t$ corresponds to $-\infty$ to $\infty$: running
 the transformation (that is integrating out high energy scales) 
 corresponds to {\it increasing} $t$.
 Solving this gives
 \be
 c_{nmM}=k~\exp(-D~t)\;,
 \ee
 where $k$ is an arbitrary constant fixed by the boundary conditions. 
 The standard renormalization group
 terminology is
 \bea
 &&D < 0\longrightarrow~{\rm relevant}~{\rm coupling}\;,\nn\\
 &&D = 0\longrightarrow~{\rm marginal}~{\rm coupling}\;,\nn\\
 &&D > 0\longrightarrow~{\rm irrelevant}~{\rm coupling}
 \;.\label{380}
 \eea
 We see that at this linear order
  the relevant couplings are exponentially enhanced as the
  energy scales are integrated out whereas the marginal couplings
  remain fixed and the irrelevant couplings are exponentially suppressed---hence
  the terminology. Of course this changes when the non-linear 
  terms are  accounted for, but typically (as long as the couplings
  never get too big) this only changes the results
  logarithmically in $s$, that is linearly in $t$: the non-linear changes
  occur slowly. 
  
  The results of this initial example
  correspond to the usual equal-time
  renormalization group terminology.
  Recalling Eq.~(\ref{eq:D}) and unraveling the above results
  we see that there are a finite number of structures---only taking into account
  the transverse scales recall---that are relevant and
  marginal whereas there are an infinite number of structures that are
  irrelevant. The only allowed relevant, marginal and irrelevant structures
  according to the transverse renormalization group (with no zero-modes)
  in scalar light-front theory in 3+1-dimensions
  are respectively
  \bea
 D < 0&\longrightarrow&~(a^\da_1a_2,~a^\da_1a_2a_3,~a^\da_1a^\da_2a_3)
 \;,\nn\\
 D = 0&\longrightarrow&~(a^\da_1a_2~p_\perp^2,~
 a^\da_1a_2a_3a_4,~a^\da_1a^\da_2a_3a_4,~a^\da_1a^\da_2a^\da_3a_4)
 \;,\nn\\
 D > 0&\longrightarrow&~(a^\da_1a_2~p_\perp^4,~a^\da_1a^\da_2a_3~p_\perp^2,~
 a^\da_1a^\da_2a_3a_4~p_\perp^2,~a^\da_1a^\da_2a^\da_3a_4a_5,~\ldots)
  \;.
  \eea
 
 Now we work the previous example, scalar light-front theory
in $3+1$-dimensions, but only keep track of the longitudinal
 scales. We will {\it not} assume longitudinal locality. A 
  general interaction (with no zero mode terms
 such as $a_1 a_2 a_3 a_4$)
 has the form
 \bea
 {\overline v}_s&=&\sum_{n=-\infty}^\infty\sum_{M=2}^\infty\sum_{m=1}^{M-1}~c_{nmM}~
 \int_{1,2,\cdots,M}16\pi^3\del^3\left(
 p_1+p_2+\cdots+p_m-p_{m+1}-p_{m+2}-\cdots-p_M
 \right)\nn\\
 &&\hskip2in\times a_1^\da a_2^\da \cdots a_m^\da~ a_{m+1} a_{m+2} \cdots a_M~
(p^+)^{n}~s^D
 \;,
 \eea
 using the same notation as in Eq.~(\ref{eq:transv}), but with
 $p_\perp^{2n}\longrightarrow(p^+)^{n}$ 
 (note now $-\infty\leq n\leq\infty$). Note that
  $s\longrightarrow 1/({\rm energy})^2\longrightarrow (p^+)^2$,
  so now to make the coupling $c_{nmM}$ dimensionless we must have
 \be
 D=-\frac{n}{2}\;.
 \ee
 `$M$' does not enter (unlike  the transverse renormalization group)
 because $a_i^\dag$ are dimensionless with respect to the longitudinal scales. The
 discussion now follows as below Eq.~(\ref{eq:D}) except with this new value for $D$.
 Thus we have: Linear order longitudinal renormalization group implies 
 \be
 c_{nmM}=k\exp(-D\,t)=k\exp(n\,t/2)\;,
 \ee
 where $k$ is an arbitrary constant fixed by the boundary conditions. Recall that integrating out
 energy scales corresponds to increasing $t$. Sticking to the standard terminology of Eq.~(\ref{380})
 we have
\bea
 &&n> 0\longrightarrow~{\rm relevant}~{\rm coupling}\;,\nn\\
 &&n= 0\longrightarrow~{\rm marginal}~{\rm coupling}\;,\nn\\
 &&n< 0\longrightarrow~{\rm irrelevant}~{\rm coupling}
 \;.
 \eea
 We see that the sequence of nonlocal operators 
 $\left(\frac{1}{\der^+}\right)\phi^4$, $\left(\frac{1}{\der^+}\right)^2\phi^4$,
  $\ldots$ becomes more and more
 irrelevant in the longitudinal renormalization group
  sense, whereas recall that in the transverse renormalization group
 sense the sequence $\left(\der^\perp\right)^2\phi^4$, $\left(\der^\perp\right)^4\phi^4$, 
 $\ldots$ becomes more and more irrelevant.
 The same type of simplification that occurs in the transverse renormalization
  group when {\it locality} is
 assumed may occur in the longitudinal renormalization group when 
 {\it non-locality} is assumed. This may be a deep conclusion, with
 perhaps far-reaching consequences, but complications occur in practice
 when running a longitudinal renormalization group \cite{white161},
 and it is not clear at present how to proceed.
 It is clear however, with these last two examples as witness, that
 longitudinal and transverse scales must be treated differently in light-front
 field theory as long advertised by Wilson and collaborators \cite{longpaper}.

\subsection{Simple example to elucidate coupling coherence}\label{perry}

This example is from \cite{perrywilson}. However, we will use the G-W transformation here.
Recall the flow equation in the G-W formalism
\be
{\overline v}_\lam = v+\del v_{_{\Lam}} -\int_\lam^\Lam [ v_{\lam^\pr},T_{\lam^\pr}]
d \lam^\pr\;.
\ee
This example is massless coupled scalar theory in $3+1$ dimensions: 
\be
v=\int d^2x^\perp dx^-\left(\frac{g_1}{4!}\phi_1^4+\frac{g_2}{4!}\phi_2^4+
\frac{g_3}{2!\,2!}\phi_1^2\phi_2^2\right)
\;.
\ee
We normal order this interaction and drop zero-modes.

To elucidate the discussion on the renormalization scale $\mu$
at the beginning of this section, we show how it enters the calculation through
the running of the marginal couplings at one-loop. The counterterms
for these above three marginal couplings
are defined by
\be
\del v_{_{\Lam}}=-\int_\Lam^\mu [ v_{\lam^\pr},T_{\lam^\pr}]
d \lam^\pr\;,\label{eq:3.84}
\ee
where $\Lam$ is the initial bare cutoff and $\mu$ is the renormalization scale.
Given this, the above effective interaction at the final cutoff $\lam$ 
for these three marginal couplings
is 
\be
{\overline v}_\lam = v -\int_\lam^\mu [ v_{\lam^\pr},T_{\lam^\pr}]
d \lam^\pr\;,
\ee
where $v$ has implicit dependence on $\mu$ through
 the couplings $g_1$, $g_2$ and $g_3$.
Note that the bare cutoff dependence is gone.

Now we calculate the two-particle goes to two-particle matrix elements
of ${\overline v}_\lam$ through one loop, setting the transverse Jacobi
momenta to zero---which isolates the running of
the marginal couplings---and derive how the marginal couplings
$g_1$, $g_2$ and $g_3$ must
run so that the respective matrix elements
of ${\overline v}_\lam$ are independent of the renormalization scale $\mu$. We will not
go through the explicit calculation, but will just present the results.
Through one loop we obtain
\bea
\frac{dg_1}{dt}&=&\frac{3}{16\pi^2}\left(g_1^2+g_3^2\right)\;,\nn\\
\frac{dg_2}{dt}&=&\frac{3}{16\pi^2}\left(g_2^2+g_3^2\right)\;,\nn\\
\frac{dg_3}{dt}&=&\frac{1}{16\pi^2}\left(g_1 \,g_3+g_2\,g_3+4\,g_3^2\right)
\;,\label{eq:THEONE}
\eea
where $t\equiv \ln(\mu)$. 
$\mu$ has dimension mass and arose from 
a light-front energy renormalization scale
$\frac{\mu^2}{{\cal P}^+}$; now integrating out energy scales corresponds to {\it decreasing}
$t$. We see that 
the theory is not asymptotically free. 

Coupling coherence enters when we search for solutions to these
flow equations. The initial theory has three independent marginal
couplings. Under what conditions will this number be reduced? Also, the initial
theory is not $O(2)$ symmetric with respect to $\phi_1^2+\phi_2^2$. Is it possible
to somehow restore this symmetry? Coupling coherence is one way. The first
postulate (the second one comes below)
of coupling coherence in this example is
 that $g_2$ and $g_3$ do not depend explicitly on
 the renormalization scale $\mu$ but rather  only implicitly
through their dependence on $g_1$---one can say they run coherently with $g_1$.
In equations this postulate is that $\frac{\der g_2}{\der t}=0$ and 
$\frac{\der g_3}{\der t}=0$. Thus, through one loop we have
\bea
\frac{dg_1}{dt}&=&\frac{\der g_1}{\der t}=\frac{3}{16\pi^2}\left(g_1^2+g_3^2\right)
\;,
\nn\\
\frac{dg_2}{dt}&=&\frac{\der g_2}{\der g_1}\frac{\der g_1}{\der t}=
\frac{3}{16\pi^2}\left(g_2^2+g_3^2\right)\;,\nn\\
\frac{dg_3}{dt}&=&\frac{\der g_3}{\der g_1}\frac{\der g_1}{\der t}=
\frac{1}{16\pi^2}\left(g_1 \,g_3+g_2\,g_3+4\,g_3^2\right)
\;,
\eea
where we inserted the results from Eq.~(\ref{eq:THEONE}) on the right-hand-side of these equations.
Inserting the top equation into the other two gives
\bea
\frac{\der g_2}{\der g_1}\left(g_1^2+g_3^2\right)&=&\left(g_2^2+g_3^2\right)
\;,\nn\\
\frac{\der g_3}{\der g_1}\left(g_1^2+g_3^2\right)&=&\frac{1}{3}\left(g_1
\,g_3+g_2\,g_3+4\,g_3^2\right)
\;,\label{eq:3.88}
\eea
once again valid through one loop.

Two non-trivial solutions to these two coupled equations given the boundary
condition of coupling coherence ($\left.g_2\right|_{g_1\longrightarrow 0}=0$ and 
$\left.g_3\right|_{g_1\longrightarrow 0}=0$) are (i)
a decoupled theory:
\beaa
(g_2=g_1,~g_3=g_1)&\longrightarrow&v=\frac{g_1}{4!}
\left\{\left(\frac{\phi_1+\phi_2}{2^{\frac{1}{4}}}\right)^4+
\left(\frac{\phi_1-\phi_2}{2^{\frac{1}{4}}}\right)^4\right\}
\;,
\eeaa
and (ii) an interesting $O(2)$ symmetric theory:
\beaa
\left(g_2=g_1,~g_3=\frac{g_1}{3}\right)
&\longrightarrow&v=\frac{g_1}{4!}\left(\phi_1^2+\phi_2^2\right)^2
\;.
\eeaa
As promised, coupling coherence is a way to reduce the number of independent
couplings and restore symmetries not manifest in the flow equations. This
choice of which marginal coupling is to be the independent coupling is where
all the physics lies. In QED it is clear that $e$ is the coupling of choice. 

 For further simple one-loop
examples which show how coupling coherence leads naturally
to massless gauge bosons when explicit gauge invariance is broken by the regulator, 
to massless fermions
when explicit chiral invariance\footnote{Chiral
symmetry is very interesting on the light-front. Its charge 
 $Q^5_{LF}$ measures  helicity.
See \cite{chiral} and references within for further discussion.}
 is broken, and to the hidden $\phi\longrightarrow -\phi$ 
symmetry in scalar theory when working in
the broken phase (of particular interest for light-front calculations without 
zero-modes), see \cite{perrywilson}.
    
\chapter{Positronium's ground state spin splitting}\label{ch:pos}

Now we apply the procedure outlined in Sections~\ref{G-W} and \ref{3.3}
 to obtain positronium's  
ground state spin splitting   through order $\alpha^4$. 
First, we derive $H_\lam$ to second order in $e$.
  This includes a discussion of the effective fermion
self-energy, but the effective photon self-energy (for $\lam <2m$)
and electromagnetic coupling do not run
at this order.   
Then we move on to the diagonalization of $H_\lam$. 
This starts with 
a discussion of our zeroth order Hamiltonian, ${\cal H}_0$, which
will be treated analytically. We introduce 
a coordinate
change that takes $(x \in [0,1]) \longrightarrow (\kap_z \in [-\infty,\infty])$,
 which allows easier identification of ${\cal H}_0$. 
 We solve for the spectrum of ${\cal H}_0$ exactly, which among other things,
fixes  the
 $\alpha$-scaling of the momenta in the matrix elements in BSPT.
 Then we move on to 
a derivation of the perturbative effects coming from 
low-energy (energy transfer below $\lam$) photon
emission, absorption and annihilation
 at order $e^2$, which includes a discussion of the full electron and positron
 self-energies
 and a derivation to order $e^2$ of the complete exchange and annihilation
 interactions.
 Given this, we determine the 
  range of $\lam$
  that allows the effects of low-energy (energy transfer below $\lam$)
   photon emission and absorption 
 to be  transferred to the effective interactions in the $|e {\overline
 e}\rangle$
 sector alone, and at the same time,
 does not cut into the nonperturbative features of the
  solutions of ${\cal H}_0$.
 Finally, we  proceed with  BSPT in ${\cal V}$
 noting that all shifts appear in the few-body 
 sector, $|e {\overline e}\rangle$, alone.

We present the well-known result from equal-time calculations to compare
with our results later.
Through order $\al^4$ an energy level in positronium with quantum
numbers $(n,l,S,J)$\footnote{$n$ is the 
principal and $l$ is the orbital
 angular momentum quantum number; 
 $S$ is the total electron plus positron spin quantum number and 
 ${\bf J}$ = ${\bf l}$+${\bf S}$ with $J$ the corresponding quantum number.} 
 according to QED is given by 
 \cite{bstext}
\beaa
E_{nlSJ}=-\frac{Ryd}{2 n^2}+\left[
\frac{11}{32 n^4}+\left(\veps_{lSJ}-\frac{1}{2 l+1}\right)\frac{1}{n^3}
\right] \al^2 ~{\rm Ryd}
\;,
\eeaa
where ${\rm Ryd}=\frac{m \al^2}{2}\simeq 13.6~{\rm eV}$,
\beaa
\veps_{l1J}=\frac{7}{6} \del_{l,0} +
\frac{1-\del_{l,0}}{2 (2 l+1)}\left(
\del_{J,l+1} \frac{3l+4}{(l+1)(2l+3)}
-\del_{J,l}\frac{1}{l(l+1)}-\del_{J,l-1} \frac{3l-1}{l(2l-1)}
\right)
\eeaa
and $\veps_{l0J}=0$.
This is a well established result first derived in 1947 \cite{oldpos}.
For the ground state spin splitting this gives
\bea
E_{1^3\!S}-E_{1^1\!S}=\frac{7}{6}\,\al^2 \,{\rm Ryd}
\eea
through order $\al^4$
in the standard spectroscopic notation, $n^{2S+1}\!l$. In passing it is 
interesting to note that in positronium (since fine and hyperfine structure are at 
the same order) no degeneracy with respect
to $J$ remains through order
$\al^4$---unlike hydrogen where through order $\al^4$
there is the famous $2 S_{\frac{1}{2}}$ and $2 P_{\frac{1}{2}}$ 
degeneracy [and it is of course this splitting
at order $\al^5\ln(1/\al)$ that is the famous Lamb shift, the
dominant part of which is calculated in Chapter~\ref{ch:lamb}].
\section{Derivation of $H_\lam$ through second order}\label{4.1}

From Eqs.~(\ref{eq:hap3})--(\ref{eq:bigtheta}), the final renormalized Hamiltonian to second order 
is given by
\bea
\langle i|H_\lam|j\rangle&=& f_{\lam ij} \left\{
h_{ij}+
v_{ij}+
\sum_{k} {v}_{ik} 
  {v}_{kj} \left(
  \frac{g^{(\lam \Lam)}_{ikj}}{\Delta_{ik}}\right.\right.\nn\\
  &&\left.\left.~~~~~~~~~~~~~~~~~~~~~~~~~~~~~~~~~ +
  \frac{g^{(\lam \Lam)}_{jki}}{\Delta_{jk}}
  \right)+
  \del v_{\Lam ij}^{^{(2)}}+{\cal O}(e^3)
\right\}\;.
\label{eq:44}
\eea
$g^{(\lam \Lam)}_{ikj}$ is written in Eq.~(\ref{eq:abg}) and $v$ is given by
Eq.~(\ref{eq:46}) with all the proton fields set to zero.
\subsection{Renormalization issues in positronium}\label{4.1.1}

The form of $\del v_{_{\Lam}}^{^{(2)}}$
follows from the constraint that $H_\lam$ satisfies 
coupling coherence.
To order $e^2$  the fermion and photon masses run, but the coupling does not. 
First, we  discuss
 the result for the electron
  self-energy coming from the second-order effective interactions in $H_\lam$.
 Consult Appendix~\ref{brentrules} for a listing of the free matrix elements
 of the canonical interaction $v$
  used in this
 dissertation.
 Specifically,
 we calculate a free matrix element of $H_\lam$ given in Eq.~(\ref{eq:44}) in
 the electron self-energy channel.
 
 At second-order, an electron self-energy effective interaction
  arises because photon emission in $\hr$ is restricted to be below $\lam$.
 That is, the energy scales  $\Lam$ down to $\lam$ from photon emission have
 been `integrated out' and placed in effective interactions.
 At second-order the explicit form of this electron self-energy effective
 interaction is (note $\veps_i=\veps_j$ so 
 $f_{\lam ij}=1$)
 \beaa
&&\frac{ \del \Sigma_e^{^{(2)}}}{p_1^+}\equiv
\left.\frac{\langle i|v_\lam^{^{(2)}}|j\rangle}
{\langle i | j \rangle}\right|_{{\rm electron~ self-energy}}\nn\\
&&~~~=\frac{1}{\langle 3 | 1 \rangle}
\sum_{s_e s_\gam}\int_{p_e p_\gam}\theta(p_e^+-\eps {\cal P}^+)\theta(p_\gam^+-\eps {\cal P}^+)
 (p_3s_3|v_{ee\gam}|p_es_ep_\gam s_\gam )(p_es_ep_\gam s_\gam |v_{ee\gam}|p_1s_1)\nn\\
 &&~~~\times\left[ 16 \pi^3 \del^3(p_e+p_\gam-p_3)\right]\left[ 16 \pi^3
 \del^3(p_e+p_\gam-p_1)\right]\left(f_{\Lam jk}-f_{\lam jk}\right)/\Del_{jk}
 +\frac{\del v_\Lam^{^{(2)}}}{p_1^+}\;,
 \eeaa
 where $v_{ee\gam}\equiv\int d^2x^\perp dx^- {\cal H}_{ee\gam}$ is the canonical
 emission and absorption interaction, and
  the initial, intermediate and final free states are labeled $|j\rangle=|1,2\rangle$,
  $|k\rangle=|p_es_e,p_\gam s_\gam ,p_2s_2\rangle$ and $|i\rangle=|3,4\rangle$ respectively.
 \begin{figure}[t!]\hskip0in
\centerline{\epsfig{figure=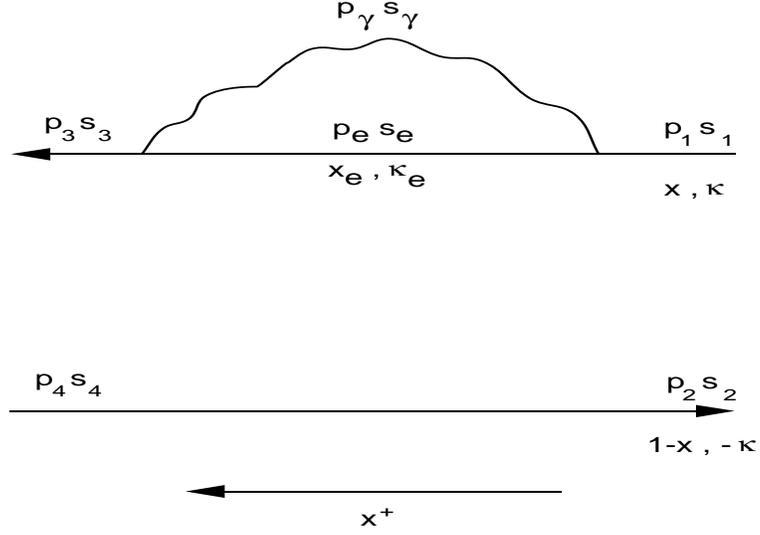,height=3in,width=4.5in}}
\caption{Momenta and spin label conventions for the second-order effective electron
self-energy
interaction.} 
\label{selfenergy}
\end{figure}
 See Figure~\ref{selfenergy} for the momenta and spin label conventions.
 $\del v_\Lam^{^{(2)}}$ are the aforementioned
 second-order counterterms to be determined through coupling
 coherence below. `$\eps$' is the infrared regulator discussed in the paragraph containing
 Eq.~(\ref{eq:2.67}) that we are forced to introduce.
 We define our Jacobi variables by
 \beaa
 {\cal P}&=&p_1+p_2\;,\\
 p_1&=&\left(x{\cal P}^+~,~\kap + x {\cal P}^\perp \right)\;,\\
 p_e&=&\left(x_ep_1^+~,~\kap_e + x_e p_1^\perp \right)\;,\\
 p_\gam&=&\left((1-x_e)p_1^+~,~-\kap_e + (1-x_e) p_1^\perp \right)
 \;.
 \eeaa
 Note that
 \be
 p_e^-+p_\gam^-=\frac{{p_1^\perp}^2+\left(\frac{\kap_e^2+m^2}{x_e}+\frac{\kap_e^2}{1-x_e}\right)}
 {p_1^+}
 \;.
 \ee
 The above becomes (see Appendix~\ref{brentrules} for the matrix elements of 
 $v_{ee\gam}$ in 
 the free basis)
 \bea
 \del \Sigma_e^{^{(2)}}&=&
 -e^2 \int d^2\kap_e\int_{\eps/x}^{1-\eps/x}dx_e\frac{2x_e}{16\pi^3x_e(1-x_e)}\nn\\
 &&~~~~~\times
 \frac{\left(f_{\Lam jk}-f_{\lam jk}\right)
 \left[\kap_e^2\left(\frac{2}{(1-x_e)^2}+\frac{1}{x_e^2}+\frac{2}{x_e(1-x_e)
 }\right)+\frac{m^2(1-x_e)^2}{x_e^2}\right]}
 {\frac{\kap_e^2+m^2}{x_e}+\frac{\kap_e^2}{1-x_e}-m^2}\nn\\
 &&~~~~~~~~~~+~\del v_\Lam^{^{(2)}}\;,
 \eea
 where
 \be
 f_{\lam^\pr jk}=\theta\left({\lam^\pr}^2x+m^2-\frac{m^2}{x_e}-\frac{\kap_e^2}{x_e(1-x_e)}\right)
 =\theta\left(G_{\lam^\pr}[x_e]_{}^{}\right)\theta\left(G_{\lam^\pr}[x_e]-\kap_e^2\right)
 \;,\ee
 with
 \be
 G_{\lam^\pr}[x_e]\equiv x_e(1-x_e)\left({\lam^\pr}^2x+m^2-\frac{m^2}{x_e}\right)
 \;.
 \ee
 Including these constraints we have\footnote{We use $\eps/x
 \ll\frac{m^2}{m^2+{\lam^\pr}^2x}$. We see that a non-zero renormalized electron mass $m$
 ``regulates" the infrared electron momentum [$p_e^+\longrightarrow 0$]
 divergence.}
 \bea
 \del \Sigma_e^{^{(2)}}&=&-e^2\pi\left[
 \int_{\frac{m^2}{m^2+\Lam^2 x}}^{1-\eps/x}dx_e\int_0^{G_{\Lam}[x_e]}d(\kap_e^2)-
 \int_{\frac{m^2}{m^2+\lam^2 x}}^{1-\eps/x}dx_e\int_0^{G_{\lam}[x_e]}d(\kap_e^2)\right]
   \frac{2x_e}{16\pi^3x_e(1-x_e)}\nn\\
 &&~~~~~\times
 \frac{
 \left[\kap_e^2\left(\frac{2}{(1-x_e)^2}+\frac{1}{x_e^2}+\frac{2}{x_e(1-x_e)
 }\right)+\frac{m^2(1-x_e)^2}{x_e^2}\right]}
 {\frac{\kap_e^2+m^2}{x_e}+\frac{\kap_e^2}{1-x_e}-m^2}+~\del v_\Lam^{^{(2)}}\;.
 \eea
 Performing this integral gives
 \be
  \del \Sigma_e^{^{(2)}}=-(\del\Sigma_\Lam^{^{(2)}}
-\del\Sigma_\lam^{^{(2)}})+\del v_{\Lam}^{^{(2)}}
\label{eq:72}
 \;,
 \ee
 where we have defined
 \bea
 \del\Sigma_{\lam^\pr}^{^{(2)}}&\equiv& \frac{\alpha}{2 \pi} 
 \left[
 -\frac{3 {\lam^\pr}^2 x+m^2}{2}
 +\frac{m^2 \left(\frac{m^2}{2}+{\lam^\pr}^2 x
 \right)}{m^2+{\lam^\pr}^2 x}
 -3 m^2 \ln \left(\frac{m^2+{\lam^\pr}^2 x}{m^2}
 \right)\right. \nonumber\\
&&~~~~~~~~~~~~~~~~~~~~ 
 \left.-2 {\lam^\pr}^2 x\ln \left(\frac{m^2+{\lam^\pr}^2 x}{{\lam^\pr}^2 x}
 \right)
 +2 {\lam^\pr}^2 x \ln\left(\frac{ x}{\eps}
 \right)
 \right]
 \;.
 \label{eq:fermionmass}
\eea
For respective $\del\Sigma_{\lam^\pr}^{^{(2)}}$ terms of Eq.~(\ref{eq:72}),
 ${\lam^\pr}=\Lam$ and $\lam$.
 Note that the energy dependence on the electron's relative transverse Jacobi momentum, $\kap$,
 does not change.
Also, note that the result is infrared singular.

Now we determine $\del v_\Lam^{^{(2)}}$ through coupling coherence given the above results.
Constraining the electron mass to run coherently with the cutoff according to
Eq.~(\ref{eq:alot})  amounts to the requirement
\be
\left[m^2-(\del\Sigma_\Lam^{^{(2)}}
-\del\Sigma_\lam^{^{(2)}})+\del v_{\Lam}^{^{(2)}} +{\cal
O}\left(e^4\right)\right]
=\left[m^2+\del v_{\Lam}^{^{(2)}}+{\cal O}\left(e^4\right)\right]_{\Lam
\rightarrow \lam}
\label{eq:1945}
\ee
This fixes the  counterterm\footnote{Any finite ${\cal O}(e^2)$
running cutoff 
 independent
term can be added to the counterterm, and Eq.~(\ref{eq:1945}) would still be
satisfied. This term can only depend on the renormalization scale $\mu$.
Setting this term to zero is our choice of renormalization prescription. \label{fn:16}}
\bea
\del v_{\Lam}^{^{(2)}}&=&\del\Sigma_{\Lam}^{^{(2)}}+{\cal O}(e^4)\;,
\label{eq:1000}
\eea
and to second order the fermion mass renormalization is complete. 

In summary, through second-order the coherent electron mass-squared
is
\be
m_\lam^2\equiv 
\left[m^2-(\del\Sigma_\Lam^{^{(2)}}
-\del\Sigma_\lam^{^{(2)}})+\del v_{\Lam}^{^{(2)}} \right]
=m^2+\del\Sigma_\lam^{^{(2)}}\;,\label{coherentelectron}
\ee
where $\del\Sigma_\lam^{^{(2)}}$ is written in Eq.~(\ref{eq:fermionmass}).
In $\hr$ there is 
of course still the photon emission interaction below $\lam$ 
that must be considered. The
form of this interaction is $f_\lam\int d^2x^\perp dx^-{\cal H}_{ee\gam}$. 
  This is considered below in Subsection~{4.2.2},
   and the resulting combination of $m_\lam^2$ and these
 effects from ``low-energy emission" 
 add to the physical electron mass-squared $m_{phys}^2$, which as will
be shown is infrared finite and  with our choice of renormalization prescription
(see footnote~\ref{fn:16}), through second order,
is given by $m^2$, the renormalized electron mass-squared in
the free Hamiltonian $h$.\footnote{So this is a ``physical" renormalization prescription;
 not that it matters, because physical results are of course independent of the
 choice of renormalization prescription.}
 
For arbitrary $\lam$, 
the photon mass also runs at order $e^2$. The discussion follows that of the
electron mass except 
for the fact that the running photon mass is infrared finite. 
For $\lam^2 < (2 m)^2$, the resulting coherent photon mass vanishes because pair production is
no longer possible. In this chapter we choose
 $\lam^2 \ll m^2$, thus the photon mass is zero to all orders in perturbation
theory, as required by gauge invariance.
There are additional difficulties with dependent marginal couplings that are encountered
at ${\cal O}(e^3)$, but this is beyond the focus of this dissertation.
\subsection{$H_\lam$ through order $e^2$: exchange and annihilation
channels}\label{4.1.2}

To complete the derivation of $H_\lam$ through second-order, we need to write the
coherent interactions for the exchange and annihilation channels in the $|e
{\overline e} \rangle$
sector.
At second order, these come from
 tree level diagrams, with no divergences or running couplings. Thus, the coherent
 results
follow from
\bea
\del v_{\Lam}^{^{(2)}}&\equiv& -\int_\Lam^\infty [
v_{\lam^\pr}^{^{(1)}},T_{\lam^\pr}^{^{(1)}}]
d \lam^\pr
\;.
\label{eq:iiii}
\eea
To show that 
$\del v_{\Lam}^{^{(2)}}$ produces a coherent interaction recall
Eq.~(\ref{eq:23}). We have
\bea
{\overline v}_\lam &=& v-\int_\Lam^\infty [
v_{\lam^\pr}^{^{(1)}},T_{\lam^\pr}^{^{(1)}}]
d \lam^\pr -\int_\lam^\Lam [ v_{\lam^\pr}^{^{(1)}},T_{\lam^\pr}^{^{(1)}}]
d \lam^\pr\nonumber +{\cal O}(e^3)\\
&=&v-\int_\lam^\infty [ v_{\lam^\pr}^{^{(1)}},T_{\lam^\pr}^{^{(1)}}]
d \lam^\pr+{\cal O}(e^3)\\
&=&\left.\left[v-\int_\Lam^\infty [
v_{\lam^\pr}^{^{(1)}},T_{\lam^\pr}^{^{(1)}}]
d \lam^\pr+{\cal O}(e^3)
\right]\right|_{\Lam \rightarrow \lam}
\;,
\eea
which satisfies the coupling coherence constraint, Eq.~(\ref{eq:alot}). 
At second order this seems trivial, but at higher orders the constraint that
only
$e$ and $m$ run independently with the cutoff places severe constraints on the
Hamiltonian.
\begin{figure}[t]\vskip-.5in\hskip-.5in
\centerline{\epsfig{figure=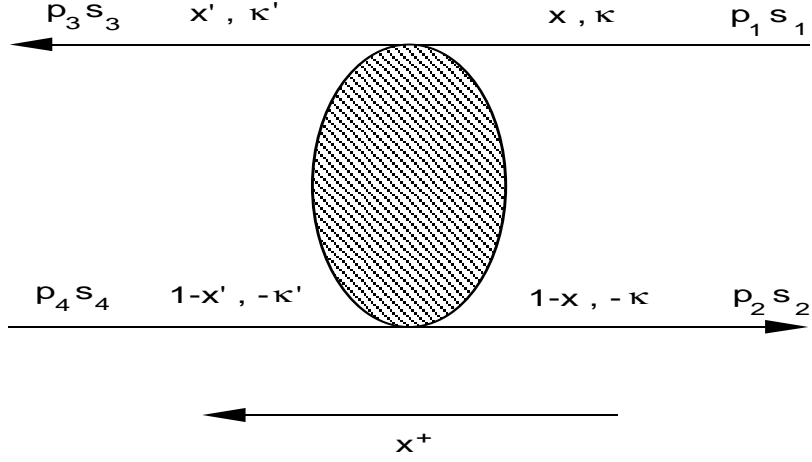,height=3in,width=5in}}
\caption{This illustrates the spin and momenta label conventions 
used for this positronium calculation.} 
\label{posfig1}
\end{figure}

Given this second-order interaction, the free matrix elements of $H_\lam$,
shown  
 in Eq.~(\ref{eq:44}), in the exchange and annihilation channels 
are

\noindent
\underline{Exchange Channel}:
 \bea
 V_{_{\lam,exchange}}&\equiv&\frac{\langle e (3) {\overline e} (4)|
 H_\lam|e(1){\overline e}(2)\rangle|_{_{exchange}}}
 {16\pi^3\del^3(p_1+p_2-p_3-p_4) \sqrt{x x^\prime (1-x)(1-x^\prime)}}\nonumber\\
 &\equiv&
 V_1+V_2
 +{\cal O}(e^4)\;,
 \label{eq:saeko3}
 \eea 
 where
 \bea
V_1&=&- \;e^2 \;N_1 \;\;\theta\left(
 \lam^2-\left|{\cal M}_0^2-
 {{\cal M}_0^\pr}^2\right|\right)\nonumber\\
&&\times \left(
  \frac{\theta\left(
 \left|\Delta_1\right|-\left|\Delta_2\right|\right)
 \theta\left(
 \left|\Delta_1\right|-\lam^2\right)}{DEN_1}
 +
  \frac{\theta\left(
 \left|\Delta_2\right|-\left|\Delta_1\right|\right)
 \theta\left(
 \left|\Delta_2\right|-\lam^2\right)}{DEN_2}
  \right)\;,~~\\
V_2&=&- \;e^2\;\theta\left(
 \lam^2-\left|{\cal M}_0^2-
 {{\cal M}_0^\pr}^2\right|\right)
 \left(
 \frac{4 }{(x-x^\prime)^2}\del_{s_1s_3}\del_{s_2s_4}
 \right)
\;.
\label{eq:76}
\eea
The variables  are defined as follows (see Figure~\ref{posfig1} also):
\beaa
\bullet\;\; p_1&=&\left(x {\cal P}^+ , \kappa+x {\cal P}^\perp\right)\;,
\;p_2=\left((1-x) {\cal P}^+ , -\kappa+(1-x) {\cal P}^\perp\right)\;\\
\bullet\;\; p_3&=&\left(x^\prime 
{\cal P}^+ , \kappa^\prime+x^\prime {\cal P}^\perp\right)\;,
\;p_4=\left((1-x^\prime) {\cal P}^+ , -\kappa^\prime+(1-x^\prime)
 {\cal P}^\perp\right)\nonumber\;
\\
\bullet\;\; N_1&=&\del_{s_1 s_3}\del_{s_2 s_4} T_1^\perp \cdot T_2^\perp-
2 m^2 \del_{s_2 {\overline s_4}}\del_{s_2 {\overline s_1}}
\del_{s_3 {\overline s_1}}\frac{(x-x^\prime)^2}{x 
x^\prime (1-x) (1-x^\prime)}\\
&+&i m \sqrt{2} (x^\prime-x)\left(
\frac{s_1}{x x^\prime}\del_{{\overline s_1} s_3}\del_{s_2 s_4} 
\eps_{s_1}^\perp \cdot T_1^\perp+
\frac{s_2}{(1-x) (1-x^\prime)}\del_{{\overline s_4} s_2}\del_{s_1 s_3} 
\eps_{s_2}^\perp \cdot T_2^\perp
\right)\;\\
\bullet\;\;T_1^i&=&-\frac{2 (\kappa^i-{\kappa^\prime}^i)}{x-x^\prime}
-\frac{{\kappa^\prime}^i({\overline s_2})}{1-x^\prime}
-\frac{\kappa^i(s_2)}{1-x}\;,\;
T_2^i=\frac{2 (\kappa^i-{\kappa^\prime}^i)}{x-x^\prime}
-\frac{{\kappa^\prime}^i({\overline s_1})}{x^\prime}
-\frac{\kappa^i(s_1)}{x}\\
\bullet\;\;\kappa^j(s)&=&\kappa^j+i \;s \;\eps_{jk}\; \kappa^k
\;\;(s\;=\;\pm\;1\;{\rm and}\;{\overline s}\equiv -s)\;\;;
\eps_{12}=-\eps_{21}=1\;,\;\eps_{11}=\eps_{22}=0\\
\bullet\;\;\Delta_1&=&\frac{DEN_1}{x^\pr-x}\;,\;\Delta_2=\frac{DEN_2}{x^\pr-x}\\
\bullet\;\;DEN_1&=&\frac{(\kap x^\prime-\kap^\prime x)^2}{x x^\prime}+
\frac{(m x-m x^\pr)^2}{x x^\pr}\;,\;
DEN_2=DEN_1|_{x \rightarrow 1-x\;,\;x^\pr \rightarrow 1-x^\pr}\\
\bullet\;\;{\cal M}_0^2&=&\frac{{\kap}^2+m^2}{x (1-x)}\;,\;
{{\cal M}_0^\pr}^2=\frac{{\kap^\pr}^2+m^2}{x^\pr (1-x^\pr)}\;.
\eeaa

\noindent
\underline{Annihilation Channel}:
\bea
 V_{_{\lam,annihil}}&\equiv&\frac{\langle e (3) {\overline e} (4)|
 H_\lam|e(1){\overline e}(2)\rangle|_{_{annihilation}}}
 {16\pi^3\del^3(p_1+p_2-p_3-p_4) \sqrt{x x^\prime (1-x)(1-x^\prime)}}\nonumber\\
 &\equiv&
 V_3+
 V_4
 +{\cal O}(e^4)\;,
 \label{eq:saeko4}
 \eea 
 where
 \bea
&&V_3= \;e^2 \;N_2 \;\;\theta\left(
 \lam^2-\left|{\cal M}_0^2
-{{\cal M}_0^\pr}^2\right|\right)
 \left(
  \frac{\theta\left({\cal M}_0^2
-{{\cal M}_0^\pr}^2
 \right)
 \theta\left(
 {\cal M}_0^2-\lam^2\right)}{{\cal M}_0^2}\right.\nonumber\\
 &&~~~~~~~~~~~~~~~~~~~~~~~~~~~~~~~~~~~~~~~~~~+\left.
  \frac{\theta\left({{\cal M}_0^\pr}^2-{\cal M}_0^2
 \right)
 \theta\left(
 {{\cal M}_0^\pr}^2-\lam^2\right)}{{{\cal M}_0^\pr}^2}
  \right)~,\\
&&V_4=\;4 e^2 \theta\left(
 \lam^2-\left|{\cal M}_0^2
-{{\cal M}_0^\pr}^2\right|\right)
\del_{s_1 {\overline s_2}} \del_{s_3 {\overline s_4}}\;,\label{eq:747}
\eea
and
\beaa
\;\; N_2&=&\del_{s_1 {\overline s_2}}\del_{s_4 {\overline s_3}} T_3^\perp 
\cdot T_4^\perp+
 \del_{s_1  s_2}\del_{s_1  s_4}
\del_{s_3  s_4}\frac{2 m^2}{x 
x^\prime (1-x) (1-x^\prime)}\\
&&~~~~+ i m \sqrt{2} \left(
\frac{s_1}{x (1-x)} \del_{s_1 s_2}\del_{ s_4 {\overline s_3}} 
\eps_{s_1}^\perp \cdot T_3^\perp-
\frac{s_4}{x^\prime (1-x^\prime)}\del_{ s_1 {\overline s_2}}\del_{ s_3 s_4} 
{\eps_{s_4}^\perp}^\ast \cdot T_4^\perp
\right)\;,\\
\;\;T_3^i&=&
\frac{{\kappa^\prime}^i( s_3)}{1-x^\prime}
-\frac{{\kappa^\pr}^i({\overline s_3})}{x^\pr}\;,\;
T_4^i=
\frac{{\kappa}^i({\overline s_1})}{1-x}-\frac{{\kappa}^i( s_1)}{x}
\;.\eeaa
$V_2$ and $V_4$ are  canonical instantaneous exchange and annihilation
interactions, respectively, with widths restricted by the
regulating function, $f_\lam$. 
$V_1$ and $V_3$ are effective  interactions that arise because 
 photon emission and annihilation have vertices with widths restricted 
by the
regulating function, $f_\lam$. 
\section{Diagonalization of $\hr$}

First we discuss the lowest order spectrum of $H_\lam$, after which we 
 discuss BSPT, renormalization and a limiting procedure which allows
 the effects of low-energy (energy transfer below $\lam$)
 emission to be transferred to the $|e {\overline e}\rangle$
 sector alone.
\subsection{${\cal H}_0$, a coordinate change and its exact spectrum}\label{4.2.1}

${\cal H}_0$, 
as motivated from the form of
our second-order effective Hamiltonian, in the $|e {\overline e}\rangle$
sector is
\bea
{\cal H}_0&=&h+{\cal V}_{_{C}}\;,
\eea
where $h$ is the free Hamiltonian given in Eq.~(\ref{eq:hap2}),
and ${\cal V}_{_{C}}$ 
is given by [using the same variables defined below Eq.~(\ref{eq:76});
note,
$\kap_z$ is defined below by Eq.~(\ref{eq:xonehalf})]
\bea
&&\frac{\langle e (3) {\overline e} (4)|
 {\cal V}_{_{C}}|e(1){\overline e}(2)\rangle}
 {16\pi^3\del^3(p_1+p_2-p_3-p_4) \sqrt{x x^\prime (1-x)(1-x^\prime)}}\equiv V_{_{C}}\;,\nn\\
 &&\hskip1.0in{\rm where}~~~~~V_{_{C}}\equiv
 -\frac{16 m^2 e^2 \del_{s_1 s_3} \del_{s_2
 s_4}}{(\kap-\kap^\pr)^2+(\kap_z-\kap_z^\pr)^2}
\;.\label{eq:ruo}
\eea

In all other sectors, we choose ${\cal H}_0 = h$. This is convenient for
 the leading order spin-splitting calculation (order $\al^4$) in positronium,
 but for example in the Lamb shift calculation it is more convenient to
 keep the Coulomb interaction between the oppositely charged valence particles
 to all orders in all sectors.
 
 As already mentioned,
${\cal H}_0$ 
 in the $|e{\overline e}\rangle$ sector is motivated from the form of our
 second order effective
  Hamiltonian
 $H_\lam$: it arises as the leading order term in a nonrelativistic limit of the
  instantaneous photon exchange interaction
 combined with the two time-orderings of the dynamical photon exchange
 interaction.
 We choose it to simplify  positronium bound-state calculations. Other choices
 are possible, and must
 be used to study problems such as photon emission.
 Later, in BSPT this choice is shown to produce the leading order contribution
 to 
 positronium's binding-energy (order $\al^2$)
 as long as $m\al^2\ll\tilde{\lam}\ll m\al$.
 
 The coordinates
 $\kap_z$ and $\kap_z^\pr$ in Eq.~(\ref{eq:ruo}) 
 follow from a standard coordinate transformation that
  takes
 the range of longitudinal momentum fraction, $x \in [0,1]$ to 
 $\kap_z \in [-\infty, \infty]$.
 This coordinate change is
 \bea
 x&\equiv&\frac{1}{2}+\frac{\kap_z}{2 \sqrt{\kap^2+\kap_z^2+m^2}}\;.
 \label{eq:xonehalf}\eea
 We introduce  
 a new three-vector 
  \bea
 {\bf p}&\equiv& (\kap,\kap_z)
 \;.
 \label{eq:nicerelation}\eea
 Note that
 \bea
 {\cal M}_0^2&\equiv&\frac{\kap^2+m^2}{x (1-x)}=4(m^2+{\bf p}^2)
 \eea
 is invariant with respect to rotations in the space of vectors ${\bf p}$.
The nonrelativistic kinematics of Eqs.~(\ref{nr1}) and (\ref{nr2}) in 
terms of this  three-vector
become
\bea
\frac{|{\bf p}|}{m} &=& {\cal O}(\alpha)\;.\label{eq:nr3}
\eea

Note
 the  simple forms that our  ``exchange channel denominators" take in the
 nonrelativistic limit
 \bea
 DEN_1&=& ({\bf p}-{\bf p}^\pr)^2
 -\frac{(\kap_z-\kap_z^\pr) ({\bf p}^2-{{\bf p}^\pr}^2)}{m}
 +{\cal O}\left[\left(\frac{{\bf p}}{m}\right)^5 m^2\right]\label{eq:ohno}~,\\
 DEN_2&=& ({\bf p}-{\bf p}^\pr)^2
 +\frac{(\kap_z-\kap_z^\pr) ({\bf p}^2-{{\bf p}^\pr}^2)}{m}
 +{\cal O}\left[\left(\frac{{\bf p}}{m}\right)^5 m^2\right]~.\label{eq:ohno2}
 \eea
 Also note the form that the longitudinal momentum fraction 
 transferred between the electron and positron
 takes
 \bea
 x-x^\pr &=& \frac{\kap_z-\kap_z^\pr}{2 m}+\frac{({{\bf p}^\pr}^2
 \kap_z^\pr-{\bf p}^2 \kap_z)}{4 m^3}
 +{\cal O}\left[\left(\frac{{\bf p}}{m}\right)^5\right]
 \;.
 \label{eq:96}\eea
 These formulas are consistently used throughout this chapter.
  
  Now we describe the leading 
  Schr{\"o}dinger equation. We seek  solutions
  of the following eigenvalue equation
\bea
{\cal H}_0 |\psi_N({\cal P})\rangle&=&{\cal E}_N |\psi_N({\cal P})\rangle\;,
\label{eq:3333}
\eea
where ${\cal E}_N\equiv \frac{{{\cal P}^\perp}^2+{\cal M}_N^2}{{\cal P}^+}$.
 {\em ${\cal H}_0$  is diagonal with respect to the different particle sectors,
 thus
we can solve Eq.~(\ref{eq:3333}) sector by sector.}
In all sectors other than  $|e {\overline e}\rangle$,
${\cal H}_0=h$, and the solution is trivial.
For the $|e {\overline e}\rangle$ sector,
a general $|\psi_N({\cal P})\rangle$ is
 \bea
 |\psi_N({\cal P})\rangle&=&\sum_{s_1 s_2} \int_{p_1 p_2} \sqrt{p_1^+ p_2^+}
 16 \pi^3 \del^3({\cal P}-p_1-p_2) \nn\\
 &&~~~~~~~~\times\;{\tilde{\phi}}_N(x \kap s_1 s_2)~
 b_{s_1}^\dagger(p_1)\; d_{s_2}^\dagger(p_2)\; |0\rangle\;,
 \eea
 with norm
 \beaa
 &&\langle \psi_{N} ({\cal P})|\psi_{N^\pr} ({\cal P}^\pr)\rangle \equiv
 \del_{N N^\pr}16\pi^3 {\cal P}^+
 \del^3\left({\cal P}-{\cal P}^\pr\right)\\
 &&\Longrightarrow\;
 \sum_{s_1 s_2}\frac{\int d^2 \kap \int_0^1 dx}{16\pi^3}
 {{\tilde{\phi}}_N}^\ast(x \kap s_1 s_2) {\tilde{\phi}}_{N^\pr}(x \kap s_1
 s_2)\;=\;\del_{N N^\pr}\;.
 \eeaa
 The tilde on ${\tilde{\phi}}_{N}$ will be notationally convenient below. 
 In the $|e {\overline e}\rangle$ sector, Eq.~(\ref{eq:3333}) becomes
 \be
 \left({\cal M}_{_{N}}^2-\frac{{\kap^\pr}^2+m^2}{x^\pr (1-x^\pr)}
 \right) {\tilde{\phi}}_N(x^\pr \kap^\pr s_3 s_4)=
 \sum_{s_1 s_2}\frac{\int d^2 \kap \int_0^1 dx}{16\pi^3}
 ~V_{_{C}}~ {\tilde{\phi}}_N(x \kap s_1 s_2)
 \;.\label{eq:sc}
 \ee
 After the above coordinate change, this becomes [J(p) is the Jacobian of the
 coordinate change written below]
 \be
 \left[{\cal M}_{_{N}}^2-4 (m^2+{{\bf p}^\pr}^2)
 \right] \phi_N({\bf p}^\pr  s_3 s_4)=
 \sum_{s_1 s_2}\int\frac{ d^3 p \sqrt{J(p)J(p^\prime)}}{16\pi^3}
 ~V_{_{C}}~
  \phi_N({\bf p} s_1 s_2)
 \;,~~~~\label{eq:full}
 \ee
 where the tilde on the wavefunction has been removed by redefining the norm
 in a convenient fashion
 \bea
\del_{N N^\pr}&=&\sum_{s_1 s_2}\frac{\int d^2 \kap \int_0^1 dx}{16\pi^3}
 {\tp_N}^\ast(x \kap s_1 s_2) \tp_{N^\pr}(x \kap s_1 s_2) =
 \sum_{s_1 s_2} \int d^3 p \frac{J(p)}{16\pi^3}
 {\tp_N}^\ast({\bf p} s_1 s_2) \tp_{N^\pr}({\bf p} s_1 s_2)
 \nonumber\\
 &\equiv&\sum_{s_1 s_2} \int d^3 p \;\phi_N^{ \ast}({\bf p} s_1 s_2)
 \phi_{N^\pr}({\bf p} s_1 s_2)
 \;,\label{eq:norm}\eea
 and the Jacobian of the coordinate change is 
 \bea
 J(p)&\equiv&\frac{dx}{d\kap_z}=\frac{\kap^2+m^2}{2({\bf
 p}^2+m^2)^{\frac{3}{2}}}
 \;.\eea
 Note that the Jacobian factor in Eq.~(\ref{eq:full}) satisfies
 \bea
 \sqrt{J(p)J(p^\prime)}&=& \frac{1}{2 m}\left[1-
\frac{{\bf p}^2+2\kap_z^2+{{\bf p}^\prime}^2+2{\kap_z^\pr}^2}{4 m^2}+
{\cal O}\left(\frac{{\bf p}^4}{m^4},\frac{{{\bf
p}^\pr}^4}{m^4},\cdots\right)\right]
 \;.\label{eq:jac}
 \eea
 
 Before defining ${\cal H}_0$ in the $|e {\overline e}\rangle$ sector 
 we mention a subtle but important point in 
 the definition of ${\cal H}_0$.  {\em
 ${\cal H}_0$ in the $|e {\overline e}\rangle$ sector 
 will not be defined by Eq.~(\ref{eq:full})}. Rather, it will be defined by
 taking the leading order
  nonrelativistic expansion of the Jacobian factor in Eq.~(\ref{eq:full}). This
  gives
   \bea
 \left[{\cal M}_{_{N}}^2-4 (m^2+{{\bf p}^\pr}^2)
 \right] \phi_N({\bf p}^\pr  s_3 s_4)&=&
 \sum_{s_1 s_2}\int\frac{ d^3 p \left(\frac{1}{2 m}\right)}{16\pi^3}
 ~V_{_{C}}~
  \phi_N({\bf p} s_1 s_2)
 \;,
 \label{eq:cccoul}
 \eea
 where $V_{_{C}}$ is defined in Eq.~(\ref{eq:ruo}). This ${\cal H}_0$ will be
 diagonalized
 exactly, and the subsequent BSPT will be set up as an expansion in
 ${\cal V}\equiv H_\lam-{\cal H}_0$,
 which will then be regrouped in terms of a consistent expansion in
 $\al$ and $\al \ln(1/\al)$ to some prescribed order. 
  First, we discuss the exact
 diagonalization of
 ${\cal H}_0$.
 
 Putting  the expression for $V_{_{C}}$ into Eq.~(\ref{eq:cccoul}) results in
 the following
 equation
 \bea
 \left(-{\cal B}_{_{N}}+\frac{{{\bf p}^\pr}^2}{m}
 \right) \phi_N({\bf p}^\pr  s_3 s_4)&=&
 \frac{\alpha}{2 \pi^2}\int\frac{ d^3 p }{({\bf p}-{\bf p}^\pr)^2}
  \phi_N({\bf p} s_3 s_4)
 \;.\label{eq:C}
\eea
 This is recognized as the familiar nonrelativistic Schr{\"o}dinger equation for
 positronium.
 Note that we have defined a leading order binding-energy, $-{\cal B}_N$, as
 \bea
 {\cal M}_N^2 &\equiv& 
 4 m^2 + 4 m {\cal B}_{_{N}}
 \;,
 \eea
 where ${\cal M}_N^2$ is the leading order mass-squared.
 Note the difference in the definition of our leading order binding-energy and
 our exact binding-energy as given by $M_{_{N}}^2\equiv(2 m+B_{_{N}})^2$ [see 
 Appendix~\ref{app:mvsb} for
 further discussion
 on this difference], where $M_N^2$ is the exact mass-squared.
 
 To proceed with the 
 solution\footnote{The solution of Eq.~(\ref{eq:C}) is of course well-known,
 but introducing hyperspherical harmonics (which may not be so well-known)
  is essential for the later analytic calculation in second order
 BSPT, so we go through some detail here.} 
 of Eq.~(\ref{eq:C}) note that there is no spin
 dependence in
 the operator so the spin part just factors out
  \bea
\phi_{\nu, s_e,s_{_{{\overline e}}}}({\bf p}^\pr s_3s_4)&\equiv&\phi_\nu({\bf
p}^\pr) 
\del_{s_e s_3} \del_{s_{_{{\overline e}}} s_4}
\;.
\eea
 We rewrote
$N$ as $(\nu, s_e,s_{_{{\overline e}}})$, where $(s_e,s_{_{{\overline e}}})$ 
label the
spin quantum numbers and $\nu$ labels all other quantum 
numbers, 
which are
discrete for the bound states
and continuous for the scattering states.

The solutions to Eq.~(\ref{eq:C}) are well known.
 For ${\cal B}_{_{N}} < 0$, following Fock~\cite{fock1935}, 
 we change coordinates according to
 \bea
 m {\cal B}_{_{N}}&\equiv&-e_n^2~,\\
u&\equiv&(u_0,{\bf u}) ~,\\
 u_0&\equiv&\cos(\omega)\equiv\frac{e_n^2-{\bf p}^2}{e_n^2+{\bf p}^2}~,\\
 {\bf u}&\equiv&\frac{{\bf p}}{p}\sin(\omega) \equiv \sin(\omega)\left[
\sin(\theta) \cos(\phi) ,sin(\theta) \sin(\phi) , \cos(\theta)\right]\nonumber\\
&\equiv& \frac{2 e_n {\bf p}}{e_n^2+{\bf p}^2}
 \;.\eea
 Useful relations implied by this coordinate change are in 
 Appendix~\ref{app:hyper}.
 Note that in our  notation we anticipate that $\nu$ will be given 
 by $(n, l, m_{_{l}})$, the usual principal and angular momentum
  quantum numbers, and that the leading order binding-energy
 will  depend only on the principal quantum number, $n$.
Given this, 
 Eq.~(\ref{eq:C}) becomes
\bea
\psi_\nu(\Omega_{ p^\pr})&\equiv&\frac{\alpha}{2 \pi^2}\frac{m}{2 e_n} 
\int  \frac{d \Omega_p}{|u-u^\pr|^2} \psi_\nu(\Omega_p)
\;,\eea
where
\bea
\psi_\nu(\Omega_p)&\equiv&\frac{(e_n^2+{\bf p}^2)^2}{4 (e_n)^{\frac{5}{2}}}
\phi_\nu({\bf p})\;.
\label{eq:110}
\eea
Using Eq.~(\ref{eq:1900}) of
 Appendix~\ref{app:hyper}, this is seen to have the following solution:
\bea
&&\psi_\nu(\Omega_p)=Y_\nu(\Omega_p)~~{\rm with}~~ 
\frac{\alpha}{2 \pi^2}\frac{m}{2 e_n} \frac{2 \pi^2}{n} = 1
\;,\eea
where
$Y_\nu(\Omega_p)$ is a hyperspherical harmonic.
Thus,
\be
e_n=\frac{m \alpha}{2 n}~\Longrightarrow~
 {\cal B}_{_{N}}=-\frac{m \alpha^2}{4 n^2}
\;,
\label{eq:112}
\ee
and
\be
\phi_\nu({\bf p})=\frac{4 (e_n)^{\frac{5}{2}}}{(e_n^2+{\bf p}^2)^2}Y_\nu(\Omega_p)\;.
\ee
  This 
is the standard nonrelativistic solution for the bound states of positronium.
This completes the solution
of ${{\cal H}}_0$ for the bound-states. 
The scattering states are also needed in our second-order BSPT calculation.
We  use Green's function techniques to include the scattering states where required.
\subsection{BSPT, renormalization and a limit}\label{4.2.2}

Here we use the BSPT formulas (appropriately generalized to the degenerate case)
of Section~\ref{3.3}
to analyze positronium's leading ground state spin splitting.
The potential to be used in BSPT is
\bea
{\cal V}&=&H_\lam-{\cal H}_0\;,
\eea
where the eigenvalue equation for ${\cal H}_0$  is given by
Eq.~(\ref{eq:cccoul}),
and $H_\lam$ to second order is given in Section~\ref{4.1}. We will be perturbing about
the nonperturbative eigenstates
 of ${\cal H}_0$.  

First, we discuss electron mass renormalization. In second-order BSPT there is
an
electron mass shift coming from the $f_\lam v$ part of $H_\lam$, with
$v$ given by $\int d^2 x^\perp d x^- {\cal H}_{e e \gamma}$ [see
Eq.~(\ref{eq:48})].
This is photon emission and absorption restricted by the regulating function,
$f_\lam$.
The calculation is similar to that of Subsection~\ref{4.1.1}. Assuming
$\langle {\cal M}_{_{N}}^2-{\cal M}_0^2\rangle = {\cal O}(e^2)$, this
electron mass-squared shift is
\bea
\del m^2&=&-\del\Sigma_\lam^{^{(2)}}+{\cal O}(e^4)\;.
\label{eq:35}
\eea
$\del\Sigma_\lam^{^{(2)}}$ is the same function that was defined in
Eq.~(\ref{eq:fermionmass}). Checking the consistency:
using this result [Eq.~(\ref{eq:35})]
one obtains $\langle {\cal M}_{_{N}}^2-{\cal M}_0^2\rangle
=\langle 4 m^2+ 4 m {\cal B}_{_{N}}-4 (m^2+{\bf p}^2)\rangle={\cal O}(e^4)$,
and our initial assumption is satisfied. Combining
this  with the coherent electron 
mass-squared $m_\lam^2$ of Eq.~(\ref{coherentelectron}), 
 the physical electron mass-squared $m_{phys}^2$ through 
second-order is
\bea
m_{phys}^2&=&m_\lam^2+\del m^2\nonumber\\
&=& \left[m^2-\left(\del\Sigma_{\Lam}^{^{(2)}}- 
\del\Sigma_{\lam}^{^{(2)}}\right)+\del v_{\Lam}^{^{(2)}}\right]+\left[-
\del\Sigma_{\lam}^{^{(2)}}\right]+{\cal O}(e^4)\nonumber\\
&=&m^2+{\cal O}(e^4)
\;.
\eea
In this last step we recalled the result from coupling coherence of
Eq.~(\ref{eq:1000}).
We see that through second order, the physical electron mass is equivalent to the renormalized
electron
mass in the free Hamiltonian $h$. 
As already mentioned, we see that our choice of renormalization prescription mentioned in
footnote~\ref{fn:16} corresponds to a physical prescription as the physical electron mass is 
equivalent to the renormalized electron mass through second order.
Of course we do not have to use this physical prescription, but it is our choice here,
and must be maintained consistently in higher order  calculations.
As promised below Eq.~(\ref{coherentelectron}), we see that treating photons
perturbatively
has led to an exact cancellation of the infrared divergence in the
coherent electron mass-squared $m_\lam^2$, and  the physical electron mass
through second-order is infrared finite.

Now we move on to the discussion of BSPT.
The only  channels to order $e^2$ are exchange and annihilation. Parts of these
effective
interactions are given in Subsection~\ref{4.1.2}. We also need to include the perturbative
mixing of the $|e {\overline e} \gamma \rangle$ 
and $|\gamma\rangle$ sectors with the 
$|e {\overline e}  \rangle$ sector arising from $f_\lam v$, with 
$v=\int d^2 x^\perp d x^- {\cal H}_{e e \gamma}$. In second-order 
BSPT this gives rise to the following effective interactions that must be
added to 
$V_{_{\lam,exchange}}$ and $V_{_{\lam,annihil}}$ 
of Eqs.~(\ref{eq:saeko3}) and (\ref{eq:saeko4})
respectively.

\noindent
\underline{Exchange Channel}:
\bea
V_5&=& \frac{- e^2 N_1
\theta\left(\lam^2-\left|\Delta_1\right|\right)
\theta\left(\lam^2-\left|\Delta_2\right|\right)}{DEN_3}
\;,
\eea
with 
\beaa
DEN_3 &=& (\kap -\kap^\pr)^2+\frac{1}{2} (x-x^\pr) A +|x-x^\pr|
\left( \frac{1}{2} \left( {\cal M}_0^2+{{\cal M}_0^\pr}^2 \right) -{\cal
M}_{_{N}}^2\right)
\;,\eeaa
and
\beaa
A&=&
\frac{\kap^2+m^2}{1-x}-\frac{{\kap^\pr}^2+m^2}{1-x^\pr}+\frac{{\kap^\pr}^2+m^2}{x^\pr}
-\frac{\kap^2+m^2}{x}
\;.
\eeaa

\noindent
\underline{Annihilation Channel}:
\bea
V_6&=&\;e^2 \;N_2 \;\;\frac{\theta\left(\lam^2-{\cal M}_0^2\right)
\theta\left(\lam^2-{{\cal M}_0^\pr}^2\right)}{{\cal M}_{_{N}}^2}
\;.
\eea
Note that in a nonrelativistic expansion [after the coordinate change of
Eq.~(\ref{eq:xonehalf})]
the above ``exchange channel denominator" becomes
\be
DEN_3=({\bf p}-{\bf p}^\pr)^2+|x-x^\pr|
\left( \frac{1}{2} \left( {\cal M}_0^2+{{\cal M}_0^\pr}^2 \right) -{\cal
M}_{_{N}}^2\right)+
 {\cal O}\left[\left(\frac{{\bf p}}{m}\right)^6 m^2\right]
\;.~~~
\ee

The full exchange and annihilation channel interactions to order $e^2$ are
\bea
V_{_{exchange}}&\equiv&V_{_{\lam,exchange}}+V_5\label{eq:saeko1}\;,\\
V_{_{annihil}}&\equiv&V_{_{\lam,annihil}}+V_6\label{eq:saeko2}
\;,
\eea
where
 Eqs.~(\ref{eq:saeko3}) and (\ref{eq:saeko4}) give $V_{_{\lam,exchange}}$ and 
$V_{_{\lam,annihil}}$ respectively.

One way to summarize the results, recalling the form of Eq.~(\ref{eq:full})
and the norm in Eq.~(\ref{eq:norm}), is to state:
The full order $e^2$ effective interactions give rise to the following {\it first}
order BSPT shift of the 
 bound-state mass-squared spectrum of ${\cal H}_0$
\bea
&&\del^{^{(1)}}M^2(s_3,s_4;s_1,s_2)
\equiv\langle
\phi_{n,l,m_l,s_3,s_4}|V|\phi_{n,l,m_l,s_1,s_2}\rangle\nonumber\\
&& ~~~~~~=\int d^3 p~ d^3 p^\pr \phi_{n,l,m_l}^\ast({\bf p}^\pr)
 V({\bf p}^\pr,s_3,s_4;{\bf p},s_1,s_2)
\phi_{n,l,m_l}({\bf p})\;,
\label{eq:dirac}\eea
where
\bea
V({\bf p}^\pr,s_3,s_4;{\bf p},s_1,s_2)&=&
 \sqrt{
 \frac{J(p)}{16 \pi^3}\frac{J(p^\pr)}{16 \pi^3}} \left(
 V_{_{exchange}}+V_{_{annihil}}\right)\nn\\
 &&~~~~~~~~~~~~~~~~~~~
 - \sqrt{
 \frac{
 \left(\frac{1}{2 m}\right)}{16 \pi^3}\frac{\left(\frac{1}{2 m}\right)}{16
 \pi^3}
 } 
 \left(V_{_{C}}\right)
\;.\label{eq:fermi}
\eea
The Dirac notation in Eq.~(\ref{eq:dirac})
will be used in the remainder of this chapter.
See Eqs.~(\ref{eq:ruo}), (\ref{eq:saeko1}) and (\ref{eq:saeko2}) for $V_{_{C}}$,
$V_{_{exchange}}$ and $V_{_{annihil}}$ respectively. The interaction $V$ must
be diagonalized in the degenerate spin space following the standard rules 
of degenerate BSPT. Note that 
in order to get all the shifts through a consistent order in $\al$ (in this
case through order $\al^4$),
$V$ needs to be considered in {\it second} order
BSPT  also.

The diagonalization of $V$ 
in the degenerate spin space
 follows shortly, but first
 we determine the range of $\lam$ that allows 
 the effects of low-energy (energy transfer below $\lam$)
   photon emission and absorption 
 to be  transferred to the effective interactions in the $|e {\overline
 e}\rangle$
 sector alone, 
 and at the same time  does not
remove the nonperturbative bound-state physics of interest. This range is
\bea
&&\frac{\left|M_N^2-(2 m)^2\right|}{{\cal P}^+} \ll \frac{\lam^2}{{\cal P}^+}
\ll q^{-}_{_{photon}}
\;,
\label{eq:theone}
\eea
where $M_N^2$ is the bound-state mass-squared and $q^{-}_{_{photon}}$ is
 the dominant energy of an emitted photon.
After the solutions of ${\cal H}_0$ are known the $\alpha$-scaling in all BSPT
matrix elements
is known and the bounds in Eq.~(\ref{eq:theone}) become
\bea
&&m^2 \alpha^2 \ll \lam^2 \ll m^2 \alpha
\;.\label{eq:4.58}
\eea
This is satisfied  under the following limit
\bea
\lam^2\;&\longrightarrow&{\rm a\;fixed\;number}~,\label{eq:limit2}\\
\frac{ m^2\alpha^2}{\lam^2}  &\longrightarrow& 0~,\\
 \frac{m^2\alpha }{\lam^2} &\longrightarrow& \infty
\;.\label{eq:limit3}
\eea
Given the nonrelativistic limit
\bea
\alpha &\longrightarrow& 0~,\\
\frac{m^2}{\lam^2}&\longrightarrow&\infty
\;,
\eea
this implies 
\bea
&&\frac{m^2}{\lam^2}\propto
 \alpha^{-\frac{k}{2}} 
\;,\label{eq:2000}
\eea
with
\bea
&&2 < k <4~~.
\eea
Note that this ``window of opportunity" is available to us because, (i) we have
an adjustable effective cutoff $\frac{\lam^2}{{\cal P}^+}$
in the theory, and (ii) QED is 
a theory with two dynamical energy scales, $\frac{m^2 \alpha^2}{{\cal P}^+}$ and
$\frac{m^2 \alpha}{{\cal P}^+}$, a fact known for a long time, and the reason
that QED
calculations have been so successful over the years.

Given the above limit [Eqs. (\ref{eq:limit2})--(\ref{eq:limit3})],
\bea
&\bullet&\theta\left(\lam^2-4|{\bf p}^2-{{\bf p}^\pr}^2|\right)
 , \theta\left(|\Delta_1|-\lam^2\right) ,
\theta\left(|\Delta_2|-\lam^2\right)\longrightarrow 1
\label{eq:w}\\
&\bullet&\theta\left(4({\bf p}^2+m^2)-\lam^2\right) ,
\theta\left(4 ({{\bf p}^\pr}^2+m^2)-\lam^2\right) \longrightarrow 1
\label{eq:ww}\\
&\bullet&
  \theta\left(\lam^2-|\Delta_1|\right) ,
\theta\left(\lam^2-|\Delta_2|\right)\longrightarrow 0\\
&\bullet&\theta\left(\lam^2-4({\bf p}^2+m^2)\right) ,
\theta\left(\lam^2-4({{\bf p}^\pr}^2+m^2)\right) \longrightarrow 0
\;.\label{eq:1888}
\eea
Now we proceed with the diagonalization of $V$
 in the degenerate spin space [see Eqs.~(\ref{eq:dirac}) and (\ref{eq:fermi})].
 In BSPT with $V$
 we will calculate all  corrections through order $\alpha^4$ that give rise to a
spin splitting structure in the ground state of ${\cal H}_0$.
First, we write the general $V$ more explicitly given the above limits
 in Eqs.~(\ref{eq:w})--(\ref{eq:1888}):
\bea
V({\bf p}^\pr, s_3,s_4; {\bf p}, s_1, s_2) &=&
\frac{1}{16 \pi^3} \frac{1}{2 m} \left[
1-\frac{{\bf p}^2+2 \kap_z^2+{{\bf p}^\pr}^2+2 {\kap_z^\pr}^2}
{4 m^2}+{\cal O}\left(\frac{{\bf p}^4}{m^4}
\right)
\right]\nonumber\\
&\times& \left(
-\frac{e^2 N_1}{DEN_4}-\frac{4 e^2}{(x-x^\pr)^2}\del_{s_1 s_3} \del_{s_2 s_4}
 +\frac{e^2 N_2}{DEN_5}+4 e^2 \del_{s_1 {\overline s}_2} \del_{s_3 {\overline
 s}_4}
\right)\nonumber\\
&-&\frac{1}{16 \pi^3}\frac{1}{2 m} V_{_{C}}
\;,\label{eq:4.70}
\eea
where
\bea
&\bullet& \frac{1}{DEN_4} \equiv \frac{\theta_{12}}{DEN_1}+
\frac{\theta_{21}}{DEN_2}\;\;,\;\theta_{12}\equiv
\theta\left(DEN_1-DEN_2\right)\\
&\bullet& \frac{1}{DEN_5}\equiv \frac{\theta\left({\cal M}_0^2-{{\cal
M}_0^\pr}^2\right)}{{\cal M}_0^2}
+\frac{\theta\left({{\cal M}_0^\pr}^2-{{\cal M}_0}^2\right)}{{{\cal M}_0^\pr}^2}
\;.
\eea
Note that we have expanded out the Jacobian factors according to
Eq.~(\ref{eq:jac}).
Also, $DEN_1$ and $DEN_2$ are defined below Eq.~(\ref{eq:76}) and written in
their expanded
version in Eqs.~(\ref{eq:ohno}) and (\ref{eq:ohno2}) respectively.
Finally, $N_1$ and $N_2$ are written below Eqs.~(\ref{eq:76}) and (\ref{eq:747})
respectively.

Since the
eigenstate wavefunctions of 
${\cal H}_0$  force 
${\bf p}$ to scale as ${\bf p} \sim m \alpha$, it is useful
to note the  $\alpha$-scaling 
of the matrix elements of $V$ in momentum space in a nonrelativistic expansion. Recalling  that 
we are always 
 assuming $\alpha \longrightarrow 0$
 (without which our matrix elements would 
 not have a well-defined $\alpha$ scaling), we see the following structure
 arising
\bea
V&=&V^{^{(0)}}+V^{^{(1)}} +V^{^{(2)}}+ \cdots\;,\label{eq:61897}
\eea
where a momentum space matrix element of $V^{^{(S)}}$ scales as $V^{^{(S)}}\!\!\!\sim\alpha^S$.
Thus in first-order BSPT these respective terms scale as
(this follows noting  $\int d^3p\left | \phi_N(p)\right |^2~=~1$ and 
 $V$ is not diagonal in
momentum space)
\bea
\del^{^{(1)}}M_{N N^\pr}^2 =
\langle \phi_N|V^{^{(S)}}|\phi_{N^\pr} \rangle 
 \sim \alpha^{3 + S}
\;.\label{eq:80}
\eea
To be consistent through order $\alpha^4$ in first-order BSPT 
we need to look at all the matrix 
 elements of
 $V^{^{(S)}}$ with $S \leq 1$;\footnote{
For example,
 $ \frac{e^2 {\bf p}}
 {({\bf p}-{\bf p}^\pr)^2} 
 \sim\frac{\alpha^2}{\alpha^2} 
 \Rightarrow S = 0\;$.} and in second-order BSPT, since $\left |{\cal M}_1^2-{\cal M}_n^2\right
 |\sim m^2 \al^2$, we need to consider $V^{^{(0)}}$ in second-order BSPT (since
 $\al^3\al^3/\al^2=\al^4$).
 
 A final discussion that we must have,
  before we  write out these expressions for $V^{^{(S)}}$, is how we are going
  to deal with $DEN_4$ and 
  $DEN_5$ defined above.\footnote{Actually the $DEN_5$ term is handled with
  analogous 
  techniques as the $DEN_4$ term, and has even smaller corrections than those of
  $DEN_4$.
   Thus, we will just discuss the $DEN_4$ term in what follows and here state
   the result
   for the $DEN_5$ term: Take $DEN_5 \longrightarrow 4 m^2$; 
   the corrections to this start shifting the bound state mass
   at order $\alpha^6$.}
  These denominators are dealt with by noting
  the following formulas
  \bea
\frac{\theta(a-b)}{a}+\frac{\theta(b-a)}{b}&=&
\frac{1}{2} \frac{\theta(a-b)+\theta(b-a)}{a}+
\frac{1}{2} \frac{\theta(a-b)+\theta(b-a)}{b} \nonumber\\
&+&\frac{1}{2} \frac{\theta(a-b)-\theta(b-a)}{a}-
\frac{1}{2} \frac{\theta(a-b)-\theta(b-a)}{b}\nonumber\\
&=&\frac{1}{2}\left( \frac{1}{a}+\frac{1}{b}\right)
+\frac{1}{2} \left[\theta(a-b)-\theta(b-a)\right]\left(
\frac{1}{a}-\frac{1}{b}\right)  \nonumber \\
&=&\frac{1}{2}\left( \frac{1}{a}+\frac{1}{b}\right)+\frac{1}{2} \frac{\left| a-b
\right|}
{a-b} \left(
\frac{1}{a}-\frac{1}{b}\right)\nonumber\\
&=&\frac{1}{2}\left( \frac{1}{a}+\frac{1}{b}\right)-\frac{1}{2} \frac{\left| a-b
\right|}
{a~b}
\;.\label{eq:seco}
  \eea
 To proceed it is useful to  note
 \bea
 && DEN_1= ({\bf p}-{\bf p}^\pr)^2
 -\frac{(\kap_z-\kap_z^\pr) ({\bf p}^2-{{\bf p}^\pr}^2)}{m}
 +{\cal O}\left[\left(\frac{{\bf p}}{m}\right)^5 m^2\right]~,\\
&& DEN_1=DEN_2-\frac{2 (\kap_z-\kap_z^\pr) ({\bf p}^2-{{\bf p}^\pr}^2)}{m}
 +{\cal O}\left[\left(\frac{{\bf p}}{m}\right)^5 m^2\right]~,\\
 &&\frac{1}{2}\left(\frac{1}{DEN_1}+\frac{1}{DEN_2}
 \right)=\frac{1}{\pp}+\nn\\
 &&~~~~~~~~~~~~~~~~~~~~~~~~~~~~~~~~~~+\frac{(\kap_z-\kap_z^\pr)^2({\bf p}^2-{{\bf
 p}^\pr}^2)^2}
 {m^2 ({\bf p}-{\bf p}^\pr)^6}
 +{\cal O}\left[\left(\frac{{\bf p}}{m}\right)^2\frac{1}{ m^2}\right]
 \;.\label{eq:rrr}
 \eea
 Especially note that this last equation scales as: $\frac{1}{\alpha^2}
 +1+\alpha^2+\cdots$, i.e. the corrections start at order $1$ (not order
 $\frac{1}{\alpha}$);
 this implies that only the $\frac{1}{\pp}$ term 
 of Eq.~(\ref{eq:rrr}) contributes to the spin splittings to 
 order $\alpha^4$. But we still have to discuss the second term that arises in
 Eq.~(\ref{eq:seco}).
 This term is given by 
 \be
 \left.\frac{1}{DEN_4}\right|_{_{second~term}} =
 -\frac{1}{2} \frac{\left| DEN_1-DEN_2 \right|}
{DEN_1 DEN_2}=-\frac{1}{2}\frac{\left|\frac{2 (\kap_z-\kap_z^\pr) ({\bf
p}^2-{{\bf p}^\pr}^2)}{m}\right|}
{\left({\bf p}-{\bf p}^\pr\right)^4}+{\cal O}\left(\alpha^{0}\right)
 \;.~~~~\label{eq:152}
 \ee
 Including $N_1$, this starts out as an ${\cal O}(\alpha^3)$ spin conserving
 contribution, which does not contribute to the splitting. 
 The next order contribution is ${\cal O}(\alpha^4)$ with spin
 structure,
 but is odd under ${\bf p} \leftrightarrow
 {\bf p}^\pr$, and thus integrates to zero in first-order BSPT (in second-order BSPT it
 contributes $\al^4 \al^4/\al^2 \sim \al^6$). However, the ${\cal O}(\alpha^3)$
 spin conserving term 
  appears to lead to an order $\alpha^4$ shift to the spin splittings
 in second-order BSPT when the cross terms with $V^{^{(0)}}$ of
 Eq.~(\ref{eq:rzero})
 are considered;\footnote{$V^{^{(0)}}$ of Eq.~(\ref{eq:rzero}) comes from the
 first
 term on the right-hand-side of Eq.~(\ref{eq:rrr}) combined with the complete
 next to leading order term of
 $N_1$.} however,
 these cross term contributions add to  zero due to the facts that
 the ${\cal O}(\alpha^3)$ term including Eq.~(\ref{eq:152}) conserves spin and
 the ${\cal O}(\alpha^3)$ term including Eq.~(\ref{eq:152}) is even while the
 term from
 Eq.~(\ref{eq:rzero}) is odd under ${\bf p} \leftrightarrow
 {\bf p}^\pr$. 
 
 To summarize the  preceding discussion
 of $DEN_4$ and $DEN_5$, we can say that through order $\alpha^4$, for the spin
 splittings
 of  positronium,
 there are no relativistic corrections to the following replacements: 
 \bea
 &&DEN_4\longrightarrow \pp~~{\rm and}~~DEN_5\longrightarrow 4 m^2
 \;.
 \eea
  This is valid for the ground and excited states, but in what
  follows we specialize to the ground state for simplicity.
 
 Given this general conclusion about $DEN_4$ and $DEN_5$, 
  we list the pieces of $V$ that contribute to positronium's ground state
  spin splittings
 through order $\alpha^4$.
  Explicitly, as far as the $\alpha$-scaling goes, we need to consider
 \bea
 V^{^{(0)}}({\bf p}^\pr s_3 s_4;
 {\bf p} s_1 s_2)&=& \frac{- c_{_{ex}} e^2}{4  \pi^3 ({\bf p}-{\bf p}^\pr)^2}
  { v}^{(0)}({\bf p}^\pr s_3 s_4;
 {\bf p} s_1 s_2)
 \;,\label{eq:rzero}
 \eea
 where
 \bea
 &&{ v}^{(0)}({\bf p}^\pr s_3 s_4;
 {\bf p} s_1 s_2)\equiv\left[\del_{s_1 {\overline s}_3} \del_{s_2 s_4}
 f_1({\bf p}^\pr s_3 s_4;
 {\bf p} s_1 s_2)+
 \del_{s_1 s_3} \del_{s_2 {\overline s}_4} f_2({\bf p}^\pr s_3 s_4;
 {\bf p} s_1 s_2)\right]\label{eq:132},~~~~~~~\\
 &&f_1({\bf p}^\pr s_3 s_4;
 {\bf p} s_1 s_2) \equiv
 s_1 (\kap_y-\kap_y^\pr)-i (\kap_x-\kap_x^\pr) ~,\\
 &&f_2({\bf p}^\pr s_3 s_4;
 {\bf p} s_1 s_2) \equiv s_4 (\kap_y-\kap_y^\pr)+i (\kap_x-\kap_x^\pr) 
 \;.
 \eea
 Recall that $s_i = \pm 1~(i=1,2,3,4)$ only. The only other interaction we need
 to consider is
 \bea
 V^{^{(1)}}({\bf p}^\pr s_3 s_4;
 {\bf p} s_1 s_2)&=& \frac{e^2}{4 m \pi^3}\left[c_{_{an}} \del_{s_1 s_2}
 \del_{s_1 s_4}
  \del_{s_3 s_4}+ c_{_{ex}} \del_{s_2 {\overline s}_4} \del_{s_2 {\overline
  s}_1}
  \del_{s_3 {\overline s}_1}+\frac{}{}\right.\nonumber\\
 &&~~~~+ \left.\left(c_{_{an}} \frac{1}{2} - c_{_{ex}}
 \frac{(\kap-\kap^\pr)^2}{({\bf p}-{\bf p}^\pr)^2}\right)
  \del_{s_1 {\overline s}_2} \del_{s_3 {\overline s}_4}\right]\;.\label{eq:61898}
 \eea
 The constants $c_{_{ex}}$ and $c_{_{an}}$
  were introduced only to distinguish the terms that arise from the `exchange'
  and 
 `annihilation' channels respectively, and $c_{_{ex}} = c_{_{an}} = 1$ (as given by the theory)
 will be
 used in the
 remainder of this calculation. 
 
 Two simplifications were made in 
 deriving  $V^{^{(1)}}$.  
   First,  
 we did not include terms that are a constant along the diagonal in
 spin space, because these do not contribute to the spin {\it splittings}.
  Second,
  we noted that
 terms of the following type integrate to zero
 \bea
&& \langle \phi_{_{1,0,0,s_3,s_4}}|\frac{e^2 (\kap_x \kap_y^\pr,\kap_z
\kap_x,\kap \times 
\kap^\pr)}{({\bf p}-
{\bf p}^\pr)^2}|\phi_{_{1,0,0,s_1,s_2}} \rangle=(0,0,0)
 \;,
 \eea
 and thus were not included in the definition of
 $V^{^{(1)}}$.

The ground state spin splitting through order $\alpha^4$ contains
 contributions from $V^{^{(1)}}$ in
first-order BSPT ($V^{^{(0)}}$ vanishes in first-order BSPT) and $V^{^{(0)}}$ 
in second-order BSPT.
We  begin with the first-order BSPT calculation.

 \noindent
 \underline{First-Order BSPT}:
 
The lowest order wavefunctions are given near the end of Subsection~\ref{4.2.1} (see
Appendix~\ref{app:hyper}
for the hyperspherical harmonics). $V^{^{(1)}}$ in first-order BSPT contributes
the following to positronium's ground state mass-squared
 \bea
\del M_{_{1}}^2&\equiv&
 \del^{^{(1)}}M^2(s_3,s_4;s_1,s_2)\nonumber\\
 &=& N \int d^3 p~ d^3 p^\pr \frac{1}{(e_1^2+{\bf p}^2)^2} 
 \frac{1}{(e_1^2+{{\bf p}^\pr}^2)^2} V^{^{(1)}}({\bf p}^\pr,s_3,s_4;{\bf
 p},s_1,s_2)
 \;,
 \label{eq:178}
 \eea
 where
 \bea
 N&=&\frac{8 e_1^5}{\pi^2}\;\;{\rm and}\;\;e_1 = \left.\frac{m \alpha}{2
 n}\right|_{n=1}
 \;.
 \eea
 Using the rotational symmetry of the integrand we can make the substitution
 \bea
 \frac{(\kap-\kap^\pr)^2}{\pp}&\longrightarrow&\frac{\frac{2}{3}\left[
 (\kap_x-\kap_x^\pr)^2+(\kap_y-\kap_y^\pr)^2+(\kap_z-\kap_z^\pr)^2\right]}{\pp}
 =\frac{2}{3}
 \;.
 \eea 
\begin{figure}[t]\vskip0in\hskip-.2in
\centerline{\epsfig{figure=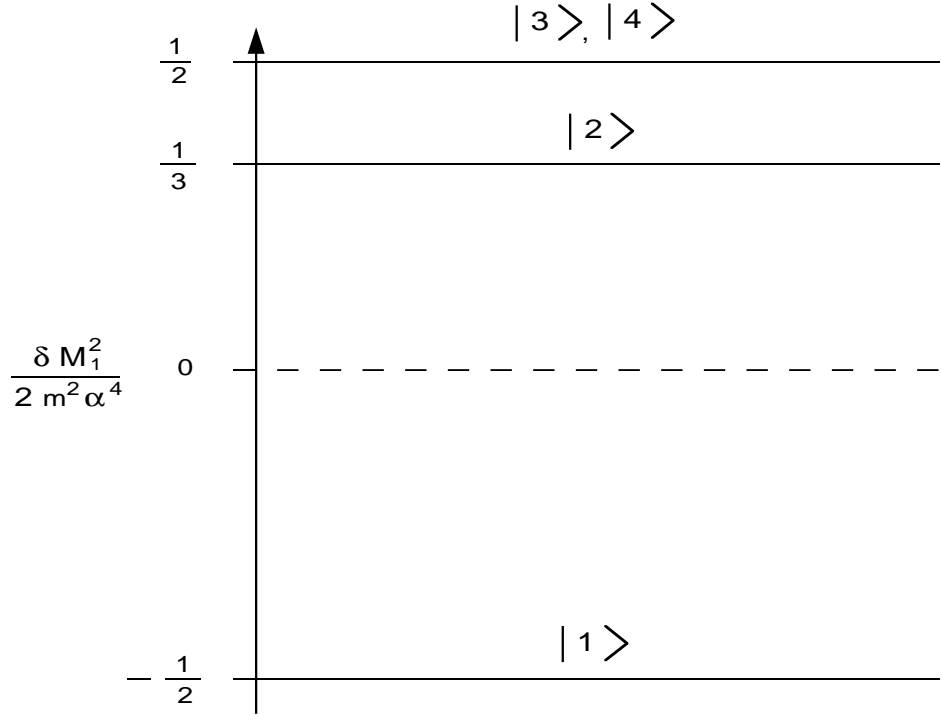,height=4in,width=5in}}
\caption{$\del M_{_{1}}^2$ is the part of the ground state spin splittings 
from first-order BSPT given by
Eq.~(\ref{eq:178}). $m$ is the electron mass and $\alpha$ is the fine structure constant.
The state labels 1, 2, 3 and 4 are explained in Eq.~(\ref{eq:187}). The two upper most
levels should coincide in a rotationally invariant theory; and after including
second-order BSPT, they do.} 
\label{posfig2}
\end{figure}
 After this,  the remaining integrals are trivial and the splittings that arise
 from diagonalization
 of the $\easy$ matrix in spin space are
 \bea
 \left\langle 1\left |\easy\right| 1 \right\rangle&=&- m^2 \alpha^4~,\\
  \left\langle 2\left |\easy\right| 2 \right\rangle&=& \frac{2}{3} m^2
  \alpha^4~,\\
  \left\langle 3\left |\easy\right|3 \right\rangle&=& m^2 \alpha^4~,\\
  \left\langle 4\left |\easy\right| 4 \right\rangle&=& m^2 \alpha^4
 \;,
 \eea
 where
 \bea
&&\left\{ |1\rangle\equiv\frac{|+-\rangle-|-+\rangle}{\sqrt{2}}\;,\;\right.\nn\\
&&\left.~~~~~~~~~
|2\rangle\equiv\frac{|+-\rangle+|-+\rangle}{\sqrt{2}}\;,\;
|3\rangle\equiv|--\rangle\;,\;
|4\rangle\equiv|++\rangle\right\}
 \;.\label{eq:187}
 \eea
Figure~\ref{posfig2} shows these results, which taken alone do not produce the degeneracies
required by
rotational invariance. 
 
Now we perform the second-order BSPT calculation. 
 
  \noindent
 \underline{Second-Order BSPT}:
 
 $V^{^{(0)}}$ gives rise to the following contribution 
 to positronium's ground state mass squared
 in second-order BSPT
 \bea
 \hard&\equiv&
\del^{^{(2)}}M^2(s_3,s_4;s_1,s_2)\nonumber\\
&=&\sum_{s_e,s_{{\overline e}}} {\sum_{\nu \neq (1,0,0)}}^{\!\!\!\!\!\!\!c}
\frac{\langle \phi_{1,0,0,s_3,s_4} | V^{^{(0)}} | \phi_{\nu,s_e,s_{{\overline
e}}}
\rangle \langle \phi_{\nu,s_e,s_{{\overline e}}}|  V^{^{(0)}}|
\phi_{1,0,0,s_1,s_2} \rangle}
{{\cal M}_1^2-{\cal M}_n^2}
  \;.
  \label{eq:happyman}
  \eea
  Recall that $\nu = (n,l,m_l)$, the usual principal and angular momentum
  quantum 
  numbers of nonrelativistic positronium (the ``$c$" on the sum emphasizes the fact that the
  continuum states must be
included also).
  The calculation of $\hard$ is tedious, but can be done analytically.
  This calculation is performed in
  Subsection~\ref{hard}. The result is [see Eq.~(\ref{eq:whatthehell})]
  \bea
 \hard &=& - \frac{m^2 \alpha^4}{24} (3 g_1+ g_2) 
 \;,\label{eq:4.96}
 \eea
where $g_1$ and $g_2$ are given  in Eqs.~(\ref{eq:g1}) and
(\ref{eq:g2}) respectively.
 
\begin{figure}[t!]\vskip0in\hskip-.2in
\centerline{\epsfig{figure=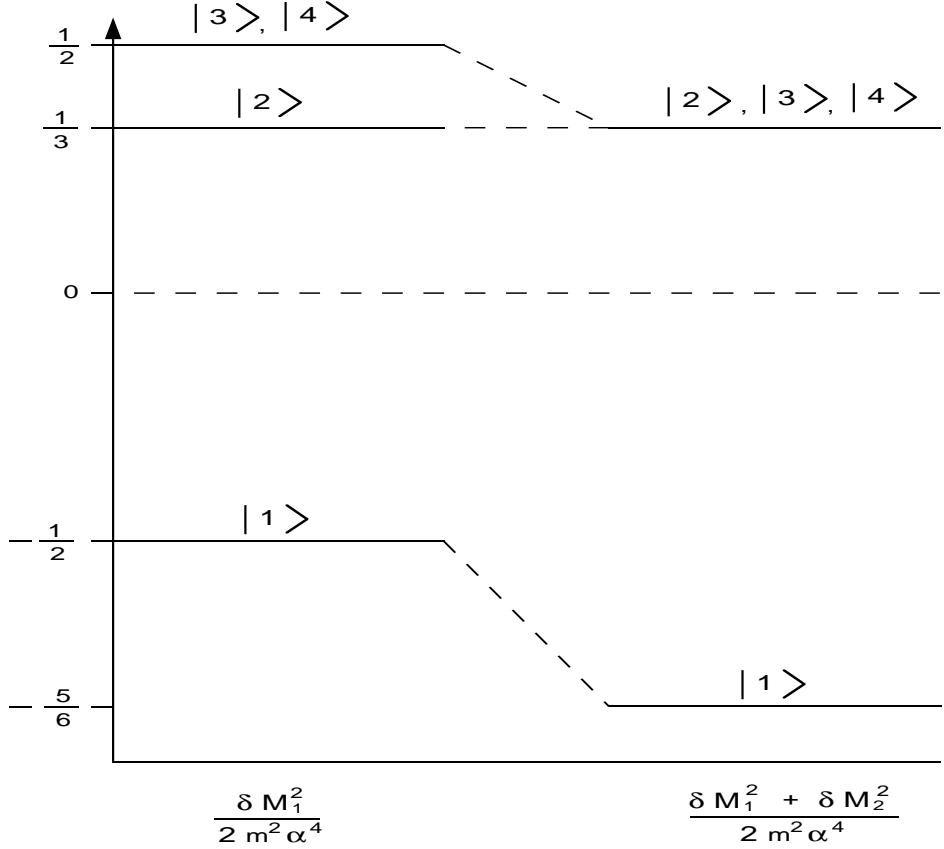,height=4.5in,width=5in}}
\caption{The combined ground state spin splitting from first- and
second-order BSPT in positronium 
through order $\alpha^4$ is illustrated 
using the same notation as in Figure~\ref{posfig2}. 
$\del M_{_{2}}^2$ is given by Eq.~(\ref{eq:happyman})
and is 
calculated in Subsection~\ref{hard}. The final combined result (on the right) corresponds
to a rotationally invariant theory.} 
\label{posfig3}
\end{figure}
 
 Now we  combine the $\easy$ and $\hard$ matrices and diagonalize the result.
 The combined matrix is
 given by
 \bea
 \frac{\easy+\hard}{2 m^2 \alpha^4}&=& \frac{1}{2} \del_{s_1 s_2} \del_{s_1 s_4}
 \del_{s_3 s_4} 
 -\frac{1}{12} \del_{s_1 {\overline s}_2} \del_{s_3 {\overline s}_4}+\frac{1}{2}
 \del_{s_2 {\overline s}_4} \del_{s_1 {\overline s}_2} \del_{s_1
  {\overline s}_3}\nonumber\\
  &&~~~~~-\frac{1}{48} (3 g_1+g_2)\;.\label{eq:combined}
 \eea
 The eigenvalues are
  \bea
 \left\langle 1 \left|\easy+\hard\right| 1 \right\rangle&=&- \frac{5}{3} m^2
 \alpha^4\label{eq:1776}~,\\
 \left\langle 2 \left|\easy+\hard\right| 2 \right\rangle&=& \frac{2}{3} m^2
 \alpha^4~,\\
  \left\langle 3 \left|\easy+\hard\right| 3 \right\rangle&=&  \frac{2}{3} m^2
  \alpha^4~,\\
 \left\langle 4 \left|\easy+\hard\right| 4 \right\rangle&=&  \frac{2}{3} m^2
 \alpha^4
 \;,\label{eq:1493}
 \eea
 and the corresponding eigenvectors are the same as in Eq.~(\ref{eq:187}).
  Figure~\ref{posfig3} displays these results. 
  
  These results translate to the well known result as detailed in Appendix~\ref{app:mvsb}:
 \bea
 B_{triplet}-B_{singlet}&=&\frac{7}{6} \alpha^2 Ryd+{\cal O}\left(m \alpha^5\right)
 \;.
 \eea
We see rotational invariance in the degeneracies of the 
ground state $n=1$ triplet
levels
exactly maintained through order $\alpha^4$.
\subsection{Calculation of $\hard$}\label{hard}

In this subsection we perform the following sum 
analytically
\bea
\hard&=&\sum_{s_e,s_{{\overline e}}} {\sum_{\nu \neq (1,0,0)}}^{\!\!\!\!\!\!\!c}
\frac{\langle \phi_{1,0,0,s_3,s_4} | V^{^{(0)}} | \phi_{\nu,s_e,s_{{\overline
e}}}
\rangle \langle \phi_{\nu,s_e,s_{{\overline e}}}| V^{^{(0)}}|
\phi_{1,0,0,s_1,s_2} \rangle}
{{\cal M}_1^2-{\cal M}_\nu^2}
  \;.
  \eea 
  Recall that for the bound states,
  $\nu=(n,l,m_l)$ the usual principal and angular momentum quantum
  numbers of 
  nonrelativistic positronium. We must also include the scattering states of 
  course. We do this with Green's function techniques as explained below.
 Recall that the spin factored completely out of our lowest-order
 Schr{\"o}dinger equation, so 
  the
 following notation is useful
 \bea
 |\phi_{\nu,s_e,s_{{\overline e}}}
\rangle&=& |\phi_\nu\rangle \otimes |s_e s_{{\overline e}}\rangle
 \;,\\
 1&=& 
{\sum_{s_e,s_{\overline e},\nu}}^{\!\!\!\!c} |\phi_\nu\rangle\langle\phi_\nu|
  \otimes |s_e s_{{\overline e}}\rangle\langle s_e s_{\overline e}|\nonumber\\
  &=&\sum_{s_e,s_{\overline e}} \int d^3 p ~|{\bf p}\rangle\langle {\bf p} |
  \otimes |s_e s_{\overline e}\rangle\langle s_e s_{\overline e}|
  \;.
 \eea
 To proceed, define the following Green's function for  arbitrary $E$
 \bea
 \frac{{G}_E}{4 m}&\equiv& {\sum_\nu}^c \frac{|\phi_\nu \rangle
 \langle \phi_\nu |}{E-{\cal M}_\nu^2}
 \;.
 \eea
 The factor $\frac{1}{4 m}$ will turn out to be useful. This Green's function 
  satisfies
 the familiar Coulomb Green's function equation
 \bea
 \del^3({\bf p}-{\bf p}^\pr) &=&
 (\tilde{E}-\frac{{{\bf p}^\pr}^2}{m}) G_E({\bf p}^\pr,{\bf p})+
 \frac{\alpha}{2 \pi^2} \int d^3 p^{\pr \pr} \frac{G_E({\bf p}^{\pr \pr},{\bf
 p})}
 {({\bf p}^\pr-{\bf p}^{\pr \pr})^2}
 \;,
 \eea
 where
 \bea
 &&\langle {\bf p}^\pr | {G}_E |{\bf p}\rangle\equiv G_E({\bf p}^\pr,{\bf
 p})~,\\
 {\rm and}&&\;\tilde{E}\equiv\frac{E-4 m^2}{4 m}\;.
 \eea
  Hostler and Schwinger
   independently obtained the solution for this Coulomb Green's function 
 in 1964~\cite{schwinger}. We find Schwinger's form useful;
   the equation  he solves is exactly the above equation 
  with the following shifts in notation
 \bea
 \left(Z e^2\right)_{Schwinger} &\longrightarrow& \alpha~,\\
 m_{Schwinger}&\longrightarrow& \frac{m}{2}~,\\
 E_{Schwinger}&\longrightarrow& \tilde{E}
 \;.
 \eea
His result is amended because the  sum we need has  $E =
{\cal M}_1^2=4m^2-4e_1^2$ ($e_1=m\al/2$ recall) and does not include $\nu = (1,0,0)$,
as dictated by the usual rules of second-order BSPT. 
This subtraction of the $\nu = (1,0,0)$ term amounts to the term
``$-\frac{1}{C}$" in $G_{III}$
below. The details of how this arises can be seen in
Eqs.~(\ref{eq:billy1})--(\ref{eq:billy2}) below.
With
this amendment, Schwinger's result is
\bea
G_{{\cal M}_1^2}^\pr({\bf p},{\bf
p}^\pr)&\equiv&G_I+G_{II}+G_{III}~,\label{eq:G}\\
G_I&=&\frac{\del^3({\bf p}-{\bf p}^\pr)}{\tilde{E}-T}~,\\
G_{II}&=& -\frac{\alpha}{2 \pi^2} \frac{1}{\tilde{E}-T}\frac{1}{\pp}
\frac{1}{\tilde{E}-T^\pr}~,\\
G_{III}&=&-\frac{\alpha}{2 \pi^2} \frac{4 e_1^2}{\tilde{E}-T}
\left[\int_{0}^{1}\frac{ d \rho }{\rho} \left(
\frac{1}{4 e_1^2 \rho \pp+C (1-\rho)^2}\right.\right.\nn\\
&&\left.\left.~~~~~~~~~~~~~~~~~~~~~~~~~~~~~~~~~-
\frac{1}{C}
\right)
\right]\frac{1}{\tilde{E}-T^\pr}~,
\eea
where
\bea
 T &=& \frac {{\bf p}^2}{m}\;,\;T^\pr =\frac{{{\bf p}^\pr}^2}{m}\;,\;e_1=\frac{m
 \alpha}{2}~,\\
C&=&(e_1^2+{\bf p}^2)(e_1^2+{{\bf p}^\pr}^2)~,\\
\tilde{E}&=&\frac{{\cal M}_1^2-4 m^2}{4 m}=-\frac{e_1^2}{m}
\;.
\eea
The prime on $G_{{\cal M}_1^2}^\pr$ denotes the fact that we
 have subtracted  the $\nu = (1,0,0)$ part of $G_{{\cal M}_1^2}$ as required by the 
usual rules of second-order BSPT. Note that this Green's function is symmetric under
${\bf p} 
\leftrightarrow {\bf p}^\pr$ and also $(p_x,p_x^\pr) \leftrightarrow
(p_y,p_y^\pr)$, symmetries
that will be used in later simplifications of the integrand of $\hard$.

 $\hard$  is now
\bea
\hard&=& \sum_{s_e,s_{{\overline e}}} \int d^3 p\, d^3 k \,d^3 p^\pr \,d^3 k^\pr\,
\langle \phi_{1,0,0} | {\bf k} \rangle V^{^{(0)}} ({\bf k},s_3,s_4; {\bf
p},s_e,s_{{\overline e}})
\nonumber\\
&&~~~~\times\left(\frac{G_I+G_{II}+G_{III}}{4 m}\right) V^{^{(0)}} ({\bf
p}^\pr,s_e,s_{{\overline e}}; 
{\bf k}^\pr,s_1,s_2)
\langle {\bf k}^\pr | \phi_{1,0,0} \rangle \\
&\equiv&\hard(I)+\hard(II)+\hard(III)\;\;~~{\rm respectively}
\;.
\eea
Now we rewrite this in terms of hyperspherical harmonics and perform the
integrations analytically.  The variables are defined as
 \bea
 &\bullet& \overbrace{[u\equiv(u_o,{\bf u})]}^{\Omega_p}\leftrightarrow[e_1,{\bf
 p}]\;,\;
 \overbrace{[u^\pr\equiv(u_o^\pr,{\bf
 u}^\pr)]}^{\Omega_{p^\pr}}\leftrightarrow[e_1,{\bf p}^\pr]
 \label{eq:333}\\
  &\bullet& \underbrace{[v\equiv(v_o,{\bf
  v})]}_{\Omega_k}\leftrightarrow[e_1,{\bf k}]\;,\;
 \underbrace{[v^\pr\equiv(v_o^\pr,{\bf
 v}^\pr)]}_{\Omega_{k^\pr}}\leftrightarrow[e_1,{\bf k}^\pr]
 \;.
 \label{eq:444}
 \eea
 See Appendix~\ref{app:hyper} for a summary of the
 mathematical relations we use. The symbols appearing in
 Eqs.~(\ref{eq:333}) and (\ref{eq:444})
 are explained there.
 Note that we use $e_1$ in these variable definitions, a choice that is
 completely general
 and turns out to be useful because we are taking expectation values of $n=1$
 states ($e_n=\frac{m\al}{2n}$ recall).
 The relations  we use are
\bea
&\bullet&\langle{\bf k}^\pr|\phi_{1,0,0}\rangle=\frac{4 e_1^{\frac{5}{2}}}
{(e_1^2+{{\bf k}^\pr}^2)^2} \frac{1}{\sqrt{2 \pi^2}}\\
&\bullet&\frac{1}{(e_1^2+{{\bf k}^\pr}^2)^2} = \frac{(1+v_o^\pr)^2}{4 e_1^4}\\
&\bullet& d^3 k^\pr=\frac{(e_1^2+{{\bf k}^\pr}^2)^3}{8 e_1^3} d \Omega_{k^\pr}=
\frac{e_1^3}{(1+v_o^\pr)^3} d \Omega_{k^\pr}
\;.
\eea
Given these,  $\hard$ becomes
\bea
\hard&=& -\frac{m^3 \alpha^5}{32 \pi^2}\int 
\frac{d\Omega_p d\Omega_{p^\pr} d\Omega_k d\Omega_{k^\pr} }
{(1+u_o)(1+u_o^\pr)^2} 
 \left[(\tilde{E}-T)
(G_I+G_{II}+G_{III}) \right]\nonumber\\
&&~~~~~\times{\cal S}\sum_{\nu \nu^\pr} \frac{1}{n n^\pr}Y_\nu(\Omega_p)
Y_{\nu^\pr}(\Omega_{p^\pr}) 
Y_\nu^\ast(\Omega_k) Y_{\nu^\pr}^\ast(\Omega_{k^\pr})
\;,
\eea
where
\bea
{\cal S}&\equiv&\sum_{s_e s_{{\overline e}}} v^{(0)}({\bf k},s_3,s_4; {\bf
p},s_e,s_{{\overline e}})
v^{(0)}({\bf p}^\pr,s_e,s_{{\overline e}}; 
{\bf k}^\pr,s_1,s_2)
\;.
\eea
Recall Eq.~(\ref{eq:132}) for the definition of $v^{(0)}$.
Using the symmetries of the integrand, the sum over spins
$s_e$ and $s_{{\overline e}}$ can be performed and a  simplification
is seen to arise. The spin completely factors out of the momenta integrations.
In other words, 
 we have
\bea
{\cal S}&=& \frac{1}{6} (3 g_1+ g_2) ({\bf p} \cdot {\bf p}^\pr+{\bf k} \cdot
{\bf k}^\pr
-2 {\bf p} \cdot {\bf k}^\pr)
\;,
\label{eq:spring}
\eea 
where
\bea
g_1&\equiv& s_1 s_3+s_2 s_4~,
\label{eq:g1}\\
g_2&\equiv& 1+s_1 s_2-s_2 s_3-s_1 s_4+s_3 s_4+s_1 s_2 s_3 s_4
\;.\label{eq:g2}
\eea
Recall that $s_i = \pm 1$, $(i=1,2,3,4)$; i.e., the `$\frac{1}{2}$' has been
factored out of these 
spins.\footnote{
In order to get these simple forms for $g_1$ and $g_2$ and to see this
spin/momentum decoupling
it was useful to note the following simple relation: 
$\del_{s s^\pr} = \frac{1}{2} s (s+s^\pr)$ 
 [true because $s^2=1$].} So, in other 
words, instead of having to do sixteen twelve dimensional integrals because the
spin and
momenta are coupled together, we just have to do one 
twelve dimensional integral that is independent of spin and then diagonalize the
result in the 
$4 \times 4$ dimensional spin space with the spin dependence given by
Eq.~(\ref{eq:spring}).

We define the following integral
\bea
\chi&\equiv&\frac{m \alpha}{8 \pi^2} \xi
\;,
\eea
where
\bea
\xi&\equiv&\int 
\frac{d\Omega_p d\Omega_{p^\pr} d\Omega_k d\Omega_{k^\pr} }
{(1+u_o)(1+u_o^\pr)^2} 
 \left[(\tilde{E}-T)
(G_I+G_{II}+G_{III}) \right]\nonumber\\
&&\times(\underbrace{{\bf p} \cdot {\bf p}^\pr}_{a}+\underbrace{{\bf k}
\cdot {\bf k}^\pr}_{b}
\underbrace{-2 {\bf p} \cdot {\bf k}^\pr}_{c})\sum_{\nu \nu^\pr}
 \frac{1}{n n^\pr}Y_\nu(\Omega_p) Y_{\nu^\pr}(\Omega_{p^\pr}) 
Y_\nu^\ast(\Omega_k) Y_{\nu^\pr}^\ast(\Omega_{k^\pr})
\;,~~
\eea
and
\bea
\hard &=& - \frac{m^2 \alpha^4}{24} (3 g_1+ g_2) \chi
\;.
\label{eq:yoyo}
\eea
For the quantities 
$\xi$, $\chi$ and $\hard$, the  labels $I$, $II$ and $III$ imply
the respective terms with 
$G_I$, $G_{II}$ and  $G_{III}$ above  [see Eq.~(\ref{eq:G})]. Also,  the
 terms $a$, $b$ and $c$ above  correspond to the respective superscripts in what
 follows.
This integration will now be performed analytically. 

First the three $G_I$ pieces.  The mathematical relations used here are
\bea
\del^3(p-p^\pr)&=&\frac{8 e_1^3}{(e_1^2+{\bf p}^2)^3}
 \del(\Omega_p-\Omega_{p^\pr})=\frac{(1+u_o)^3}{e_1^3} \del(\p-\ppr)~, \\
 {\bf p}^2&=&\frac{e_1^2}{1+u_o} (1-u_o)~,\\
 {\bf k} \cdot {\bf k}^\pr &\longrightarrow& 3 k_z k_{z}^\pr~,\\
{\bf p} \cdot {\bf k}^\pr &\longrightarrow& 3 p_z k_{z}^\pr
\;.
\eea
Note that these last two relations are possible due to the rotational symmetry
of the integrand.  Then
we expand these z-components of momenta upon the hyperspherical harmonic basis
using the following
simple relation (e.g.  the $p_z$ case)
\bea
p_z&=&\frac{e_1}{1+u_o}\left( \frac{\pi i}{\sqrt{2}} Y_{2,1,0} (\p)\right)
\;.\label{eq:252}
\eea
Now we recall the form of the
hyperspherical harmonics (see the appendix on hyperspherical
harmonics for details), and their orthonormality and phase relationships
\bea
Y_\nu(\Omega)&\equiv&Y_{n,l,m}(\Omega)\equiv f_{n,l}(\omega)
Y_{l,m}(\theta,\phi)~,\\
Y_{n,l,m}&=& (-1)^{l+m} Y_{n,l,-m}^\ast\;,\;Y_{l,m}=(-1)^m
Y_{l,-m}^\ast\;,\;f_{n,l}=
(-1)^lf_{n,l}^\ast ~,\\
d\Omega^{^{(4)}}&\equiv& d\Omega \equiv  d\Omega^{^{(3)}} d\omega \sin^2\omega
~,\\
\int d\Omega Y_\nu^\ast Y_{\nu^\pr}&=&\del_{\nu \nu^\pr}\;,\;\int
d\omega\sin^2\omega f_{n,l}^\ast
 f_{n^\pr,l}=\del_{n n^\pr}\;,\;\nn\\
 &&~~~~~~~~~~~~~\int d\Omega^{^{(3)}} Y_{l,m}^\ast
 Y_{l^\pr,m^\pr}=
 \del_{l l^\pr} \del_{m m^\pr}
\;.
\eea
After straight-forward application of these relations we obtain
\bea
&\bullet&\xi_I^a=\frac{4 \pi}{e_1} \int_0^\pi d\omega \sin^2\omega 
\frac{(1-\cos\omega)}{(1+\cos\omega)}=\frac{6 \pi^2}{e_1}
\\&\bullet&\xi_I^b=\frac{3 \pi^2}{2 e_1}\sum_{n=2}^\infty \frac{1}{n^2}
\pig
\\&\bullet&\xi_I^c=-\frac{3 \pi^2}{ e_1}\sum_{n=2}^\infty \frac{1}{n}
\pig
\;.
\eea

For the $G_{II}$ terms, we
 use the following  relations
\bea
\frac{1}{\tilde{E}-T^\pr}&=&\frac{1}{-\frac{e_1^2}{m}-\frac{{{\bf
p}^\pr}^2}{m}}=
-\frac{m}{2 e_1^2} (1+u_o^\pr)~,\\
\frac{1}{\pp}&=&\frac{(1+u_o)(1+u_o^\pr)}{e_1^2}\sum_\nu\frac{2
\pi^2}{n}Y_\nu(\p) Y_\nu^\ast (\ppr)
\;.
\eea
These give
\bea
(\tilde{E}-T) G_{II}&=&\frac{\alpha m}{2 e_1^4} (1+u_o^\pr)^2(1+u_o)\sum_\nu
 \frac{1}{n} Y_\nu(\p) Y_\nu^\ast(\ppr)
\;.
\eea
We use the rotational symmetry of the integrand and expand the integrand on the
hyperspherical harmonic
basis as was done for the three $G_I$ terms. Then we have
\bea
&\bullet&
\xi_{II}^a = \frac{3 \pi^2}{2 e_1}\sum_{n=2}^\infty\frac{1}{n}\pig
\\&\bullet&
\xi_{II}^b = \frac{3 \pi^2}{2 e_1}\sum_{n=2}^\infty\frac{1}{n^3}\pig
\\&\bullet&
\xi_{II}^c = -\frac{3 \pi^2}{ e_1}\sum_{n=2}^\infty\frac{1}{n^2}\pig
\;.
\eea

 For the three $G_{III}$ terms we use the same relations used for the three 
$G_{II}$ terms, and we  use the rotational symmetry of the integrand to rewrite
the appropriate pieces
of the integrand in terms
of $Y_{2,1,0}$ as we did for the $G_I$ and $G_{II}$ terms. 
However,  we need to discuss one additional relation that allows the 
remaining $\hard(III)$
calculation to
be done analytically. In Schwinger's 1964 paper~\cite{schwinger}
he gives the following formula
\bea
\frac{1}{2 \pi^2} \frac{1}{(1-\rho)^2+\rho(u-u^\pr)^2}&=&
\sum_{n=1}^\infty \rho^{n-1} 
\frac{1}{ n} \sum_{l,m} Y_{n,l,m}(\Omega) Y_{n,l,m}^\ast (\Omega^\pr)
\;,\label{eq:billy1}
\eea
where $u$ and $u^\pr$ are of unit length and $0 < \rho 
< 1$.\footnote{This is easily derivable from a more general standard
formula that 
Schwinger gives
\beaa
\frac{1}{4 \pi^2}\frac{1}{(u-u^\pr)^2}
=\sum_{n=1}^\infty \frac{\rho_{<}^{n-1}}{\rho_{>}^{n+1}} 
\frac{1}{2 n} \sum_{l,m} Y_{n,l,m}(\Omega) Y_{n,l,m}^\ast (\Omega^\pr)
\;.\eeaa}
Inside the brackets in $G_{III}$ we have
\bea
&&\left[\int_{0}^{1}\frac{ d \rho }{\rho} \left(
\frac{1}{4 e_1^2 \rho \pp+C (1-\rho)^2}-
\frac{1}{C}
\right)
\right] \nn\\
&&~~~~~~~~~~~~~~~~~= \left[\int_{0}^{1}\frac{ d \rho}
{\rho}\frac{1}{C}\left(\frac{1}{{(1-\rho)^2+
\rho(u-u^\pr)^2}}-1\right)\right]
\;.\label{eq:247}
\eea
Recall $C\equiv (e_1^2+{\bf p}^2)(e_1^2+{{\bf p}^\pr}^2)$.
Also recall that we are using the coordinate change of Eqs.~(\ref{eq:333}) and
(\ref{eq:444}).
Eq.~(\ref{eq:1492}) with $e_n = e_1$ then applies and was used.
In  Eq.~(\ref{eq:247}), $0 < \rho < 1$ and $u$ and $u^\pr$ are of unit length,
thus 
 Schwinger's formula [Eq.~(\ref{eq:billy1})]
can be used and we have
\bea
(\tilde{E}-T) G_{III}&=&\frac{\alpha m (1+u_o)(1+u_o^\pr)^2}{2 e_1^4}
\nn\\
&&~~~~~\times\int_0^1 d\rho\sum_{\nu \neq (1,0,0)} 
\frac{\rho^{n-2}}{n}Y_\nu(\p) Y_\nu^\ast
(\ppr)\;.
\label{eq:billy2}
\eea
Now, since $n \geq 2$ in this sum we can do the integral over $\rho$ easily,
\bea
\int_0^1 d\rho \rho^{n-2}=\left.\frac{\rho^{n-1}}{n-1}\right|^1_0 =
\frac{1}{n-1}
\;,
\eea
 and we obtain
\bea
(\tilde{E}-T) G_{III}&=&\frac{\alpha m (1+u_o)(1+u_o^\pr)^2}{2 e_1^4}\nn\\
&&~~~~~\times
 \sum_{\nu \neq (1,0,0)} \frac{1}{n (n-1)}Y_\nu(\p) Y_\nu^\ast (\ppr)\;.
\eea
For terms in $\xi$ which contain $G_{III}$, one obtains
\bea
&\bullet&
\xi_{III}^a = \frac{3 \pi^2}{2 e_1}\sum_{n=2}^\infty\frac{1}{n(n-1)}\pig
\\&\bullet&
\xi_{III}^b = \frac{3 \pi^2}{2 e_1}\sum_{n=2}^\infty\frac{1}{n^3 (n-1)}\pig
\\&\bullet&
\xi_{III}^c = -\frac{3 \pi^2}{ e_1}\sum_{n=2}^\infty\frac{1}{n^2 (n-1)}\pig
\;.
\eea

Now recall  $\chi\equiv\frac{m \alpha}{8 \pi^2} \xi$, and
 also notice  that all the summands are the same, thus putting it all together
 we have
 \bea
 \chi&=&\frac{3}{2}+\frac{3}{8}\sum_{n=2}^\infty
 \left(
 \frac{1}{n}+\frac{1}{n (n-1)}+\frac{1}{n^2}
 +\frac{1}{n^3}+\frac{1}{n^3 (n-1)}\right.\nonumber\\
 &&~~~~~- \left.\frac{2}{n}
 -\frac{2}{n^2}-\frac{2}{n^2 (n-1)}
\right) \pig\\
&=&\frac{3}{2}-\frac{3}{8}\sum_{n=2}^\infty
 \left(
 \frac{1}{n}
\right) \pig
 \;.
 \label{eq:kkk}
 \eea
 The remaining sum can be done analytically.
 To see this, first define two integrals
 \bea
 I_1&\equiv&\int \frac{d\Omega_p}{1+u_o} \frac{d\Omega_{p^\pr}}{1+u_o^\pr}=
 (4 \pi)^2\left(\int_0^\pi \frac{d\omega \sin^2 \omega}{1+\cos\omega}\right)^2=
 16 \pi^4~,\label{eq:eeooe}\\
 I_2&\equiv&\int \frac{d\Omega_p}{1+u_o} \frac{d\Omega_{p^\pr}}{1+u_o^\pr} 
 \frac{(\kap-\kap^\pr)^2}{\pp}=
 \frac{2}{3}I_1
 \;.\label{eq:sss}
 \eea
 The last equality followed from rotational symmetry of the integrand.
 Thus $I_2=32 \pi^4/3$.
 We can also calculate $I_2$ a hard way which gives\footnote{
 We use  $ \kap_x\equiv p_x=\frac{e_1}{1+u_o} \frac{\pi i}{2}
 \left[Y_{2,1,-1}(\Omega_p)-Y_{2,1,1}(\Omega_p)\right]$, 
 $\kap_y\equiv p_y=-\frac{e_1}{1+u_o}
 \frac{\pi}{2}\left[Y_{2,1,-1}(\Omega_p)+\right.$\linebreak$\left.
 Y_{2,1,1}(\Omega_p)\right]$, Eq.~(\ref{eq:1492}) and Eq.~(\ref{eq:1900}), 
  explicitly write the integrand in
  Eq.~(\ref{eq:sss}) out, and then use the phase and orthonormality
  relations of the hyperspherical harmonics to simplify the remaining expression.}
 \bea
 I_2&=&16 \pi^4-4\pi^4 \sum_{n=2}^\infty\frac{1}{n} \pig
 \;.
 \eea
 Thus, we have
 \bea
&&\sum_{n=2}^\infty\frac{1}{n} \pig=\frac{4}{3}
 \;.
 \eea
 
  Combining this result with Eq.~(\ref{eq:kkk}) gives
\bea
 \chi&=&\frac{3}{2}-\frac{3}{8}\left( \frac{4}{3}\right)=1
 \;.
 \eea
 Thus, recalling Eq.~(\ref{eq:yoyo}), we have
 \bea
 \hard &=& - \frac{m^2 \alpha^4}{24} (3 g_1+ g_2) 
 \;,
 \label{eq:whatthehell}
 \eea
 where $g_1$ and $g_2$ are given in Eqs.~(\ref{eq:g1}) and (\ref{eq:g2})
 respectively. This is the promised result written in Eq.~(\ref{eq:4.96}).
\section{Singlet-triplet splitting: A simpler method}

Perhaps the most straightforward approach to calculate the singlet-triplet
splitting is to just get busy and calculate,
since the nonrelativistic Coulomb spectrum and states are so well known. This is
exactly what is done in the previous section; however, as seen by the complexity
of Subsection~\ref{hard}, the calculation is complicated and at the
level of a ``Lamb shift calculation." Here we present a simpler method
to calculate this shift.\footnote{The idea behind
  this simpler method originated
  with Brisudov\'{a} and Perry \cite{m1}. To the consistent order in momenta required
  for fine structure,
  their transformation is equivalent to a Melosh rotation \cite{melosh}.} 
  This simpler method uses a unitary transformation
to ``remove" ${{ V}}^{^{(0)}}$ much in the spirit of Schwinger's early QED
calculations \cite{schwinger2}. 

First, we set up the notation. The exact eigenvalue equation is
\be
\left(H_0+V\right)|\Phi_N\rangle = M_N^2 |\Phi_N\rangle
\;,\label{eq:sadd}
\ee
where $M_N$ is the mass of the state and
\bea
&\bullet& \langle \Phi_N |\Phi_{N^\pr}\rangle = \del_{N N^\pr}\\
&\bullet& 1 = \sum_{s_1 s_2} \int d^3p \;|{\bf p} s_1 s_2 \rangle \langle {\bf p} s_1 s_2 | =
\sum_{s_1 s_2} \int d^3x\; |{\bf x} s_1 s_2 \rangle \langle {\bf x} s_1 s_2 |\nn\\
&&~~~~~~~~~~~~~~~~~~~~~~~~~~~~~~~~~~~~~~~~~~~~~~=
\sum_{N} |\Phi_N \rangle \langle \Phi_N|\\
&\bullet& \langle {\bf p}^\pr s_3 s_4 | {{ V}} 
| {\bf p} s_1 s_2 \rangle = { V}({\bf p}^\pr s_3 s_4;
 {\bf p} s_1 s_2)\;\;\\
&\bullet&\langle {\bf p}^\pr s_3 s_4 | {{ H}}_0 | {\bf p} s_1 s_2 \rangle =
4 (m^2+{\bf p}^2)\del^3(p-p^\pr) \del_{s_1 s_3} \del_{s_2 s_4}\nn\\
&&~~~~~~~~~~~~~~~~~~~~~~~~~~~~~~~~ -(4 m) \frac{\alpha}{2 \pi^2}
\frac{ \del_{s_1s_3}\del_{s_2s_4}
  }{({\bf p}-{\bf p}^\pr)^2}\\
&\bullet& M_N^2= (2 m+B_N)^2
\;\label{eq:101}.
\eea
$m$ is the electron mass,
$-B_N$ is the binding energy, and
$N$ labels all the quantum numbers of the state. For notational purposes note that
we label the final relative three-momentum with a prime, and that the initial and final
electrons
are labeled by ``1"  and ``3" respectively, and the initial and final positrons are labeled by
``2" and ``4" respectively. $V$ is given by Eq.~(\ref{eq:4.70}), 
and its leading order and
next to leading order matrix elements are given in Eqs.~(\ref{eq:rzero}) and 
(\ref{eq:61898}) respectively.
In zeroth order ${{ V}}$ is neglected and Eq.(\ref{eq:sadd}) becomes
\bea
&&{{ H}}_0
 |\phi_N\rangle = {\cal M}_N^2 |\phi_N\rangle = \left(
 4 m^2 + 4 m {\cal B}_N\right) |\phi_N\rangle \label{eq:zero}~.
 \eea
 This last equality defines our zeroth order binding energy, $-{\cal B}_N$.
 Projecting this eigenvalue equation into momentum space gives
 \bea
 \left(-{\cal B}_N+\frac{{{\bf p}^\pr}^2}{m}
 \right) \phi_N({\bf p}^\pr  s_3 s_4)&=&
 \frac{\alpha}{2 \pi^2}\int\frac{ d^3 p }{({\bf p}-{\bf p}^\pr)^2}
  \phi_N({\bf p} s_3 s_4)
 \;,\label{eq:C2}
\eea
the familiar nonrelativistic Schr{\"o}dinger equation of positronium.

After the simplification discussed below, 
as in the Coulomb gauge equal-time calculation, to obtain
the ground state singlet-triplet 
splitting through order $\al^4$, 
only the wave function at the origin is required, which we
thus record
\bea
\left[\phi_N\left({\bf x}=0\right)\right]^2&=& \frac{1}{(2 \pi)^3}
\left(\int d^3 p ~\phi_N\left({\bf p}\right)\right)^2=\frac{1}{\pi} \left(
\frac{m \alpha}{2 n}\right)^3 \del_{l,0}
\;.\label{eq:hy}
\eea
$n$ is the principal quantum number, and $l$ is the orbital angular momentum quantum number.
This sets up our notation and we  proceed with the simpler method.

The simpler method begins by acting on the Hamiltonian
 with a general unitary transformation with hermitian generator ${Q}$:
\bea
  {H}&=& {{ H}}_0+{{ V}}^{^{(0)}}+
 {{ V}}^{^{(1)}}+{{ V}}^{^{(2)}}+\cdots~,\\
 {H}^\pr&=& e^{i{Q}} {H} e^{-i{Q}}\nonumber\\
 &=&{H}+i\left[{Q}, {H}
 \right]+\frac{i^2}{2!}\left[{Q},\left[{Q},{H}\right]
 \right]+\cdots \;.\label{eq:hprime}
 \eea
 Now define ${Q}$ by requiring its commutator with ${{ H}}_0$ to cancel 
 ${{ V}}^{^{(0)}}$:
 \bea
 {{ V}}^{^{(0)}}+ i \left[
 {Q},{{ H}}_0
 \right]&\equiv&0
 \;.
 \label{eq:elimination}
 \eea
 Putting this into Eq.~(\ref{eq:hprime}) gives
  \bea
  {H}^\pr&=& {{ H}}_0+\left(1-\frac{1}{2!}\right)
   \left[ i {Q},{{ V}}^{^{(0)}}\right]+
  e^{i{Q}} \left({{ V}}^{^{(1)}}+{{ V}}^{^{(2)}}+\cdots\right)
  e^{-i{Q}}
  \nonumber\\
 && ~~~~~~~~~+\left\{\left(\frac{1}{2!}-\frac{1}{3!}\right)\left[i {Q},
 \left[i {Q},{{ V}}^{^{(0)}}\right]
 \right]\right.\nn\\
 &&\left.~~~~~~~~~~~~~~~~~~~~~~+
 \left(\frac{1}{3!}-\frac{1}{4!}\right) 
 \left[ i {Q},\left[ i  {Q},\left[ i {Q}, {{ V}}^{^{(0)}}\right]
 \right]\right]
 +\cdots\right\}
  \;.
  \eea
  Note that ${H}$ and ${H}^\pr$  have equivalent lowest order spectrums
  given by
  ${{ H}}_0$ (this can be seen easily by 
  looking at matrix elements of the equations in Coulomb states,
  that is in states of ${{ H}}_0$). 
  However, the corrections to $H_0$ in $H$ start at order $\al^3$, whereas they start
  at order $\al^4$ in ${H}^\pr$.
  To summarize,  we must diagonalize the following
  matrix element in spin space to consistently obtain the   
  ground state singlet-triplet splitting in positronium through order $\alpha^4$:
  \be
 \langle\phi_{1,0,0,s_3,s_4}|
 {{ V}}^{^{(1)}}+\frac{1}{2} \left[ i {Q},{{ V}}^{^{(0)}}\right]
 |\phi_{1,0,0,s_1,s_2}\rangle
 \;,\label{eq:FFF}
 \ee
 where ${Q}$ is a solution to Eq.~(\ref{eq:elimination}).
 Note that this is a {\it first} order
 bound-state perturbation theory shift.
 The quantum numbers are 
 $N=(n,l,m_l,s_e,s_{\overline{e}})\longrightarrow (1,0,0,s_e,s_{\overline{e}})$
 for the ground state.
 
 In what follows we will solve Eq.~(\ref{eq:elimination}) for ${Q}$ in the free
 basis in momentum space,\footnote{This is the simplification: to solve for ${Q}$ in the
 free basis; if ${Q}$ is solved for in the Coulomb basis the $\hard$ calculation follows
 the one carried out in Subsection~\ref{hard}.} and  then calculate the shift defined by 
 Eq.~(\ref{eq:FFF}). 
 
 From the form of ${{ V}}^{^{(0)}}$ and ${{ H}}_0$ we 
 see that ${Q}$ has the following general form
 \bea
 \langle {\bf p}^\pr s_3 s_4|i {Q}|{\bf p} s_1 s_2 \rangle&=&
 \del^3(p-p^\pr) \langle {\bf p}^\pr s_3 s_4|i {R}|{\bf p} s_1 s_2 \rangle\;,
 \eea
 where from Eq.~(\ref{eq:elimination}), 
 ${R}$ satisfies
 \bea
 \frac{{ v}^{(0)}({\bf p}^\pr s_3 s_4;
 {\bf p} s_1 s_2)}{2 m}&=&
 \langle {\bf p} s_3 s_4|i {R}|{\bf p} s_1 s_2 \rangle-
 \langle {\bf p}^\pr s_3 s_4|i {R}|{\bf p}^\pr s_1 s_2 \rangle\;.
 \label{eq:hu}
 \eea
 Recall Eq.~(\ref{eq:132}) for the form of ${ v}^{(0)}$. Thus, 
  ${R}$ is given by
 \bea
 \langle {\bf p} s_3 s_4|i {R}|{\bf p} s_1 s_2 \rangle&=&
\frac{\del_{s_1 {\overline s}_3} \del_{s_2 s_4}}{2m}\left(s_1 p_y-i p_x\right)+ 
 \frac{\del_{s_1 s_3} \del_{s_2 {\overline s}_4}}{2m}\left(s_4 p_y+i p_x\right)\;.\eea
 
 Since ${Q}$ is diagonal in momentum space it is a simple matter to calculate
 the contributions from Eq.~(\ref{eq:FFF}). Define
 \bea
 \del M_{_{1}}^2&=&\langle\phi_{1,0,0,s_3,s_4}|
 {{ V}}^{^{(1)}}
 |\phi_{1,0,0,s_1,s_2}\rangle~,\label{eq:1782}\\
 \del M_{_{2}}^2&=&\langle\phi_{1,0,0,s_3,s_4}|
 \frac{1}{2} \left[ i {Q},{{ V}}^{^{(0)}}\right]
 |\phi_{1,0,0,s_1,s_2}\rangle
 \;.\label{eq:179}
 \eea
  
  First, $\easy$:  
 \bea
 \del M_{_{1}}^2&=&  \int d^3 p d^3 p^\pr \langle \phi_{100}|{\bf p}^\pr\rangle\langle{\bf
 p}|\phi_{100}\rangle 
  { V}^{^{(1)}}({\bf p}^\pr s_3 s_4;
 {\bf p} s_1 s_2)
 \;.
 \eea
 Using the rotational symmetry of the integrand, we can replace
 \bea
 \frac{(p_\perp-p_\perp^\pr)^2}{\pp}&\longrightarrow&\frac{\frac{2}{3}\left[
 (p_x-p_x^\pr)^2+(p_y-p_y^\pr)^2+(p_z-p_z^\pr)^2\right]}{\pp}
 =\frac{2}{3}
 \;.
 \eea 
 After this,  the remaining integrals are trivial [recall Eq.~(\ref{eq:hy})] and we have
 \bea
 \frac{\easy}{2 m^2 \alpha^4}&=& \frac{1}{2} \del_{s_1 s_2} \del_{s_1 s_4} \del_{s_3 s_4} 
 -\frac{1}{12} \del_{s_1 {\overline s}_2} \del_{s_3 {\overline s}_4}+\frac{1}{2} 
 \del_{s_2 {\overline s}_4} \del_{s_1 {\overline s}_2} \del_{s_1
  {\overline s}_3}
  \;.
 \eea
 
 Next,  $\hard$: 
 \bea
 \hard&=&\langle\phi_{1,0,0,s_3,s_4}|
 \frac{1}{2} \left[ i {Q},{{ V}}^{^{(0)}}\right]
 |\phi_{1,0,0,s_1,s_2}\rangle\\
 &=&\frac{1}{2}\sum_{s_e s_{\overline{e}}}
 \int d^3 p d^3 p^\pr \langle \phi_{100}|{\bf p}^\pr\rangle\langle{\bf p}|\phi_{100}\rangle
 \left(\langle {\bf p}^\pr s_3 s_4 |
 i {R}
 |{\bf p}^\pr s_e s_{\overline{e}}\rangle
 \langle {\bf p}^\pr s_e s_{\overline e}| {{ V}}^{^{(0)}}|{\bf p} s_1 s_2
 \rangle\right.
 \nonumber\\
&&~~~~~ \left.-\langle {\bf p}^\pr s_3 s_4 |
 {{ V}}^{^{(0)}}
 |{\bf p} s_e s_{\overline{e}}\rangle
 \langle {\bf p} s_e s_{\overline e}|i {R} |{\bf p} s_1 s_2 \rangle\right)\;.
 \eea
 Recalling Eq.~(\ref{eq:rzero}) and using Eq.~(\ref{eq:hu}) we have
 \bea
 \hard&=&\frac{\alpha}{\pi^2}\int d^3p d^3 p^\pr 
 \langle \phi_{100}|{\bf p}^\pr\rangle\langle{\bf p}|\phi_{100}\rangle
 \frac{F}{\pp}
 \;,
 \eea
 where
 \bea
 F&=&\sum_{s_e s_{\overline{e}}}\langle {\bf p} s_e s_{\overline{e}} |
 i {R}
 |{\bf p} s_1 s_2\rangle
 \langle {\bf p}^\pr s_3 s_4| {v}^{(0)}|{\bf p} s_e s_{\overline{e}} \rangle\\
 &=&
 \frac{1}{2} \sum_{s_e s_{\overline{e}}}\left(
 \langle {\bf p} s_e s_{\overline{e}} |
 i {R}
 |{\bf p} s_1 s_2\rangle-\langle {\bf p}^\pr s_e s_{\overline{e}} |
 i {R}
 |{\bf p}^\pr s_1 s_2\rangle
 \right)\langle {\bf p}^\pr s_3 s_4| {v}^{(0)}|{\bf p} s_e s_{\overline{e}} \rangle
 \;,~~
 \eea
 using the fact that ${v}^{(0)}$ is odd under ${\bf p} \longleftrightarrow {\bf p}^\pr$ 
 in this last step.
 Using Eq.~(\ref{eq:hu}) this becomes
 \bea
 F&=&\frac{1}{4m}\sum_{s_e s_{\overline{e}}}{ v}^{(0)}({\bf p}^\pr s_3 s_4;
 {\bf p} s_e s_{\overline{e}}){ v}^{(0)}({\bf p}^\pr s_e s_{\overline{e}};
 {\bf p} s_1 s_2)
 \;.
 \eea
 Using the even symmetry of the rest of the integrand under the operations
 $(p_x \longrightarrow - p_x,p_x^\pr \longrightarrow - p_x^\pr)$ and 
 $(p_x \longleftrightarrow p_y, p_x^\pr \longleftrightarrow p_y^\pr)$ this sum
 can be simplified with result
 \bea
 F&=&-\frac{1}{24 m} \left(3 g_1+g_2\right) \pp
 \;,
 \eea
 where
 \bea
g_1&=& s_1 s_3+s_2 s_4~,
\label{eq:g12}\\
g_2&=& 1+s_1 s_2-s_2 s_3-s_1 s_4+s_3 s_4+s_1 s_2 s_3 s_4
\;.\label{eq:g22}
\eea
Recall that $s_i = \pm 1$, $(i=1,2,3,4)$;  the `$\frac{1}{2}$' has been factored out of these 
spins.\footnote{
In order to get these simple forms for $g_1$ and $g_2$ 
it was useful to note the following simple relation: 
$\del_{s s^\pr} = \frac{1}{2} s (s+s^\pr)$ 
 [true because $s^2=1$].}
 The result was written in this form to show the equivalence with Subsection~\ref{hard}.
 Note how much simpler the $\hard$ calculation is in the present section.
 Putting it all together we have
 \bea
 \hard&=&
 -\frac{\alpha}{24 \pi^2 m} \left(3 g_1+g_2\right) \int d^3 p d^3 p^\pr \langle
 \phi_{100}|{\bf
 p}^\pr\rangle\langle{\bf
 p}|\phi_{100}\rangle \\
 &=&-\frac{m^2 \alpha^4}{24}\left(3 g_1+g_2\right)\;,
 \eea
 using Eq.~(\ref{eq:hy}) in this last step.
 
 Combining the results we have
 \bea
 \frac{\easy+\hard}{2 m^2 \alpha^4}&=& \frac{1}{2} \del_{s_1 s_2} \del_{s_1 s_4} \del_{s_3
 s_4} 
 -\frac{1}{12} \del_{s_1 {\overline s}_2} \del_{s_3 {\overline s}_4}+\frac{1}{2} 
 \del_{s_2 {\overline s}_4} \del_{s_1 {\overline s}_2} \del_{s_1
  {\overline s}_3}\nonumber\\
  &&~~~~~-\frac{1}{48} (3 g_1+g_2)\;,
 \eea
 which is the same as Eq.~(\ref{eq:combined}) as was to be shown.
       
\chapter{Lamb shift of hydrogen}\label{ch:lamb}

Experimentally the Lamb shift was discovered by Lamb and Retherford
in 1947 \cite{lamb}.  Later that year, Bethe submitted
his seminal theoretical paper 
\cite{bethe}. And  quantum field theory began to look complete. 
 For a review of some of
these early calculations see Sections~19 and 21 of
Bethe and Salpeter's classic text \cite{bstext} and references within. 
Some selected early
papers can be found in \cite{schreview}. The agreed upon result---that has
stood the test of time---that arose
from this very active period, including QED effects through one loop, for
the $n=2$ and $j=1/2$ levels of hydrogen is 
\cite{bstextp103}
\bea
\del E_{2S_{\frac{1}{2}}}&=&\frac{\al^3 Ryd}{3 \pi}\left[
\ln\left(\frac{m}{{\overline \beta}(2,0)}\right)-\ln 2+\frac{5}{6}-\frac{3}{8}
-\frac{1}{5}+\frac{3}{8}
\right]\;,\label{eq:5.1}\\
\del E_{2P_{\frac{1}{2}}}&=&\frac{\al^3 Ryd}{3 \pi}\left[
\ln\left(\frac{Ryd}{{\overline \beta}(2,1)}\right)-\frac{1}{8}
\right]\label{eq:5.2}
\;,
\eea
where the last terms ($+\frac{3}{8}$ and $-\frac{1}{8}$ respectively) 
are the anomalous magnetic moment of the electron
contributions and
the $-\frac{1}{5}$ is the vacuum polarization contribution.
Each of these terms has an interesting history \cite{history}.
The notation ${\overline \beta}(n,l)$ will be explained later in this chapter;
it is a particular average excitation energy of hydrogen that comes up
in the calculation. Putting in
the experimental parameters \cite{expt} gives the following for the 
theoretical Lamb shift of hydrogen through
one loop in covariant\footnote{Actually, this ``covariant" restriction can
be removed as Kroll and Lamb \cite{krolllamb}, and French and Weisskopf 
\cite{french} showed.} 
QED equal-time calculations
\be
E_{2S_{\frac{1}{2}}}-E_{2P_{\frac{1}{2}}}=  1052.19  \,{\rm MHz}\, (2 \pi \hbar)
\;.
\ee
As motivated in the Introduction, we
 will only calculate the dominant part of this shift.
 Two modern experimental results are \cite{lambe1}
 \be
 E_{2S_{\frac{1}{2}}}-E_{2P_{\frac{1}{2}}}=  1057.845 \pm 0.009\, {\rm MHz}\,
  (2 \pi \hbar)
 \ee
 and \cite{lambe2}
 \be
E_{2S_{\frac{1}{2}}}-E_{2P_{\frac{1}{2}}}=  1057.851 \pm 0.002 \,{\rm MHz}\,
 (2 \pi \hbar)
 \;.
 \ee
For a modern status report on the theory of the hydrogen Lamb shift 
see \cite{statusgrotch} and references within. 
For a modern review and text see \cite{kinoshita} and references
within. 
For some further selected references on the subject over the years see \cite{lamball}.
Here we must apologize for the inadequate references to the many papers
on this subject.

We proceed with an overview
 of our Lamb shift calculation.  In hydrogen there
is a small amplitude for a bound electron to emit and re-absorb a photon, which
leads to a small shift in the binding-energy.  This is the dominant source of
the Lamb shift, and the only part of this shift we compute in this chapter.
This requires electron self-energy renormalization, but 
removal of all
the bare cutoff $\tilde{\Lam}$ dependence requires a complete 4th order
calculation,
which is beyond the scope of 
this dissertation. We work with a finite bare cutoff: $\tilde{\Lam}=m\sqrt{2}$.
However, we do
show that our
results are independent of the effective 
cutoff, $\tilde{\lam}$,\footnote{$\tilde{\lam}=\lam-m-m_p$ and $\tilde{\Lam}=\Lam-m-m_p$: convenient
definitions for discussing restrictions on typical binding-energy scales.}
which validates our adjusting the effective cutoff  into the range
$m\al^2\ll \tilde{\lam}\ll m\al$, which is necessary to obtain the results in the few-body sector alone.

The energy scale for the electron binding energy is $m\alpha^2$, while the
scale for photons that couple to the bound-states is $m\alpha$.  This energy
gap makes the theory amenable to the use of effective Hamiltonian techniques. 
For simplicity, we use a Bloch transformation \cite{bloch}
to remove the high energy scale
({\it i.e.}, $m\alpha $) from the states, and an effective Hamiltonian is
derived which acts in the low-energy space alone.  This effective Hamiltonian
is treated in BSPT, as outlined in Chapter~\ref{ch3}.  The
difference between the $2 S_{1 \over 2}$ and the $2 P_{1 \over 2}$ energy
levels, which are degenerate to lowest order, is calculated.

We divide the calculation into two parts, low- and high-energy intermediate
photon contributions.  The low-energy photons satisfy $|{\bf k}| <
\tilde{\lambda}$, while the high-energy photons satisfy
$\tilde{\lambda} < |{\bf k}| < m$.  $\tilde{\lambda}$ is the effective cutoff
for
the theory, which is chosen to lie in the range $m\al^2\ll\tilde{\lam}\ll m\al$.
This choice
lies between the two dominant energy scales in the problem and allows us to
avoid near degeneracy problems.  When an actual number is required we use

\begin{equation}
\tilde{\lambda} = \alpha\sqrt{\alpha}\; m \sim 6 \times 10^{-4}\; m \;.
\label{eq:32415}\end{equation}

\noindent Note that the spectrum of the exact effective Hamiltonian is
independent of $\tilde{\lambda}$, but our approximations introduce
$\tilde{\lambda}$-dependence.  The range for $\tilde{\lambda}$ is chosen so
that this independence can be derived consistently order by order in the few-body
sector alone.

One further introductory comment, the high photon energy 
($\tilde{\lam}<|{\bf
k}| < m$)  part of the shift is further divided into two regions,
$\tilde{\lam}<|{\bf k}| < b$ and $b<|{\bf k}| < m$, where $b$ is an arbitrary
parameter chosen in the range $m \alpha \ll b \ll m$. This  simplifies the
calculation with appropriate approximations being used in the respective
regions. The result must obviously be independent of this arbitrary division
point $b$, and is, unless ``non-matching" approximations are used in the
respective regions.

We now outline this chapter. 
In Section~\ref{lamb1} we review the basic theoretical framework of this
light-front Hamiltonian approach for this Lamb shift calculation, and then
in Section~\ref{lamb2}
 we proceed with a  discussion on  the
origin of the Coulomb interaction in the electron-proton system. Section~\ref{lamb3}
 contains the heart of
the Lamb shift calculation. In the final section, Section~\ref{lamb4}, 
we summarize and
discuss our results.

\section{Theoretical framework}\label{lamb1}

The proton will be treated as a point
particle. 
The Lagrangian for the electron, proton, and photon system is ($e > 0$)
\bea
{\cal L}&=&-\frac{1}{4}F_{\mu \nu} F^{\mu \nu} +
\overline{\psi}_e(i \not\! {\partial } +e \not\!\! {A }- 
 m )\psi_e+
 \overline{\psi}_p(i \not\! {\partial } -e \not\!\! {A }- 
 m_p )\psi_p~.
\eea
The reduced mass of
the system is defined in the standard way
\bea
m_r&=&\frac{m m_p}{m+m_p}=m \left(1-m/m_p+{\cal O}(1/m_p^2)\right)
\;.\label{eq:reduced}
\eea
Note that  we take the limit $m_p/m \longrightarrow \infty$
because we are only interested in the dominant part of the Lamb shift, but
we will keep the reduced mass with a finite proton mass
for the derivation of the Coulomb potential.
The Lagrangian leads to the canonical Hamiltonian $H$ in the light-cone
gauge $A^+ =0$ that was derived in Section~\ref{canonical}. The final
form of the canonical Hamiltonian is written in 
Eqs.~(\ref{eq:can1})--(\ref{eq:instant}).
For a summary of the light-front conventions used for 
hydrogen see Appendix~\ref{appendixlf}.

Given the canonical Hamiltonian $H$ we cut off the theory by 
requiring the free energies of all states to satisfy
\bea
&& \veps_i \leq
 \frac{{{\cal P}^\perp}^2+\Lam^2}{{\cal P}^+}
 \;,
 \eea
 where $\Lam$ is the bare cutoff, and ${\cal P}=
 \left({\cal P}^+,{\cal P}^\perp\right)$ is the total momentum of
 the hydrogen state.
 Then, with a Bloch transformation
  we remove the states with free energies satisfying
 \bea
&&\frac{{{\cal P}^\perp}^2+\lam^2}{{\cal P}^+}\leq  \veps_i \leq
 \frac{{{\cal P}^\perp}^2+\Lam^2}{{\cal P}^+}
 \;,
 \eea
 where $\lam$ is the effective cutoff.
 The result is an effective Hamiltonian, $H_\lam$, acting in 
 the low-energy ($ \veps_i \leq
 \frac{{{\cal P}^\perp}^2+\lam^2}{{\cal P}^+}$) space.
 We do not discuss the derivation of $H_\lam$ any further, but 
 instead refer the interested reader to Section~\ref{blochsec}.
 
 Given $H_\lam$, we then  make the following division
 \bea
 &&H_\lam={\cal H}_0 + \left(H_\lam-{\cal H}_0\right)~\equiv~{\cal H}_0+{\cal
 V}~,
 \eea
 where ${\cal H}_0$ is an approximation that can be solved nonperturbatively
 (and analytically for this QED calculation) and ${\cal V}$ is treated in BSPT. For this Lamb
 shift calculation, we
 treat the Coulomb interaction between the electron and proton to 
all orders in all Fock sectors. The 
 test of ${\cal H}_0$ is whether BSPT converges or not and closely related: Is 
 the $\lam$-dependence of the spectrum weakened by higher orders of BSPT?
\section{Lowest order Schr\"{o}dinger equation}\label{lamb2}

The 
primary {\em assumption} we make in this QED bound-state calculation is
that the Coulomb interaction dominates all other physics. 
After this assumption,
 the kinematic length scale of our system is fixed,
\beaa
&&a_0 \sim \frac{1}{p} \sim \frac{1}{m \alpha}\sim \frac{137}{m}
\;,
\eeaa
which then fixes our dynamic time and length scale,
\beaa
&&t\sim\frac{1}{p^2/(2m)}\sim\frac{1}{m \alpha^2}\sim \frac{137^2}{m}
\;:
\eeaa
as is well known,
dynamic changes occur very slowly in this system.
Note that in this QED calculation we will treat photons as free since they 
carry no charge and interact very weakly at low energies. 
After choosing ${\cal H}_0$,
 the $\alpha$-scaling of our BSPT is fixed, and the spectrum is then
calculated to some desired order in $\alpha$ and $\al \,\ln(1/\al)$.

In the Coulomb gauge, the Coulomb interaction appears directly in the canonical
Hamiltonian,
which 
of course is not true in the light-cone gauge.
In the light-cone gauge,
the Coulomb interaction arises from the leading terms in a nonrelativistic expansion
of a combination of two types of interactions
in our effective Hamiltonian: instantaneous photon exchange and the two time
orderings of dynamical photon exchange. This is illustrated in 
Figure~5.1.
These interactions arise from first and second-order effective interactions
respectively. See Eq.~(\ref{eq:heff})  for 
the form of the effective Hamiltonian, $H_\lam$. Now we discuss the derivation of the Coulomb
interaction in the light-cone gauge for the electron-proton system.

\begin{figure}[t]\hskip.5in
\centerline{\epsfig{figure=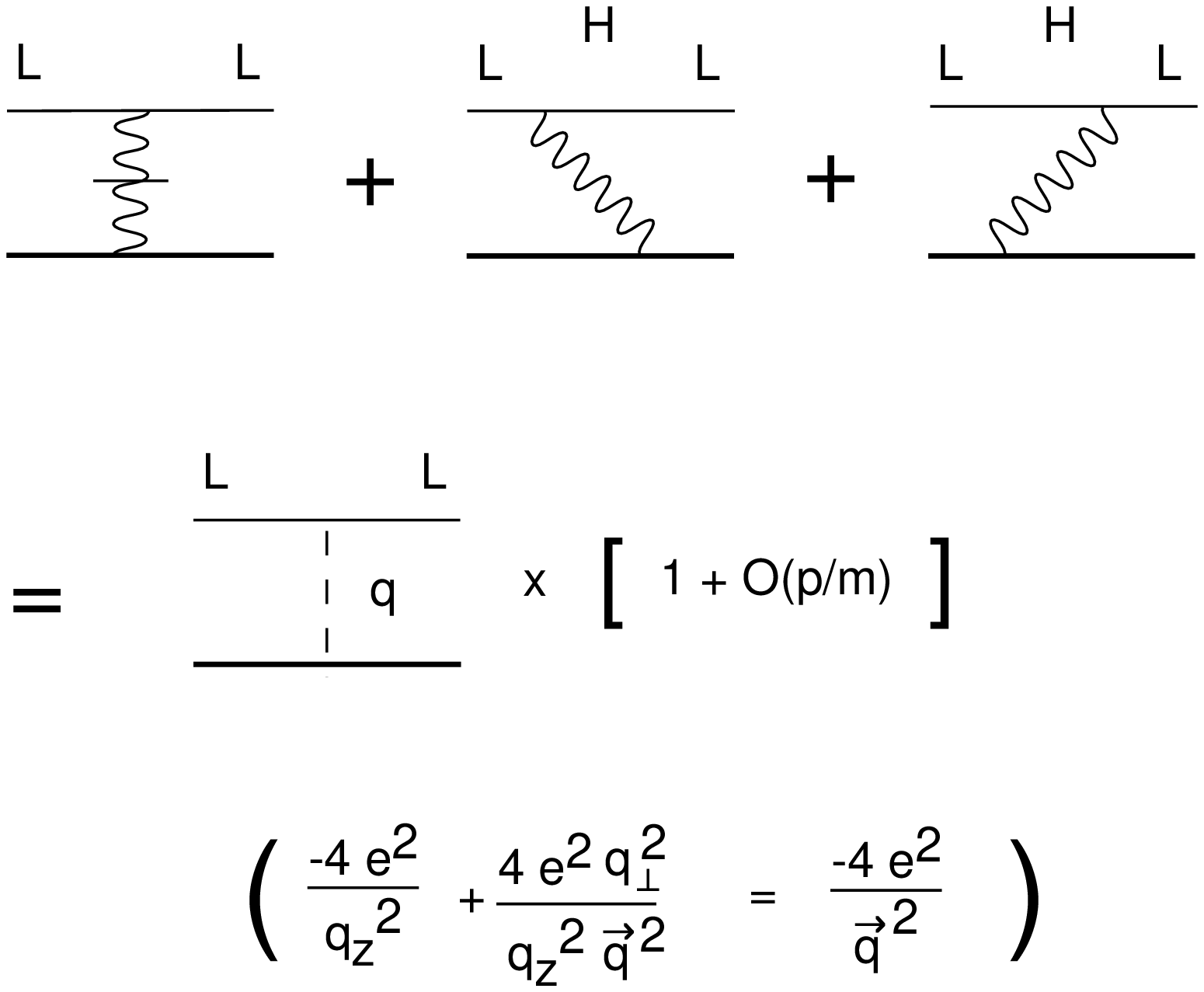,height=4in,width=5.0in}}
\caption{The effective interactions that add to give the Coulomb potential. ``H"
implies that the photon energy is greater than $\tilde{\lam}$. ``L" implies that
the
electron kinetic energy is less than $\tilde{\lam}$. 
We choose $m \alpha^2 \ll
\tilde{\lam}
\ll m \alpha$; these ``H" and ``L" constraints 
can thus be removed to leading order, and we are left with $-4 e^2/
{\bf q}^2$, the Coulomb
potential.} 
\label{lambfig1}
\end{figure}

The time-independent
Schr{\"o}dinger equation in light-front coordinates 
that the sum of the three time-ordered diagrams in
Figure~5.1 satisfies is\footnote{For a derivation of Eq.~(\ref{eq:lowsch}) 
 from the Schr{\"o}dinger equation in Fock space 
see Subsection~\ref{4.2.1} where this was done for the equal-mass case.}
\bea
 \left({\cal M}_{_{N}}^2-\frac{{\kap^\pr}^2+m^2}{x^\pr}-
 \frac{{\kap^\pr}^2+m_p^2}{1-x^\pr}
 \right) {\tilde{\phi}}_N(x^\pr \kap^\pr s_e^\pr s_p^\pr)
 &=&
 \sum_{s_e s_p}\int d^2 \kap/(2 \pi)^2 \int_0^1 dx/(4 \pi)\nn\\
 &&~~~~~~\times~
 {\tilde V}_c~ {\tilde{\phi}}_N(x \kap s_e s_p)
 \;.\label{eq:lowsch}
 \eea
 ${\cal M}_{_{N}}^2$ is the  mass squared eigenvalue of the state ${\tilde
 \phi}_N$,
where ``$N$" labels
 all the quantum numbers of this state.
 The tildes will be notationally convenient below. 
We have introduced the following Jacobi variables
\bea
p_e&=&(x {\cal P}^+ , \kappa+ x {\cal P}^\perp)~,\\
p_e^\pr&=&(x^\pr {\cal P}^+ , \kappa^\pr+ x^\pr {\cal P}^\perp)
\;,
\eea
where $p_e$ and $p_e^\pr$ are the initial and final electron three-momentum
respectively, and
\bea
&&p_e+p_p=p_e^\pr+p_p^\pr={\cal P}=({\cal P}^+,{\cal P}^\perp)
\;
\eea
is the total momentum of the hydrogen state.
Note that $\kap$ is a two-vector.
 The norm is defined by
 \bea
 \sum_{s_e s_p}\int d^2 \kap/(2 \pi)^2 \int_0^1 dx/(4 \pi)~
 {{\tilde{\phi}}_N}^\ast(x \kap s_e s_p) 
 {\tilde{\phi}}_{N^\pr}(x \kap s_e s_p)\;=\;\del_{N N^\pr}\;.
 \eea
${\tilde V}_c$  is the sum of the interactions given by the
three diagrams in Figure~\ref{lambfig1}, and 
will not be written in all its gory detail.\footnote{The interested reader
should consult
Eq.~(\ref{eq:saeko3}) and the discussion leading
up to Eq.~(\ref{eq:C}).
In Chapter~\ref{ch:pos} the discussion is
for the equal mass case, but is readily generalized
to the unequal mass case using the rules in Appendix~\ref{brentrules}.
Note that in Chapter~\ref{ch:pos} we use a G-W
transformation instead of a Bloch transformation; the Bloch transformation is
chosen
for the current chapter because of its simplicity.} The leading order term of
${\tilde V}_c$ in a nonrelativistic expansion is defined as $V_c$ and is
written below.

The nonrelativistic expansion is defined in the following way.
A coordinate change which takes the range of longitudinal momentum fraction, $x
\in [0,1]$ to 
 $\kap_z \in [-\infty, \infty]$ is defined
 \bea
 x&=&\frac{\kap_z+\sqrt{\kap^2+\kap_z^2+m^2}}{\sqrt{\kap^2+\kap_z^2+m^2}+
 \sqrt{\kap^2+\kap_z^2+m_p^2}}\;.
 \label{eq:coord}\eea
 This step can be taken for relativistic kinematics, but
 there may be no advantage. 
 Note that this definition of $\kap_z$ corresponds to the equal-time z-momentum
 of an electron in the equal-time center-of-mass frame ${\bf p}_e+{\bf p}_p=0$.
 This intuition will guide us later.
Then, the nonrelativistic expansion is an expansion in $\bp/m$; i.e.,
we assume
\bea
 &&|{\bf p}|\ll m\;,\label{eq:nr}
\eea
where we have defined a
new three-vector in terms of our transverse Jacobi variable $\kap$ and our
new longitudinal momentum variable $\kap_z$ (which replaces our longitudinal
momentum
fraction $x$)
  \bea
 {\bf p}&\equiv& (\kap,\kap_z)
 \;.
 \label{eq:coord2}\eea
 Note that the free mass-squared in the Schr{\"o}dinger equation,
Eq.~(\ref{eq:lowsch}), after this coordinate change, becomes
\bea
 \frac{{\kap}^2+m^2}{x}+ \frac{{\kap}^2+m_p^2}{1-x}
 &=&\left(\sqrt{m^2+{\bf p}^2}+\sqrt{m_p^2+{\bf p}^2}~\right)^2\nn\\
 &=&
 (m+m_p)^2+2 (m+m_p) \left[
 \frac{{\bf p}^2}{2 m_r}-\frac{{\bf p}^4 (m-m_p)^2}{8\,
 m_r\,
  m^2 m_p^2}\right.\nn\\
  &&\left.\hskip2in+{\cal O}\left(\frac{\bp^6}{m^5}\right)
 \right]
 \;,\label{eq:30}
 \eea
 which is invariant under rotations in the space of vectors ${\bf p}$.
 $m_r$ is the reduced mass given in Eq.~(\ref{eq:reduced}).
 The longitudinal momentum fraction transfer in the nonrelativistic expansion becomes
 \be
 x^\pr-x\,=\,\frac{p_z^\pr-p_z}{m+m_p}
 +\frac{\left({\bf p}^2-{{\bf p}^\pr}^2\right)(m-m_p)}{2\,m\,m_p(m+m_p)}
 +{\cal O}\left[\frac{\bp^3}{m^2(m+m_p)}\right]
 \;.
 \ee

The leading-order term of ${\tilde V}_c$ in an expansion in
$\bp/m$ is contained in
\bea
{\tilde V}_c&\sim& (m+m_p)^2 
\left(-\frac{4 e^2}{q_z^2}+
\frac{4 e^2 {q^\perp}^2}{q_z^2 {\bf q}^2}\theta_{H}
\right)\del_{s_e s_e^\pr} \del_{ s_ps_p^\pr}
\theta_{L}
\;,
\eea
where
\bea
{\bf q} &=& {\bf p}^\pr-{\bf p}\\
\theta_{L}
&=&
\theta\left(\lam^2-
\left(\sqrt{m^2+{\bf p}^2}+\sqrt{m_p^2+{\bf p}^2}~\right)^2\right)\nn\\
&&~~~~~\times~\theta\left(\lam^2-
\left(\sqrt{m^2+{{\bf p}^\pr}^2}+\sqrt{m_p^2+{{\bf
p}^\pr}^2}~\right)^2\right)\;,\\
\theta_{H}&=&\theta
\left(\left[(m+m_p)^2+2 (m+m_p) \frac{{\bf q}^2}{2|q_z|}\right]-\lam^2
\right)\nn\\
&&~~~~~\times~
\theta
\left(\Lam^2-\left[(m+m_p)^2+2 (m+m_p) \frac{{\bf q}^2}{2|q_z|}\right]
\right)
\;.
\eea
Note that $\theta_L$ and $\theta_H$ 
 are the constraints that arise from the Bloch transformation.

It is convenient to define new cutoffs which subtract off the total free
constituent masses of the state
\bea
\tilde{\lam}&\equiv&\lam-(m+m_p)~,\nn\\
\tilde{\Lam}&\equiv&\Lam-(m+m_p)
\;.\label{eq:21}
\eea
In the limit $m_p \rightarrow \infty$ we require $\tilde{\lam}$ and
$\tilde{\Lam}$
to be held fixed.
Note that this implies
\bea
\frac{\lam^2-(m+m_p)^2}{2(m+m_p)}&=&\tilde{\lam}+\frac{\tilde{\lam}^2}{2(m+m_p)}
\stackrel{(m_p \rightarrow \infty)}{\longrightarrow}\tilde{\lam}~,\nn\\
\frac{\Lam^2-(m+m_p)^2}{2(m+m_p)}&=&\tilde{\Lam}+\frac{\tilde{\Lam}^2}{2(m+m_p)}
\stackrel{(m_p \rightarrow \infty)}{\longrightarrow}\tilde{\Lam}
\;.\label{eq:21p}
\eea

In terms of these new cutoffs,
 $\theta_{L}$ and $\theta_{H}$ above become
\bea
\theta_{L}&=& \theta
\left(\tilde{\lam}-\frac{{\bf p}^2}{2 m_r}+{\cal
O}\left(\frac{\bp^4}{m^3}\right)\right)
\theta
\left(\tilde{\lam}-\frac{{{\bf p}^\pr}^2}
{2 m_r}+{\cal O}\left(\frac{|{\bf p}^\pr|^4}{m^3}\right)\right)~,\\
\theta_{H}&=&\theta\left(\frac{{\bf q}^2}{2|q_z|}-\tilde{\lam}\right)
\theta\left(\tilde{\Lam}-\frac{{\bf q}^2}{2|q_z|}\right)
\;.
\eea

To see the Coulomb interaction arising from the $|e p\rangle$ sector alone,
we make the following requirements (which are motivated from the previous two
equations)
\bea
\frac{\bp^2}{2m_r}\ll \tilde{\lam} \ll \bp~~~~~~~~~~~&{\rm and}&~~~~~~~~~~~
\tilde{\Lam}\gg \bp
\label{eq:constraints}
\;,
\eea
also demanded for $|{\bf p}^\pr|$ of course.
 These constraints
 will be maintained consistently in this chapter. 
Given these restrictions we have
\bea
\theta_L &\approx& 1~,\\
\theta_H &\approx& 1
\;.
\eea
 ${\tilde V}_c$ becomes
\bea
{\tilde V}_c&\sim& V_c
\;,
\eea
where
\bea
V_c
&\equiv&
(m+m_p)^2 
\left(-\frac{4 e^2}{q_z^2}+
\frac{4 e^2 {q^\perp}^2}{q_z^2 {\bf q}^2}
\right)\del_{s_e s_e^\pr} \del_{s_p s_p^\pr}\nonumber\\
&=&-(m+m_p)^2 \left(\frac{4 e^2}{{\bf q}^2}\right)\del_{ s_es_e^\pr} \del_{s_p
s_p^\pr}~~.
\eea
This is summarized in Figure~\ref{lambfig1}.

To finish showing how the Coulomb interaction arises in a light-front 
Hamiltonian approach,
 we need to know the Jacobian of the
coordinate transformation of Eq.~(\ref{eq:coord}), 
\bea
 J(p)&=&\frac{dx}{d\kap_z}=
 \frac{\left(\kap_z+\sqrt{{\bf p}^2+m^2}\right)\left(\sqrt{{\bf
 p}^2+m_p^2}-\kap_z\right)}
 {\sqrt{{\bf p}^2+m^2}\sqrt{{\bf p}^2+m_p^2}\left(
 \sqrt{{\bf p}^2+m^2}+\sqrt{{\bf p}^2+m_p^2}~\right)}\nn
 \\
 &=&
  \frac{1}{m+m_p}\left[1+\kap_z\left(\frac{1}{m}-\frac{1}{m_p}\right)-
\frac{\left({\bf p}^2+2\kap_z^2\right)}{2 m m_p}+
{\cal O}\left(\frac{\bp^3}{m^3}\right)\right]
 \;.
 \eea
It is also convenient to redefine the norm
\bea
\del_{N N^\pr}&=&\sum_{s_e s_p}\int d^2 \kap/(2 \pi)^2 \int_0^1 dx/(4 \pi)~
 {\tp_N}^\ast(x \kap s_e s_p) \tp_{N^\pr}(x \kap s_e s_p) \nn\\
 &=&
 \sum_{s_e s_p} \int d^3 p~ J(p)/(16\pi^3)~
 {\tp_N}^\ast({\bf p} s_e s_p) \tp_{N^\pr}({\bf p} s_e s_p)
\nn \\
 &\equiv&\sum_{s_e s_p} \int d^3 p \;\phi_N^{ \ast}({\bf p} s_e s_p)
 \phi_{N^\pr}({\bf p} s_e s_p)
 \;.\label{eq:norm22}\eea
In this last line the tildes are removed from the wave functions by defining
\bea
\phi_{N}({\bf p} s_e s_p)&\equiv&\sqrt{\frac{J(p)}{16 \pi^3}}\tp_{N}({\bf p} s_e
s_p)~.
\label{eq:jacdf}
\eea

Putting it all together,  the leading order expression for Eq.~(\ref{eq:lowsch})
in an expansion in $\bp/m$ given the restrictions of Eq.~(\ref{eq:constraints})
is
\bea
 \left(-\beta_{n}+\frac{{{\bf p}^\pr}^2}{2m_r}
 \right) \phi_N({\bf p}^\pr  s_e^\pr s_p^\pr)&=&
 \frac{\alpha}{2 \pi^2}\int\frac{ d^3 p }{({\bf p}-{\bf p}^\pr)^2}
  \phi_N({\bf p} s_e^\pr s_p^\pr)
 \;,\label{eq:sch}
\eea
which we see is the  nonrelativistic Schr{\"o}dinger equation of hydrogen.
$m_r$ is the reduced 
mass\footnote{In what follows
 we set $m_r$ to its infinite proton mass
limit, $m$. We kept $m_r$ above 
to show that it arises in Eq.~(\ref{eq:sch}) as in the
equal-time case.}
 and $-\beta_n$ is the leading-order binding-energy defined by
\bea
\beta_n&\equiv&\frac{{\cal M}_{n}^2-(m+m_p)^2}{2 (m+m_p)}
\;,\label{eq:beta}
\eea
where ${\cal M}_{n}^2$ is the leading-order mass-squared.\footnote{The full mass-squared  
$M_N^2$, and
the full binding-energy, $-B_N$, are related by  $M_N^2\equiv(m+m_p+B_N)^2$.}
The well-known bound spectrum is
\bea
\beta_n&=&-\frac{Ryd}{n^2}
\;,
\eea
where $Ryd = m \alpha^2/2$ of course.
Note that Eq.~(\ref{eq:sch}) fixes the $\alpha$-scaling of $\bp$
\bea
&&\bp\sim m\alpha~.
\eea
Thus we see that the restrictions of Eq.~(\ref{eq:constraints}) become
\bea
m \alpha^2 \ll \tilde{\lam} \ll m \alpha~~~~~~~~~~~&{\rm and}&~~~~~~~~~~~
\tilde{\Lam}\gg m \alpha
~,
\eea
 which will be maintained consistently in this chapter.
\section{Lamb shift calculation}\label{lamb3}

Given our leading-order spectrum, we  
proceed with BSPT. As advertised, this will be divided into {\em low} and {\em
high} 
intermediate photon
energy  calculations. Before proceeding with these respective
calculations, we  discuss whether   Coulomb exchange  can be treated
perturbatively or 
nonperturbatively   in the
respective regions.

For the low-energy intermediate photon, the Coulomb interaction between the
intermediate
electron and proton must be treated nonperturbatively, whereas this interaction
can be treated perturbatively for the high-energy intermediate photon
contribution.
 This is seen
by noting that each additional Coulomb exchange contributes a Coulomb matrix
element and an
energy denominator which is dominated by the larger photon-energy scale. Thus
each additional
Coulomb exchange contributes
\bea
&&\frac{\langle \frac{\alpha}{|{\bf r}|} \rangle}{|{\bf k}|}\leq
\frac{m \alpha^2}{|{\bf k}|_{min}}
\;,\label{eq:ex}
\eea
where we used the virial theorem $\langle\phi_N|\al/r|\phi_N\rangle=2\left(
\frac{1}{2} m \al^2/n^2\right)$.
For the low  photon-energy contribution, in principle
$|{\bf k}|_{min}=0$,
 and each additional Coulomb exchange can
contribute ${\cal O}(1)$, and therefore must be treated nonperturbatively. 
Of course, when the Coulomb interaction is treated nonperturbatively,
low-energy
intermediate protons and electrons form bound-states from which
long-wavelength photons decouple. This nonperturbative effect leads to 
$|{\bf k}|_{min}\sim 16.64~{\rm Ryd}$; see Eq.~(\ref{eq:ttbbg}) below.
For 
the high  photon-energy contribution, $|{\bf k}|_{min}=\tilde{\lam}$ and from
Eq.~(\ref{eq:32415}) each
additional Coulomb exchange thus contributes at  most
\bea
&&\frac{m \alpha^2}{\tilde{\lam}}= \sqrt{\alpha} \sim
8.5~\times~10^{-2}\label{eq:exhigh}
\;,
\eea
and can therefore be treated perturbatively.
\begin{figure}[t]\hskip0in\vskip-1in
\centerline{\epsfig{figure=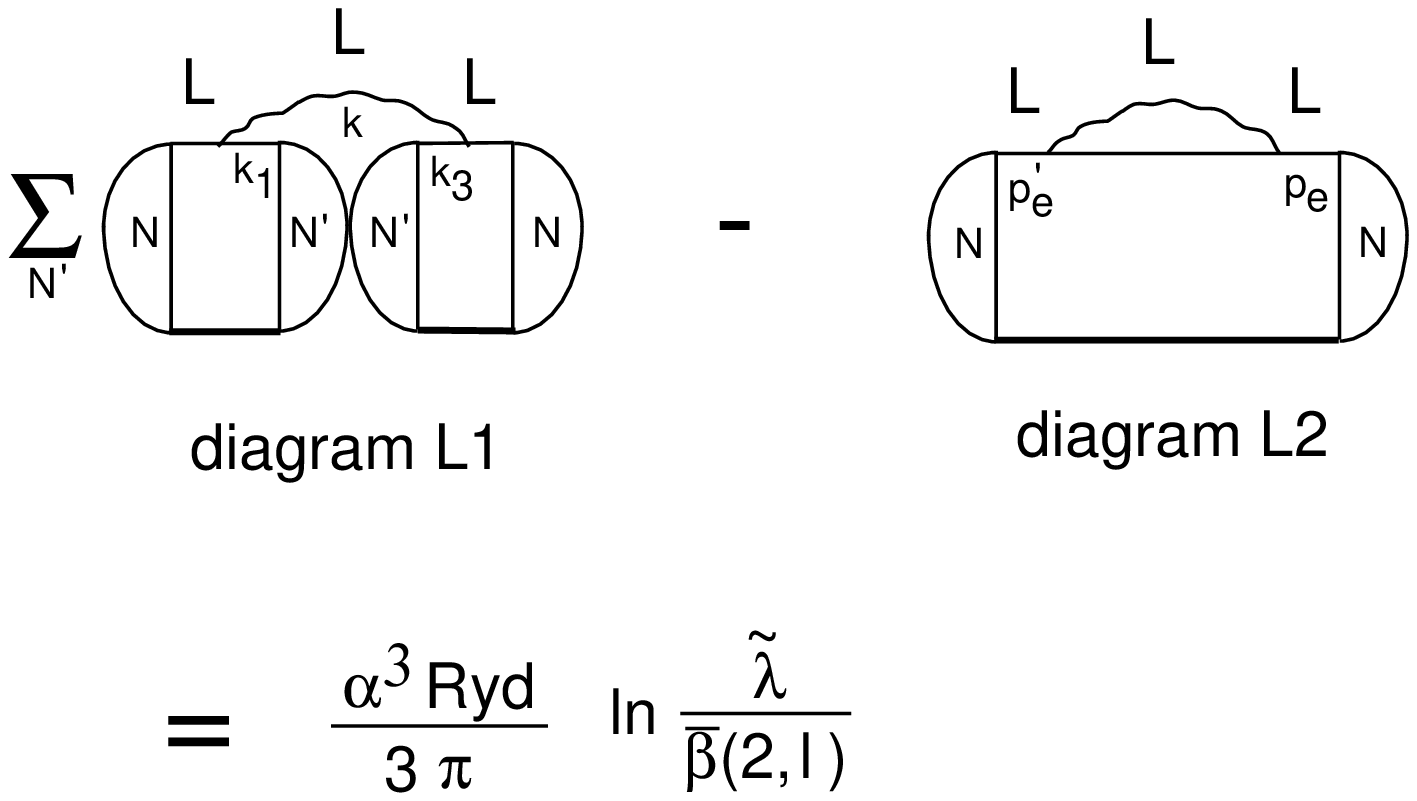,height=4in,width=5.0in}}
\caption{Low photon-energy contribution to the Lamb shift.
Diagram L1 represents the shift arising from treating photon emission below the cutoff
$\tilde{\lam}$ in second-order BSPT, where the intermediate electron-proton are bound
by the
Coulomb potential (scattering states must be included too of course).
Diagram L2 is an effective self-energy interaction plus counterterm
(shown in Figure~\ref{lambfig3}) treated in first-order BSPT.
 $\overline{\beta}(2,l)$ is the average excitation energy of the
$n=2$
levels; see Eqs.~(\ref{eq:107}) and (\ref{eq:108}) and the discussion above them for
details.} 
\label{lambfig2}
\end{figure}
\subsection{Low photon-energy contribution}\label{sec:low}

\begin{figure}[t]\hskip1in
\centerline{\epsfig{figure=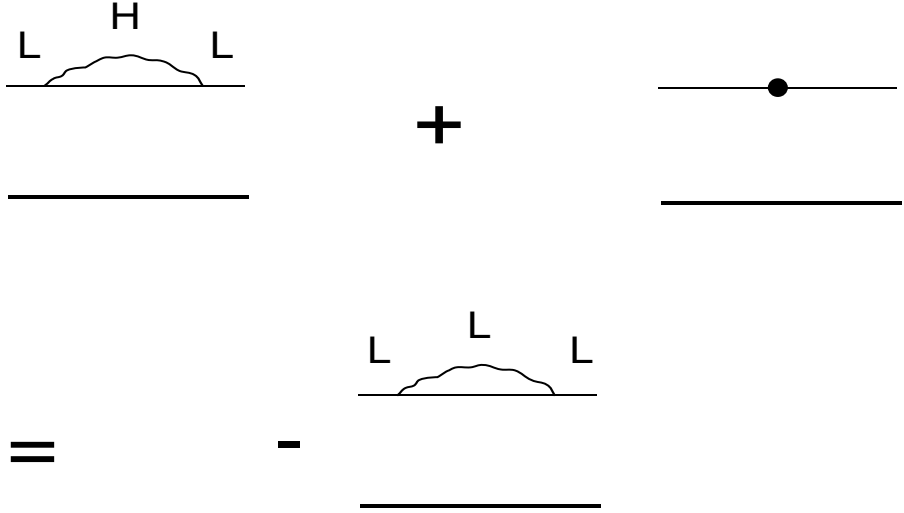,height=3in,width=5.0in}}
\caption{The sum of an effective self-energy
interaction (arising from the removal of photon emission above 
the cutoff $\tilde{\lam}$)
 and a counterterm. The counterterm is fixed by coupling coherence
 as in Subsection~\ref{4.1.1}.
  The result
is the interaction in diagram L2 of Figure~\ref{lambfig2}.} 
\label{lambfig3}
\end{figure}

The low-energy shift arises from two sources  which are shown in 
Figure~\ref{lambfig2}. The
first 
term comes from the low-energy photon emission part of the effective
Hamiltonian,
$
\langle a | \int d^2x^\perp dx^-{\cal H}_{ee\gam} | b \rangle
$,
treated in second-order BSPT, where the intermediate electron-proton are bound
by the
Coulomb potential (scattering states must be included too of course). Recall Eq.~(\ref{eq:48}) 
 for the form of ${\cal H}_{ee\gam}$.
The second term is the result of renormalizing the one-loop electron
self-energy: a counterterm
is added to the second-order self-energy effective interaction in 
$
\langle a |H_\lam | b \rangle
$, which results
in a finite (except for infrared divergences) shift to the electron self-energy.
This is shown in Figure~\ref{lambfig3}. 
 The counterterm is  fixed by requiring the electron self-energy
  to evolve coherently with
the cutoff.
The details follow those in Subsection~\ref{4.1.1} and will not be repeated here.

Note that the term where the proton emits and subsequently absorbs a photon is
down by two powers of the
proton mass with respect to the term where the electron emits and absorbs a
photon. This result
is subtle though, because it is true only after the
 light-front infrared divergences have canceled
between   two  diagrams analogous to the ones in Figure~\ref{lambfig2}.
In ${\cal H}_{pp\gam}$: the $\frac{\der^\perp}{\der^+}\cdot
A^\perp$ term leads to infrared divergences that cancel in the difference, the
$\sig\cdot\der^\perp/\der^+$ terms are the mentioned
 terms down by a power of $m_p$, and the $m_p/\der^+$ terms flip the proton's
 helicity and thus do not contribute to this self-energy shift.
 
Before proceeding with the calculation, recall that the exact binding-energy of
hydrogen, $-B_N$, in
terms of its mass-squared $M_N^2$ is
\bea
M_N^2&\equiv&(m+m_p+B_N)^2
\;.
\eea
Multiplying out the right-hand-side, note that this implies 
\bea
\frac{M_N^2-(m+m_p)^2}{2(m+m_p)}&=&B_N+\frac{B_N^2}{2(m+m_p)}
\;.
\label{eq:4444}
\eea
Recalling Eq.~(\ref{eq:beta}), which is 
the exact definition of the leading-order binding-energy, $-\beta_n$, in terms of the
leading-order mass-squared, ${\cal M}_n^2$; and also defining the mass-squared 
corrections, $\del M_N^2$,  in the form in which they appear in our calculation
\bea
 M_N^2&\equiv& {\cal M}_n^2+\del M_N^2 
\;,
\eea
combined with Eq.~(\ref{eq:4444}),
gives 
\bea
B_N&=&\beta_n+\frac{\del M_N^2}{2 (m+m_p)} -\frac{B_N^2}{2 (m+m_p)}\label{eq:robert}
\;.
\eea
 In what follows, we will show
 that as far as the fermion masses are concerned, $\del M_N^2\sim m\, m_p$
and $B_N^2\sim m^2$ consistently. Taking this as given for now, as $m_p \longrightarrow \infty$ the 
leading term of
Eq.~(\ref{eq:robert}) is
\be
B_N=\beta_n+\frac{\del M_N^2}{2 m_p}
\;.\label{eq:gG2}
\ee
Defining the binding corrections, $-\del B_N$, by
\bea
  B_N &\equiv& \beta_n+\del B_N
\;,
\eea
combined with Eq.~(\ref{eq:gG2}), gives to leading order in $1/m_p$
\bea
\del B_N&=&\frac{\del M_N^2}{2 m_p}
\;,
\label{eq:binding}
\eea
a useful formula to be used below. This formula is useful because $\del M_N^2$
is 
calculated below, but 
$\del B_N$ is the quantity that is measured.

The low-energy calculation proceeds as follows. 
The first term of Figure~\ref{lambfig2} is a second-order BSPT
shift which contributes the following to the mass-squared eigenvalue
[recall Eq.~(\ref{eq:hey1})]
\bea
\del M_{L1}^2&=&\sum_{N^\pr} \int_k \sum_{s_\gamma}\frac{
\left|\langle \psi_N\left({\cal P}\right) |
v_{ee\gam}a_{s_\gamma}^\dagger(k)|\psi_{N^\pr}\left(
{\cal P}-k\right)\rangle
\right|^2 \theta_{L1}
}
{DEN_1 (Vol)^2}
\;,\label{eq:L1}
\eea
where $k$ and $s_\gamma$ are the photon's three-momentum and spin respectively, and 
${\cal P}=\left({\cal P}^+,{\cal P}^\perp\right)$ is the total momentum of
the hydrogen state $\psi_N$, with quantum numbers $N=(n,l,m_l,s_e,s_p)\equiv
(\nu,s_e,s_p)$.
${\cal H}_{ee\gam}$ is the photon emission interaction given in Eq.~(\ref{eq:48}),
where we have defined $v_{ee\gam}\equiv \int d^2 x^\perp dx^- {\cal H}_{ee\gam}$.
$\theta_{L1}$  restricts  the
 energies of the initial, intermediate and final states to be below the
 effective
 cutoff $\frac{\lam^2+{{\cal P}^\perp}^2}{{\cal P}^+}$.
 The explicit form of these restrictions is discussed below.
Continuing the description of Eq.~(\ref{eq:L1}), 
\bea
\int_k &&= \int \frac{d^2k^\perp dk^+ \theta(k^+)}
 {16 \pi^3 k^+}=\int \frac{d^3 k}{ (2 \pi)^3 (2 |{\bf k}| )}
\;.
\eea
The last step comes from recalling that for an on mass-shell photon in the forward
light-cone $k^+=k^0+k^3=|{\bf
k}|+k^3$.
 The denominator and volume factors are
\bea
Vol&=&
\langle \psi_N({\cal P})|\psi_N({\cal P})\rangle/{\cal P}^+=
16 \pi^3 ~\del^3\left({\cal P}-{\cal P}\right)=\int d^2 x^{\perp}dx^-\;,\\
DEN_1&=&{\cal P}^+ \left(
\frac{{{\cal P}^\perp}^2+{\cal M}_n^2}{{\cal P}^+}
-\frac{{({\cal P}-k)^\perp}^2+{\cal M}_{n^\pr}^2}{({\cal P}-k)^+}
-\frac{{k^\perp}^2}{k^+}
\right)\label{eq:den1}
\;.
\eea
The two-body states  are
\bea
|\psi_N\left({\cal P}\right)\rangle&=&\int_{p_e p_p} \sqrt{p_e^+ p_p^+} 16 \pi^3
\del^3 
\left( {\cal P} -p_e-p_p\right)\nn\\
&&\hskip1in\times\tilde{\phi}_{\nu}(p_e p_p) b_{s_e}^\dagger(p_e)
B_{s_p}^\dagger
(p_p) | 0 \rangle~,\\
|\psi_{N^\pr}\left({\cal P}-k\right)\rangle&=&\int_{k_1 k_2} \sqrt{k_1^+ k_2^+}
16 \pi^3 \del^3 
\left( {\cal P}-k -k_1-k_2\right)\nn\\
&&\hskip1in\times\tilde{\phi}_{\nu^\pr}(k_1 k_2)
b_{s_e^\pr}^\dagger(k_1) 
B_{s_p^\pr}^\dagger
(k_2) | 0 \rangle
\;,
\eea
where  $\phi_N$ are solutions to Eq.~(\ref{eq:sch}), 
 the nonrelativistic Schr{\"o}dinger equation of hydrogen,
 and ${\tilde{\phi}}_N$ is related to $\phi_N$ by Eq.~(\ref{eq:jacdf}).

Straightforward algebra leads to
\bea
\del M_{L1}^2&=&{\sum_{\nu^\pr}}^c \int_k \int_{p_e} \theta\left({\cal P}^+ -
p_e^+\right)
\int_{p_e^\pr}\theta\left({\cal P}^+ - {p_e^\pr}^+\right) \int_{k_1 k_3} 
 \left(p_e^+{ p_e^\pr}^+ k_1^+ k_3^+\right) 
 \nn\\
 &\times&\!\left[16 \pi^3 \del^3 (k+k_3-p_e)\right]\left[16 \pi^3 \del^3
 (k+k_1-p_e^\pr)\right] 
 \tilde{\phi}_{\nu}^\ast\left(p_e^\pr,{\cal P}-p_e^\pr\right)~~
\nonumber\\
&\times&\!\tilde{\phi}_{\nu^\pr}\left(k_1,{\cal P}-k-k_1\right)
\tilde{\phi}_{\nu^\pr}^\ast\left(k_3,{\cal P}-k-k_3\right)
\tilde{\phi}_{\nu}\left(p_e,{\cal P}-p_e\right) \frac{N_1\theta_{L1}}{DEN_1}
,~~\label{eq:l1}
\eea
where recall that
$\nu$ and $\nu^\pr$ are shorthands for $(n,l,m_l)$ and $(n^\pr,l^\pr,m_l^\pr)$
respectively,
the usual principal and angular momentum quantum 
  numbers of nonrelativistic hydrogen
(the ``$c$" on the sum emphasizes the fact that the continuum states must be
included also).
See Eq.~(\ref{eq:den1}) for $DEN_1$.
 $N_1$ is given by 
 \be
 N_1=
 \sum_{s_e^\pr s_\gam}
 \frac{\langle 0 | b_{s_e}(p_e^\pr) ~v_{ee\gam}~ b_{s_e^\pr}^\da(k_1)
 a_{s_\gam}^\da(k)
 |0\rangle \langle 0| b_{s_e^\pr}(k_3) a_{s_\gam}(k) ~v_{ee\gam}~ b_{s_e}^\da(p_e)
 |0\rangle}
 {\sqrt{p_e^+ {p_e^\pr}^+ k_1^+ k_3^+}
 \left[16 \pi^3 \del^3 (k+k_3-p_e)\right]\left[16 \pi^3 \del^3
 (k+k_1-p_e^\pr)\right]}
 \;,
 \ee
 which after some algebra becomes
\bea
N_1&=& (4 \pi \alpha) \left[
2 m^2\left(\frac{1}{p_e^+}-\frac{1}{k_3^+}\right)
\left(\frac{1}{{p_e^\pr}^+}-\frac{1}{k_1^+}\right)
\right.\nn\\
&&~~~~~~~~~~+\left.
\left(\frac{2
k^i}{k^+}-\frac{k_1^i(s_e)}{k_1^+}-\frac{{p_e^\pr}^i(\overline{s}_e)}{{p_e^\pr}^+}\right)
\left(\frac{2
k^i}{k^+}-\frac{p_e^i(s_e)}{p_e^+}-\frac{k_3^i(\overline{s}_e)}{k_3^+}\right)
\right]\label{eq:N1}
\;,
\eea
where we have defined a new object,
\bea
p^j(s)&=&p^j+i~s~ \eps_{jk} ~ p^k
\;.
\eea
Notation:  $j,\,k= 1,2$ only, $s= \pm 1$ only, $\overline{s} = -s$,
$\eps_{12}=-\eps_{21}=1$ and 
$\eps_{11}=\eps_{22}=0$.

We now discuss $\theta_{L1}$ and then simplify $\del M_{L1}^2$ further.
Recall Eqs.~(\ref{eq:30}), (\ref{eq:21}) and
 (\ref{eq:21p}). We 
see that 
after the coordinate change defined by Eq.~(\ref{eq:coord}), in the $m_p
\rightarrow \infty$
limit, 
\be
\theta_{L1}=\theta\left(
\tilde{\lam}-T
\right)\theta\left(
\tilde{\lam}-T^\pr
\right)\theta\left(\tilde{\lam}-\frac{{\bf p}^2}{2m}+{\cal O}(\alpha^4)\right)
\theta\left(\tilde{\lam}-\frac{{{\bf p}^\pr}^2}{2m}+{\cal O}(\alpha^4)\right)
~,
\ee
where
\be
T=|{\bf k}| + \sqrt{({\bf p}-{\bf k})^2+m^2}-m 
\label{eq:t1}
\ee
and $T^\pr=T|_{{\bf p}\longrightarrow{\bf p}^\pr}$.
We have used the fact that
the wave functions  restrict $|{\bf p}| \sim m \alpha$. Recall that  we are
always assuming
$m \alpha^2 \ll \tilde{\lam}\ll m \alpha$. Thus, $\theta_{L1}$ can be
simplified 
\be
\theta_{L1}\approx\theta\left(
\tilde{\lam}-T
\right)
\theta\left(
\tilde{\lam}-T^\pr
\right)\;.
\ee
From the form of Eq.~(\ref{eq:t1}), we see that this constrains
 the photon momentum to satisfy 
\bea
|{\bf k}|&\leq& \tilde{\lam}\;,\label{eq:lowcon}
\eea
to leading order in $\alpha$.

Note that the constraints coming from $\theta_{L1}$, summarized by
Eq.~(\ref{eq:lowcon}), require the photon three-momenta in $\del M_{L1}^2$ 
of Eq.~(\ref{eq:l1})
to satisfy
\bea
k&\ll& p_e,p_e^\pr
\;.
\eea
Thus, Eq.~(\ref{eq:l1}) can be simplified further
\bea
\del M_{L1}^2&\approx&{\sum_{\nu^\pr}}^c \int_k \int_{p_e} \theta\left({\cal P}^+
- p_e^+\right)
\int_{p_e^\pr}\theta\left({\cal P}^+ - {p_e^\pr}^+\right) 
 \left(p_e^+{ p_e^\pr}^+ \right) \tilde{\phi}_\nu^\ast\left(p_e^\pr,{\cal
 P}-p_e^\pr\right)
 ~~\nn\\
 &\times&
  \tilde{\phi}_{\nu^\pr}\left(p_e^\pr,{\cal P}-p_e^\pr\right)
\tilde{\phi}_{\nu^\pr}^\ast\left(p_e,{\cal P}-p_e\right)
\tilde{\phi}_\nu\left(p_e,{\cal P}-p_e\right)
\frac{N_1}{DEN_1}
\;,~~\label{eq:l1sec}
\eea
with the constraints $k_3=p_e$, $k_1=p_e^\pr$ and
$|{\bf k}| \leq \tilde{\lam}$.

In the  $m_p \longrightarrow \infty$
limit, ${\cal P}^+\longrightarrow m_p$, and $DEN_1$ becomes
\bea
DEN_1&=&2 m_p\left(\beta_n-\beta_{n^\pr}-|{\bf k}|\right)\left[
1+{\cal O}\left(1/m_p\right)\right]
\;,
\eea
where we 
recalled ${\cal M}_n^2\equiv (m+m_p)^2+2\,(m+m_p)\,\beta_n$ and
have used $\frac{{k^\perp}^2}{k^+}+k^+=2|{\bf k}|$, 
valid for an on mass-shell photon in the forward
light-cone.
$-\beta_n$ is the binding energy of nonrelativistic hydrogen 
with numerical value $Ryd/n^2$ for the bound-states.

In the region of integration $|{\bf k}| \leq \tilde{\lam}= m \alpha
\sqrt{\alpha}\ll |{\bf p}|$, so
after the coordinate change
of Eq.~(\ref{eq:coord})  [recall Eq.~(\ref{eq:coord2})] we have
\bea
\frac{N_1}{4 \pi \alpha}&=&\frac{4 {k^\perp}^2}{{k^+}^2}+\frac{4
{k^\perp}^2}{k^+ m}+
\frac{4 ~{ p}^\perp\cdot { p}^{\pr \perp}}{m^2} \nn\\
&&~~~~~-\frac{4k^\perp}{k^+ m}\cdot\left(  { p}^\perp+{ p}^{\pr \perp}
-\frac{{ p}^\perp p_z}{m}-\frac{{ p}^{\pr \perp}p^\pr_z}{m} \right)+{\cal
O}
\left(\alpha^2 \sqrt{\alpha}\right)
\;.
\eea
The rest of the integrand is even under $k^\perp \rightarrow - k^\perp$, so
these terms in the last
line, odd in $k^\perp$, do not contribute. 

Putting it all together,
recalling Eq.~(\ref{eq:binding}),
 we have
\bea
\del B_{L1}&=&\frac{\del M_{L1}^2}{2 m_p}\approx\frac{\alpha}{4 \pi^2 }
 {\sum_{\nu^\pr}}^c \int \frac{d^3 k}{  |{\bf k}| }
~\theta\left(\tilde{\lam}-|{\bf k}|\right)
\int d^3 p \int d^3 p^\pr
\phi_\nu^\ast\left({\bf p}^\pr\right)
\phi_{\nu^\pr}\left({\bf p}^\pr\right)\nn\\
&&~~~~~~~~~~~~~~~~~~~~~~~~~~~~~~\times
\phi_{\nu^\pr}^\ast\left({\bf p}\right)\phi_\nu\left({\bf p}\right)
\frac{\frac{ {k^\perp}^2}{{k^+}^2}+\frac{ {k^\perp}^2}{k^+ m}+
\frac{{ p}^\perp\cdot\, { p}^{\pr \perp}}{m^2}}
{\beta_n-\beta_{n^\pr}-|{\bf k}|}
\;,\label{eq:bl1}
\eea
where we recalled Eq.~(\ref{eq:jacdf}), the relation between
$\phi_N$ and $\tilde{\phi}_N$.
This is infrared ($k^+ \rightarrow 0$) divergent, but we must add diagram~L2 of
Figure~\ref{lambfig2}
 to get
the total low-energy shift. 

As previously mentioned,
Diagram~L2 of Figure~\ref{lambfig2} arises from the sum of an effective second-order 
electron self-energy interaction and 
a counterterm defined such that the electron self-energy runs coherently. 
 The result of this interaction is to add the following
to the binding
\bea
\del B_{L2}&=&\frac{\del M_{L2}^2}{2 m_p}=-\frac{\alpha}{4 \pi^2 }
 {\sum_{\nu^\pr}}^c \int \frac{d^3 k}{  |{\bf k}| }
~\theta\left(\tilde{\lam}-|{\bf k}|\right)
\int d^3 p \int d^3 p^\pr
\phi_\nu^\ast\left({\bf p}^\pr\right)
\phi_{\nu^\pr}\left({\bf p}^\pr\right)\nn\\
&&~~~~~~~\times
\phi_{\nu^\pr}^\ast\left({\bf p}\right)\phi_\nu\left({\bf p}\right)
\frac{\frac{ {k^\perp}^2}{{k^+}^2}+\frac{ {k^\perp}^2}{k^+ m}+
\frac{{ p}^\perp\cdot\, { p}^{\pr \perp}}{m^2}}
{\sqrt{{\bf p}^2+m^2} - \sqrt{({\bf p}-{\bf k})^2+m^2}-|{\bf k}|}
\;.
\eea
Given the  constraint  $|{\bf k}| \leq \tilde{\lam} \ll |{\bf p}|$, this 
becomes
\bea
\del B_{L2}&\approx&~ 
\frac{\alpha}{4 \pi^2 }
 {\sum_{\nu^\pr}}^c \int \frac{d^3 k}{  |{\bf k}| }
~\theta\left(\tilde{\lam}-|{\bf k}|\right)
\int d^3 p \int d^3 p^\pr
\phi_\nu^\ast\left({\bf p}^\pr\right)
\phi_{\nu^\pr}\left({\bf p}^\pr\right)\nn\\
&&~~~~~~~~~~~~~~~~~~~~~~~~~~~~~~\times
\phi_{\nu^\pr}^\ast\left({\bf p}\right)\phi_\nu\left({\bf p}\right)
\frac{\frac{ {k^\perp}^2}{{k^+}^2}+\frac{ {k^\perp}^2}{k^+ m}+
\frac{{ p}^\perp\cdot\, { p}^{\pr \perp}}{m^2}}
{|{\bf k}|}
\;.\label{eq:bl2}
\eea
This is the famous subtraction that Bethe performed in 1947 \cite{bethe}. 
In our approach it arose as a consequence of coupling coherence. 

$\del B_{L2}$ is infrared divergent ($k^+ \longrightarrow 0$)  as is
$\del B_{L1}$. This divergence
arises from the first two terms of $N_1$ (the ones independent
of ${\bf p}$ and ${\bf p}^\pr$). Noting that
\beaa
|{\bf k}| &=& \frac{1}{2} \left(\frac{{k^\perp}^2}{k^+}+k^+\right)
\stackrel{(k^+ \rightarrow 0)}{\longrightarrow}\frac{{k^\perp}^2}{2 k^+}
\;,
\eeaa
we have
\beaa
&&\beta_n-\beta_{n^\pr}-|{\bf k}| 
\stackrel{(k^+ \rightarrow 0)}{\longrightarrow}-\frac{{k^\perp}^2}{2 k^+}
\;,
\eeaa
and  these infrared divergent contributions from the 
first two terms of $N_1$  cancel, leaving an infrared finite shift,
\bea
\del B_L&\equiv& \del B_{L1}+\del B_{L2}\nn\\
&&~~\approx\frac{\alpha}{4 \pi^2 }
 {\sum_{\nu^\pr}}^c \int \frac{d^3 k}{  |{\bf k}| }
~\theta\left(\tilde{\lam}-|{\bf k}|\right)
\int d^3 p \int d^3 p^\pr
\phi_\nu^\ast\left({\bf p}^\pr\right)
\phi_{\nu^\pr}\left({\bf p}^\pr\right)\nn\\
&&~~~~~~~\times
\phi_{\nu^\pr}^\ast\left({\bf p}\right)\phi_\nu\left({\bf p}\right)
\frac{
{ p}^\perp\cdot\, { p}^{\pr \perp}}{m^2}\left(\frac{1}
{\beta_n-\beta_{n^\pr}-|{\bf k}|}+\frac{1}{|{\bf k}|}\right)
\label{eq:bl1total}
\\\\\\
&&~~= \left(\frac{2}{3}\right)\frac{\alpha}{4 \pi^2 }
 {\sum_{\nu^\pr}}^c \int \frac{d^3 k}{  |{\bf k}| }
~\theta\left(\tilde{\lam}-|{\bf k}|\right)
\int d^3 p \int d^3 p^\pr
\phi_\nu^\ast\left({\bf p}^\pr\right)
\phi_{\nu^\pr}\left({\bf p}^\pr\right)\nn\\
&&~~~~~~~~~~\times
\phi_{\nu^\pr}^\ast\left({\bf p}\right)\phi_\nu\left({\bf p}\right)
\frac{
{\bf p}\cdot {\bf p}^{\pr }}{m^2}\left(\frac{1}
{\beta_n-\beta_{n^\pr}-|{\bf k}|}+\frac{1}{|{\bf k}|}\right)
\;.\label{eq:bl1total2}
\eea
This last step followed after averaging over directions 
 as dictated by 
rotational invariance. 

Eq.~(\ref{eq:bl1total2}) is easy to integrate, and our final result for the 
low-energy photon contribution is
\bea
\del B_L
&=&\frac{2 \alpha}{3 \pi m^2}
 {\sum_{\nu^\pr}}^c 
 \left(\beta_{n^\pr}-\beta_n\right) \ln\left|
\frac{\tilde{\lam}+\beta_{n^\pr}-\beta_n}{\beta_{n^\pr}-\beta_n}\right|
\left|\langle
\phi_\nu|\hat{{\bf p}}|
\phi_{\nu^\pr}\rangle\right|^2
\label{eq:lastl1}\\
&\approx&\frac{2 \alpha}{3 \pi m^2}
 {\sum_{\nu^\pr}}^c 
 \left(\beta_{n^\pr}-\beta_n\right) \ln\left|
\frac{\tilde{\lam}}{\beta_{n^\pr}-\beta_n}\right|
\left|\langle
\phi_\nu|\hat{{\bf p}}|
\phi_{\nu^\pr}\rangle\right|^2
\;,\label{eq:7222}
\eea
where in this last step we recalled  $\tilde{\lam} \gg m \alpha^2$.
Note the $\tilde{\lam}$-dependence in the result.
This will cancel after we correctly add the contributions coming from high-energy intermediate
photons,  which now follows.
\subsection{High photon-energy contribution}\label{sec:high}

The high-energy shift arises from three sources which are shown in Figure~5.4. 
These
are 
first-order BSPT shifts due to third and fourth order effective interactions
(see Section~\ref{blochsec}). 
The net result of these three diagrams is\footnote{In practice we calculate
the effect from the third order effective interaction with instantaneous photon
exchange and then generalize $-\frac{\alpha}{2 \pi^2 q_z^2}\,
\del V_H
\longrightarrow-\frac{\alpha}{2 \pi^2 {\bf q}^2}\,
\del V_H$. It is assumed that the leading order helicity
conserving term in a nonrelativistic expansion of 
fourth order effective interactions
with dynamical photon exchange combine with the instantaneous photon exchange resulting
in the Coulomb interaction in the same manner as what led to the original
Coulomb interaction---the result is obvious, but the algebra was not carried
out.} 
\begin{figure}[t]\hskip1.0in
\centerline{\epsfig{figure=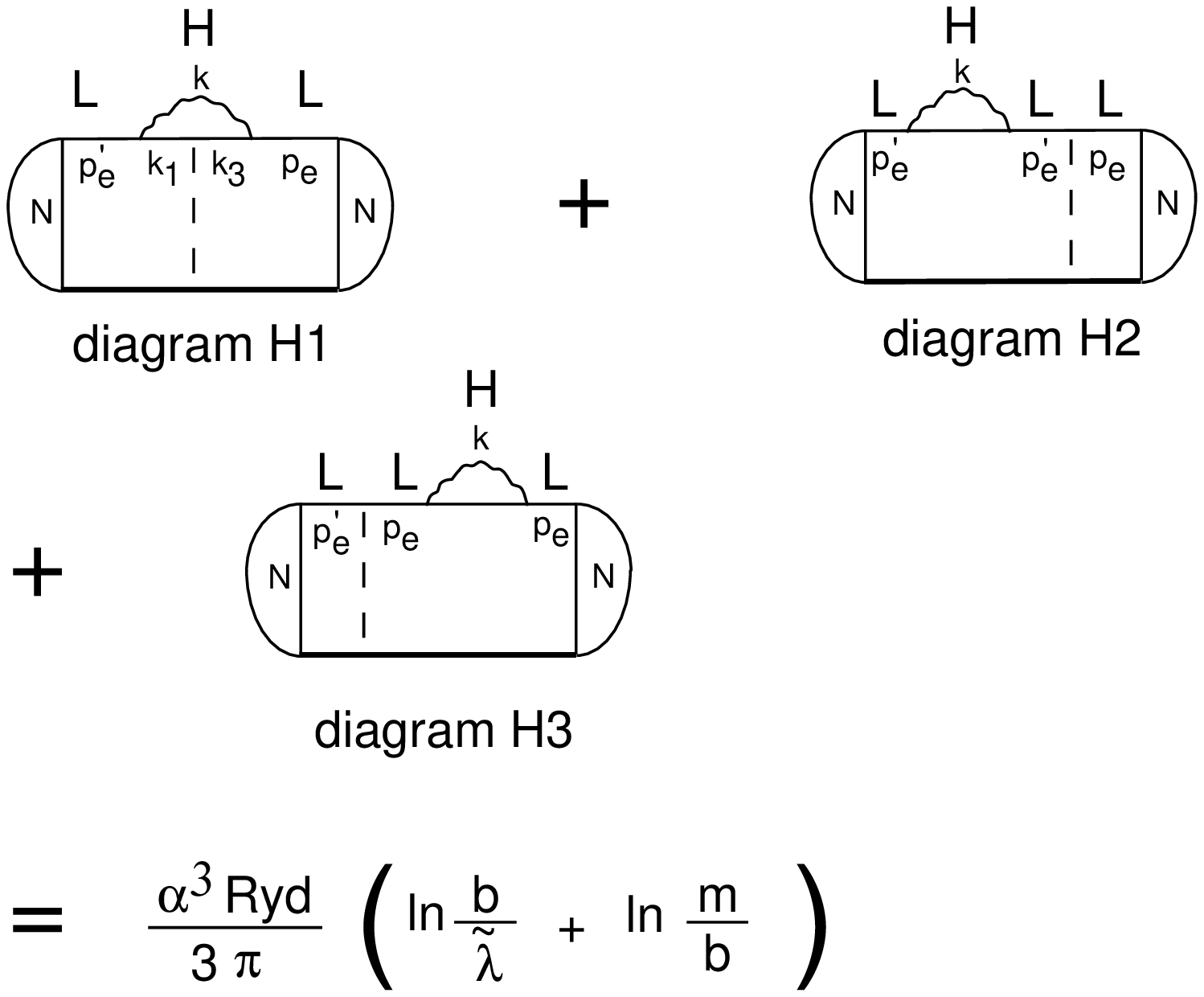,height=4.3in,width=5.0in}}
\caption{High photon-energy contribution to the Lamb shift. These are
third and fourth order effective interactions treated in 
first-order BSPT.
These effective interactions arise from the removal of photon emission above the cutoff
$\tilde{\lam}$.
`$b$' is an arbitrary scale, required to satisfy 
$m \alpha \ll b \ll m$, that was introduced to simplify the calculation. Note the 
$b$-independence of the result.
The total contribution is a sum of Figure~\ref{lambfig2} and Figure~\ref{lambfig4}. 
Note the
$\tilde{\lam}$-independence
of the combined result.} 
\label{lambfig4}
\end{figure}
\bea
-\frac{\alpha}{2 \pi^2 {\bf q}^2} &\longrightarrow&
-\frac{\alpha}{2 \pi^2 {\bf q}^2}\left(
1+\del V_H
\right)\;,\label{eq:hbaby}
\eea
where ${\bf q}$ is the exchanged momentum of the electron, and
\bea
\del V_H&=&\del V_{H1}+\del V_{H2}+\del V_{H3}
\;,
\eea
with
\bea
\del V_{H1}&=&\frac{1}{2}
\int_k\theta\left({p_e^\pr}^+ -k^+\right)\theta\left(p_e^+ -k^+\right)
N_{H1} ~\theta_{H1}\nn\\
&&~~\times \left(
\frac{\left({\cal
P}^+\right)^2}{\left(M_0^2-M^2\right)\left(M_0^2-{M^\pr}^2\right)}+
\frac{\left({\cal
P}^+\right)^2}{\left({M_0^\pr}^2-M^2\right)\left({M_0^\pr}^2-{M^\pr}^2\right)}
\right)
\label{eq:vh1}\;,~~\\
\del V_{H2}&=&-\frac{1}{2}
\int_k\theta\left({p_e^\pr}^+ -k^+\right)\theta\left(p_e^+ -k^+\right)
N_{H2} ~\theta_{H2} \nn\\
&&\hskip1.5in\times
\frac{\left({\cal
P}^+\right)^2}{\left(M_0^2-{M^\pr}^2\right)\left({M_0^\pr}^2-{M^\pr}^2\right)}
\label{eq:vh2}\;,\\
\del V_{H3}&=&-\frac{1}{2}
\int_k\theta\left({p_e^\pr}^+ -k^+\right)\theta\left(p_e^+ -k^+\right)
N_{H3}~ \theta_{H3} \nn\\
&&\hskip1.5in\times
\frac{\left({\cal
P}^+\right)^2}{\left(M_0^2-{M}^2\right)\left({M_0^\pr}^2-{M}^2\right)}
\label{eq:vh3}
\;.
\eea
The factors  $\pm \frac{1}{2}$ in front arise from the form of the Bloch
transformation
[see the third order effective interactions of
Eq.~(\ref{eq:heff})].
The vertex factors are given by
\bea
N_{H1}&=&\left(N_1\right)_{(k_1\rightarrow p_e^\pr-k,k_3 \rightarrow p_e-k)}
\label{eq:nh1}\;,\\
N_{H2}&=&\left(N_1\right)_{(k_1\rightarrow p_e^\pr-k,k_3 \rightarrow
p_e^\pr-k,p_e\rightarrow
p_e^\pr)}
\label{eq:nh2}\;,\\
N_{H3}&=&\left(N_1\right)_{(k_1\rightarrow p_e-k,k_3 \rightarrow
p_e-k,p_e^\pr\rightarrow
p_e)}
\label{eq:nh3}
\;,
\eea
 where $N_1$ was defined in Eq.~(\ref{eq:N1}). The free state masses are given
 by
 \bea
 M_0&=&\sqrt{{\bf p}^2+m^2}+\sqrt{{\bf p}^2+m_p^2}
 \;,\\
 M_0^\pr&=&\sqrt{{{\bf p}^\pr}^2+m^2}+\sqrt{{{\bf p}^\pr}^2+m_p^2}
 \;,\\
 M&=&|{\bf k}|+\sqrt{({\bf p}-{\bf k})^2+m^2}+\sqrt{{{\bf p}}^2+m_p^2}\;,
 \label{eq:ee}\\
 M^\pr&=&|{\bf k}|+\sqrt{({\bf p}^\pr-{\bf k})^2+m^2}+\sqrt{{{\bf
 p}^\pr}^2+m_p^2}\;.
 \label{eq:eee22}
 \eea
 
 The Bloch transformation constrains the free masses of the states.
 As discussed in the low photon-energy calculation, 
 the ``L" restrictions in Figure~\ref{lambfig4} can be removed given
 $\tilde{\lam} \gg m \alpha^2$. However, the ``H" restrictions lead to important
 constraints given by the $\theta_H$ factors above, which we now discuss.
 They constrain  the free masses to satisfy [recall
 Eqs.~(\ref{eq:21})-(\ref{eq:21p})]
\bea
&\tilde{\lam}\leq&M-m-m_p \leq \tilde{\Lam}\;,\label{eq:85}\\
&\tilde{\lam}\leq&M^\pr-m-m_p \leq \tilde{\Lam}\label{eq:86}
\;,
\eea
where $M$ and $M^\pr$ are defined in Eqs.~(\ref{eq:ee}) and (\ref{eq:eee22})
respectively.

As already mentioned, for convenience of calculation,
we will divide this high-energy contribution into two
regions, $\tilde{\lam} \leq |{\bf k}| \leq b$ and $b \leq |{\bf k}| \leq m$
({\em region one}
and {\em region two} respectively), with 
$m \alpha \ll b \sim m \sqrt{\alpha} \ll m$. Recall, $m \alpha^2 \ll
\tilde{\lam}
 \sim m \alpha \sqrt{\alpha} \ll m \alpha$. We now show how this division
 into these two regions arises as a result of 
  the constraints of Eqs.~(\ref{eq:85}) and (\ref{eq:86}).
 
 In this first region, $|{\bf k}| \ll m$, and  Eqs.~(\ref{eq:85}) and
 (\ref{eq:86}) become
 \bea
 \tilde{\lam}&\leq& |{\bf k}| +\frac{({\bf p}-{\bf k})^2}{2 m}\sim ~|{\bf k}|~
 \leq~ b~,\\
  \tilde{\lam}&\leq& |{\bf k}| +\frac{({\bf p}^\pr-{\bf k})^2}{2 m}\sim ~|{\bf
  k}|~ \leq ~b
 \;,
 \eea
 which is as we have already stated (recall that we always assume $m_p
 \rightarrow \infty$
 and drop the $1/m_p$ corrections since we are just after the dominant shift). 
 
  The analysis of the second region is slightly more complicated
 because $|{\bf k}| \gg m \alpha$, and near the upper limit $|{\bf k}| \sim m$.
 Since 
 $|{\bf k} |\gg m \alpha$ in this region, Eqs.~(\ref{eq:85}) and (\ref{eq:86})
 both
 become
  \bea
 b&\leq& |{\bf k}|+\sqrt{{\bf k}^2+m^2}-m ~\leq~ \tilde{\Lam}
  \;.
 \eea
 This is just a linear constraint,  
 \bea
&& b\left(\frac{2m+b}{2m+2b}\right)~\leq~|{\bf k}|~\leq~ \frac{\tilde{\Lam}}{2}
\left(\frac{\tilde{\Lam}+2 m}{\tilde{\Lam}+m}\right)
 \;,
 \eea
 which, since we choose $b \ll m$, becomes
 \bea
 && b~\leq~|{\bf k}|~\leq~ \frac{\tilde{\Lam}}{2}
\left(\frac{\tilde{\Lam}+2 m}{\tilde{\Lam}+m}\right)
 \;.\label{eq:rrr22}
 \eea
 We do not deal with removing the initial cutoff $\tilde{\Lam}$ dependence. 
 A full analysis of this dependence requires a complete 4th order calculation,
 which is beyond the scope of this dissertation. 
We cut off the photon momentum at the electron 
 mass, and proceed. 
 Note that from Eq.~(\ref{eq:rrr22}), this choice corresponds to $\tilde{\Lam}^2=2
 m^2$.
The point of calculating these high photon-energy contributions 
is to  show that our results
 are independent of the effective cutoff $\tilde{\lam}$
 by a consistent set of approximations [which is one step beyond Bethe's original
 calculation but still not at the level of calculation resulting in
 Eqs.~(\ref{eq:5.1}) and (\ref{eq:5.2})].

 Taking a sample denominator, in the {\it first} region
 \bea
 &&(M_0^2-M^2)=(M_0+M)(M_0-M)\nn\\
 &&\hskip1in\approx
 2 m_p \left(\frac{{\bf p}^2}{2m}-|{\bf k}| -\frac{({\bf p}-{\bf k})^2}{2
 m}\right)
 \approx ~-2 m_p|{\bf k}|
 \;,
 \eea
 and in  the {\it second} region
  \bea
 &&(M_0^2-M^2)=(M_0+M)(M_0-M)\nn\\
 &&\hskip1in\approx 2 m_p \left(
 m-|{\bf k}|-\sqrt{{\bf k}^2+m^2}\right)\approx-2 m_p\left(|{\bf k}|+\frac{|{\bf
 k}|^2}{2m}\right)
 \;.
 \eea
 
 Using these previous formulas, including
 ${\cal P}^+\longrightarrow m_p$ as $m_p \longrightarrow \infty$,
 Eqs.~(\ref{eq:vh1})-(\ref{eq:vh3}),
 after summing, become
 \bea
 \del V_H^\pr&=&
 -\frac{\alpha}{4 \pi} \frac{{q^\perp}^2}{m^2}
 \int_{-1}^{1} d \cos\theta \int_{\tilde{\lam}}^b \frac{d |{\bf k}|}
 {|{\bf k}|} \left[1 +
 {\cal O}\left(\frac{|{\bf k}|}{m}\right)
 \right]
 \eea
 in the first region (the ``prime" on $\del V_H^\pr$ signifies the {\em first}
 region), and
 \bea
  \del V_H^{\pr \pr}&=&
  -\frac{\alpha}{4 \pi} \frac{{q^\perp}^2}{m^2}
  \int_0^{2\pi}\frac{d\phi}{2 \pi}
  \int_{-1}^{1} d \cos\theta \int_{b}^{m} \frac{d |{\bf k}|}
 {|{\bf k}|} \left[1
 +c_n\frac{|{\bf k}|}{m}
 \left(1+\cos\theta\right)
 \right.\nn \\
 &&\hskip1.2in -c_d\frac{|{\bf k}|}{m}-2\, i \,s_e 
 \frac{{ p}^\perp\!\times{{ p}^{\pr\perp}}}{{q^\perp}^2}\,\frac{|{\bf k}|}{m}
 +\left.
 {\cal O}\left(\frac{|{\bf k}|^2}{m^2}\right)
 \right]
  \label{eq:wyatt}
 \eea
 in the second region (the ``double prime" on $\del V_H^{\pr\pr}$ signifies the {\em
 second} region). 
 In the second region since the photon momentum is not necessarily smaller
 than the electron mass, we have kept two terms in the $|{\bf k}|/m$ expansion
 of the integrand.
 In the ${\cal O}(|{\bf k}|/m)$ terms we have introduced two constants, $c_n$ and
 $c_d$, which
 denote {\it numerator} and energy {\it denominator} corrections respectively.
 Hereafter we
 set $c_n=1$ and $c_d=1$, as given by the 
 theory.
 
 This term in the last line of Eq.~(\ref{eq:wyatt})
  that is dependent on the electron
 helicity $s_e/2$ is written for completeness, but it does not contribute to
 the Lamb shift as will now be explained. 
 An expectation value of ${ p}^\perp\!\times{{ p}^{\pr\perp}}$ 
 in the $2S_{\frac{1}{2}}$ state vanishes, and in the
 $2P_{\frac{1}{2}}$ state vanishes {\it after} an average over $m_l$
 is taken as dictated by rotational invariance. Also dictated by rotational
 invariance is an average over $m_s\equiv s_e/2$, and then the term in question proportional
 to $m_s$ obviously vanishes.
 For details
 on why rotational invariance dictates these averages to be taken see the
 end of Appendix~\ref{app:avg}.
 
 Note that in combining $\del V_{H1}$, $\del V_{H2}$ and $\del V_{H3}$, many
 cancelations
 occur; most noteworthy, 
  each contribution is individually infrared divergent ($k^+ \rightarrow 0$),
  but in the sum
 the divergences cancel.
 These final equations are easily integrated, and we have
 \bea
 \del V_H^\pr&=&
 -\frac{\alpha}{2 \pi}
 \frac{{q^\perp}^2}{m^2}\ln\left(\frac{b}{\tilde{\lam}}\right)
 \;,\label{eq:1}\\
  \del V_H^{\pr \pr}&=&
  -\frac{\alpha}{2 \pi} \frac{{q^\perp}^2}{m^2}\ln\left(\frac{m}{b}\right)
  \;.\label{eq:2}
 \eea
 In the second region note that the  ${\cal O}(|{\bf k}|/m)$ terms coming
 from numerator and energy denominator corrections 
  cancel, leaving the ${\cal O}(1)$ piece alone.
 The combined high-energy contribution is
 \bea
&& \del V_H=\del V_H^\pr+\del V_H^{\pr \pr}=
-\frac{\alpha}{2 \pi}
\frac{{q^\perp}^2}{m^2}\ln\left(\frac{m}{\tilde{\lam}}\right)
\label{eq:highfinal}
 \;,
 \eea
 which is independent of $b$,
 as required for consistency. Recall that 
 ${\bf q}={\bf p}^\pr-{\bf p}$: the difference between the final and initial
 electron momenta.
 
 From the definition of $\del V_H$
 [see Eq.~(\ref{eq:hbaby})], 
 this correction shifts the energy levels an amount
 \bea
 \del B_H&=&-\frac{\alpha}{2 \pi^2}\int d^3p ~
 d^3 p^\pr \phi_\nu^\ast\left({\bf p}^\pr\right) \left(
 \frac{\del V_H}{\pp}
 \right)\phi_\nu\left({\bf p}\right)
 \;.
 \eea
 Combining this with Eq.~(\ref{eq:highfinal}) gives
\bea
 \del B_H&=&\frac{\alpha^2}{4 \pi^3 m^2}
\ln\left(\frac{m}{\tilde{\lam}}\right)\int d^3p ~
 d^3 p^\pr \phi_\nu^\ast\left({\bf p}^\pr\right) \left(
 \frac{\left({ p}^\perp-{ p}^{\pr \perp}\right)^2}{\pp}
\right)\phi_\nu\left( {\bf p} \right) \label{eq:100}\\
 &=&\left(\frac{2}{3}\right)\frac{\alpha^2}{4 \pi^3 m^2}
 \ln\left(\frac{m}{\tilde{\lam}}\right)\left(\int d^3p ~
 \phi_\nu\left({\bf p}\right)\right)^2
 \;,\label{eq:10122}
 \eea
 where in this last step we averaged over directions and noted that the wave
 function
 at the origin is real. For more details on this averaging over directions see
 Appendix~\ref{app:avg}.
\subsection{Total contribution}

In this subsection we combine the results of the last two subsections for the low and
high photon-energy contributions, and perform the required sums/integrations
to calculate the total shift between the $2S_{\frac{1}{2}}$ and
$2P_{\frac{1}{2}}$ energy levels
of hydrogen.

Adding Eqs.~(\ref{eq:7222}) and (\ref{eq:10122}) gives for the total shift
\bea
\del B&=&\del B_L+\del B_H=\frac{2 \alpha}{3 \pi m^2}
 {\sum_{\nu^\pr}}^c 
 \left(\beta_{n^\pr}-\beta_n\right) \ln\left|
\frac{\tilde{\lam}}{\beta_{n^\pr}-\beta_n}\right|
\left|\langle
\phi_\nu|\hat{{\bf p}}|
\phi_{\nu^\pr}\rangle\right|^2\nn\\
&&~~~~~~~~~~~~~~~~~~~~~~~~~~+~\frac{\alpha^2}{6 \pi^3 m^2}
 \ln\left(\frac{m}{\tilde{\lam}}\right)\left(\int d^3p ~
 \phi_\nu\left({\bf p}\right)\right)^2
\;.\label{eq:blog}\eea

For the second term we have
\bea
\left(\int d^3 p~ \phi_\nu\left({\bf p}\right)\right)^2
 &=&  \left((2 \pi)^{\frac{3}{2}}\phi_\nu\left({\bf x}=0\right)\right)^2
=\frac{(2 \pi)^{3}} {\pi}\left(\frac{m \alpha}{n}\right)^3\del_{l,0}
\;.\label{eq:103}
\eea
The $(2 \pi)^3$ factor arose because of our normalization choice [recall
Eq.~(\ref{eq:norm22})].

The first term of Eq.~(\ref{eq:blog}) 
is the famous Bethe log and must be calculated numerically, summing
over all bound and continuum states. Following standard convention we define
an average excitation energy, $\overline{\beta}(n,l)$, 
\bea
&&\ln\left(\frac{\overline{\beta}(n,l)}{Ryd}\right)
{\sum_{\nu^\pr}}^c 
 \left(\beta_{n^\pr}-\beta_n\right) 
\left|\langle
\phi_\nu|\hat{{\bf p}}|
\phi_{\nu^\pr}\rangle\right|^2\nn\\
&&\hskip1.3in
\equiv
{\sum_{\nu^\pr}}^c 
 \left(\beta_{n^\pr}-\beta_n\right) \ln\left|
\frac{\beta_{n^\pr}-\beta_n}{Ryd}\right|
\left|\langle
\phi_\nu|\hat{{\bf p}}|
\phi_{\nu^\pr}\rangle\right|^2
\;.\label{eq:4422}\eea
The sum on the left is evaluated by standard techniques [$H_c=p^2/(2m)-\alpha/r$
],
\bea
&&{\sum_{\nu^\pr}}^c 
 \left(\beta_{n^\pr}-\beta_n\right) 
\left|\langle
\phi_\nu|\hp|
\phi_{\nu^\pr}\rangle\right|^2
=\frac{1}{2}\langle
\phi_\nu\left|
\left[\hp,H_c\right]\cdot \hp+\hat{{\bf p}}
\cdot\left[H_c,\hp\right]\right|
\phi_{\nu}\rangle\nn\\
&&~~~~~~~~~~=-\frac{1}{2}\langle
\phi_\nu\left|\left[
\hp
\cdot,\left[\hp,H_c\right]\right]\right|
\phi_{\nu}\rangle=-\frac{1}{2}\left\langle
\phi_\nu\left|\left[
\hp
\cdot,-i\vec{\nabla}\left(-\alpha/r\right)\right]\right|
\phi_{\nu}\right\rangle\nn\\
&&~~~~~~~~~~=
-\frac{1}{2}\left\langle
\phi_\nu\left|(-i)^2{\vec{\nabla}}^2\left(-\alpha/r\right)\right|
\phi_{\nu}\right\rangle=
-\frac{1}{2}(-i)^2 (-\alpha) (-4\pi)\left\langle
\phi_\nu\left|\del^3\left(r\right)\right|
\phi_{\nu}\right\rangle\nn\\
&&~~~~~~~~~~=2  \alpha \left(\frac{m \alpha}{n}\right)^3\del_{l,0}\;.
\eea
This vanishes for $l \neq 0$, but the average excitation energy,
$\overline{\beta}(n,l)$,
is  defined [it is just a way to catalogue the numerical sum on the right
of Eq.~(\ref{eq:4422}), the quantity
we need to know]
with the sum on the
left-hand-side set to its value for $l=0$. In summary, $\overline{\beta}(n,l)$
for all states is defined by
\bea
\ln\left(\frac{\overline{\beta}(n,l)}{Ryd}\right)
2  \alpha \left(\frac{m \alpha}{n}\right)^3
\equiv
{\sum_{\nu^\pr}}^c 
 \left(\beta_{n^\pr}-\beta_n\right) \ln\left|
\frac{\beta_{n^\pr}-\beta_n}{Ryd}\right|
\left|\langle
\phi_\nu|\hat{{\bf p}}|
\phi_{\nu^\pr}\rangle\right|^2
\;.\label{eq:45}\eea
Without further ado, this sum has been evaluated  by R. W. Huff \cite{huff}.
His results for the $n=2$ levels are
\bea
\overline{\beta}(2,0)&=&
16.63934203(1)~
Ryd\;,\label{eq:107}\\
\overline{\beta}(2,1)&=&
0.9704293186(3)~
Ryd
\;,\label{eq:108}
\eea
where the figures in parentheses give the
number of units of estimated error in the last decimal place (R. W. Huff's
estimates).

Combining the results:
\bea
\del B_{2S_{\frac{1}{2}}}&=&\frac{2 \alpha}{3 \pi m^2} 
\ln\left(\frac{\tilde{\lam}}{\overline{\beta}(2,0)}\right)
2  \alpha \left(\frac{m \alpha}{n}\right)^3+\frac{\alpha^2}{6 \pi^3 m^2} 
\ln\left(\frac{m}{\tilde{\lam}}\right)
\frac{(2 \pi)^{3}} {\pi}\left(\frac{m \alpha}{n}\right)^3\nn\\
&=& \frac{ \alpha^3 Ryd}{3 \pi }
\ln\left(\frac{m}{\overline{\beta}(2,0)}\right)\;,\\
\del B_{2P_{\frac{1}{2}}}&=&\frac{2 \alpha}{3 \pi m^2} 
\ln\left(\frac{Ryd}{\overline{\beta}(2,1)}\right)
2  \alpha \left(\frac{m \alpha}{n}\right)^3\nn\\
&=& \frac{ \alpha^3 Ryd}{3 \pi }
\ln\left(\frac{Ryd}{\overline{\beta}(2,1)}\right)
\;.
\eea
Note the cancelation of the $\tilde{\lam}$-dependence.
Thus, the dominant part of the Lamb shift is 
\bea
\del B_{_{Lamb}}&=&\del B_{2S_{\frac{1}{2}}}-\del B_{2P_{\frac{1}{2}}}=
\frac{ \alpha^3 Ryd}{3 \pi }
\ln\left(\frac{m~\overline{\beta}(2,1)}{Ryd~\overline{\beta}(2,0)}\right)
\\
&=& (1047-4)~{\rm MHz}~ (2 \pi \hbar)= 1043~{\rm MHz}~ (2 \pi \hbar)
\;, 
\eea
where we use \cite{expt} and
the average excitation energies of Eqs.~(\ref{eq:107}) and (\ref{eq:108}).
Note that the $2P_{\frac{1}{2}}$ shift is only about one-half of a percent of
the
 $2S_{\frac{1}{2}}$ shift (but both shifts are ``up").
\section{Lamb shift summary and discussion}\label{lamb4}

\begin{figure}[t]\vskip-1in
\centerline{\epsfig{figure=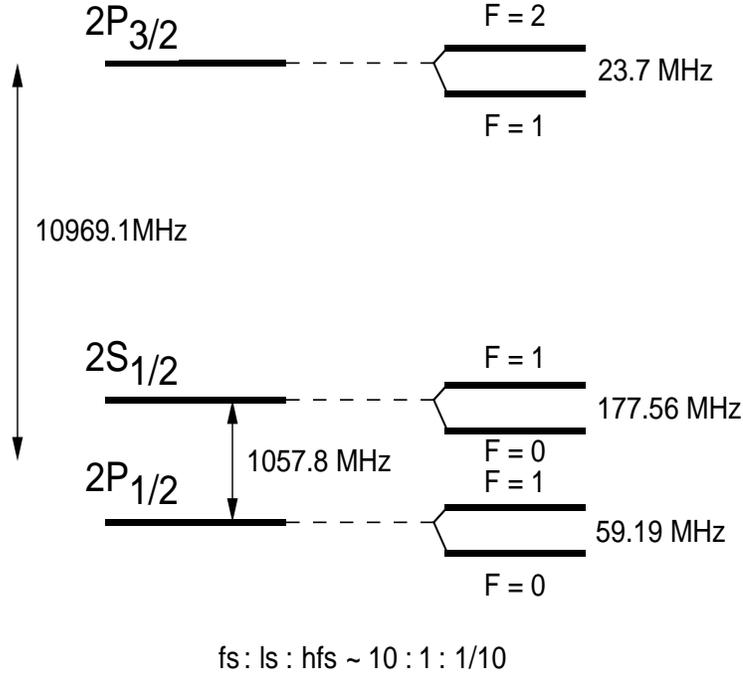,height=4.5in,width=4.0in}}
\caption{The experimental $n=2$ hydrogen spectrum: fine structure, Lamb shift
and hyperfine
structure.
${\bf F}={\bf L}+{\bf S_e}+{\bf S_p}$.} 
\label{lambfig5}
\end{figure}

In a light-front Hamiltonian approach,
 we have shown how to do a consistent Lamb shift calculation for the $n=2$,
 $j=1/2$
levels of hydrogen over the photon energy scales
\beaa
&&0\lr m \alpha^2\lr\tilde{\lam}\lr m \alpha \lr b \lr m
\;,
\eeaa
with the choices $m \alpha^2 \ll \tilde{\lam} \ll m \alpha$ and $m \alpha \ll b
\ll m$.
In a consistent set of diagrams we showed how $\tilde{\lam}$- and $b$-dependence
cancel
leaving the dominant part of the Lamb shift,
 $1043$ MHz. For completeness, the experimental
  $n=2$ spectrum of hydrogen is shown in Figure~\ref{lambfig5}.
 
 If we
 compare the three regions we see the following results (we only compare for the
 $2S_{\frac{1}{2}}$ shift since the $2P_{\frac{1}{2}}$ shift is negligible
 within our errors)
\bea
&&(0\leq|{\bf k}|\leq\tilde{\lam})~
\del B_{_{Lamb}}\sim
\frac{ \alpha^3 Ryd}{3 \pi }
\ln\left(\frac{\tilde{\lam}}{16.64 ~Ryd}\right)\sim
 46~{\rm MHz}~ (2 \pi \hbar)\;,~~
 \label{eq:ttbbg}\\ 
&&(\tilde{\lam}\leq|{\bf k}|\leq b)~
\del B_{_{Lamb}}\sim 
\frac{ \alpha^3 Ryd}{3 \pi }
\ln\left(\frac{b}{\tilde{\lam}}\right)\sim
667~{\rm MHz} ~(2 \pi \hbar)\;,\\
&&(b\leq|{\bf k}|\leq m)~\del B_{_{Lamb}}\sim 
\frac{ \alpha^3 Ryd}{3 \pi }
\ln\left(\frac{m}{b}\right)\sim
334~{\rm MHz}~ (2 \pi \hbar)
\;,
\eea
where we used $\tilde{\lam} = m \alpha \sqrt{\alpha}$ and $b= m \sqrt{\alpha}$,
consistent
choices used throughout 
 this chapter.
As expected on physical grounds, the dominant photon
 momentum that couples to hydrogen is
\bea
&&|{\bf k}| \sim 1/a_0 \sim m \alpha\;.
\eea
 That is, the photon wavelength  that couples the strongest to
 the neutral hydrogen system is of order the size of the system.
 As seen above,  the effects
of photons of this
momentum amount to about two-thirds of the Lamb shift, the dominant part of this
experimentally
observed shift.

\chapter{Summary and Discussion}

In positronium, in a consistent set of approximations,
we analytically
calculated the
 spin {\em splitting} for the $n=1$ levels through order $\alpha^4$. 
To go beyond this, and calculate the spin splitting through order $\al^5$
for example, it is clear that the effective Hamiltonian $\hr$ must be
derived through fourth order.

 Kaluza and Pirner were the first to {\it numerically} calculate the 
  ground state spin splitting of positronium (neglecting
 annihilation channel contributions)
 through order $\alpha^4$
 in a {\it light-front hamiltonian approach}~\cite{kaluzapirner},\footnote{
 A recent thesis that summarizes and extends this work is
 \cite{uwe}.}
  but they were forced to make ad hoc
 assumptions because
 their Hamiltonian depended on the full eigenvalue of the problem.
 We avoided these assumptions 
 in our approach,
 and performed the calculation analytically.
 
     The dominant part of the Lamb shift
  was calculated with
 result
 $1043$ MHz. 
 This turned out to be accurate,
 and within a consistent set of approximations
 was shown to be independent of our final cutoff $\tilde{\lam}$.
 It was important to establish this independence to validate the placement of
 our final cutoff in the range $m\al^2 \ll \tilde{\lam}\ll m \al$---which
 is necessary to obtain the result in the few-body sector alone.
 We did not show that the result was independent of the
  initial cutoff $\tilde{\Lam}$. We fixed $\tilde{\Lam}$
   such that the photon energy in the respective loop
   satisfied $|{\bf k}|_{photon}\leq m$,
 and did not derive the $\tilde{\Lam}$-independence of the result.
To obtain more precision, the full one-loop renormalization 
must be
performed
of course.\footnote{For an initial study of one-loop perturbative
renormalization of QED in a light-front Hamiltonian approach with various regulators
see \cite{pqed}. }
A final note on the Lamb shift calculation: 
Each of our five diagrams (of Figures~\ref{lambfig2} and
\ref{lambfig4}) were infrared
($k^+_{photon} \longrightarrow 0$)
divergent; however, both the sum of the two low photon-energy diagrams and the
sum of the 
three high photon-energy diagrams were infrared finite. This is an
encouraging result
for future work.

We close with a discussion of rotational invariance
in the context of light-front field theory,
and how the results of this dissertation fit into the big picture.
 The transverse
rotation generators are dynamic and our regulator
is not rotationally symmetric,\footnote{Since the generators
are dynamic, we do not even know how to regulate in a rotationally symmetric
fashion until the
dynamic solution is known.}
thus non-covariant counterterms will in general be required to obtain
a dynamic solution which adheres to the principles of 
rotational invariance.
An initial study \cite{matrot}
at the one- and two-loop level in Yukawa theory showed
 in perturbation theory that rotationally invariant results follow
 with an appropriate renormalization of the kinetic {\it and} vertex masses.
 We continued the study of rotational invariance in this dissertation from
 the perspective of the QED bound-state problem.
 In Chapter~\ref{ch:pos} we showed in a consistent set of approximations
 through order $\al^4$ in the ground state spin splitting of positronium that
 the result is rotationally symmetric and followed from the canonical
 structure of the interactions;\footnote{This necessarily
 included one loop renormalization of the fermion kinetic mass. Strictly
 speaking (from the transverse renormalization group sense), an infinite number
 of counterterms were required---these were fixed by coupling coherence.}
 however, this result could only be obtained after including the dynamic shifts
 from second order bound-state perturbation theory which necessarily included a
 sum over all intermediate bound and scattering Coulomb states. 

 As is no  surprise, the generators are dynamic, and
 rotational invariance comes about in a dynamically 
 complicated way even at the no-loop level for the bound-state problem---analytic understanding of this
 was presented in Chapter~\ref{ch:pos} and may be useful for future work.
 
 In Chapter~\ref{ch:lamb} we obtained the dominant part 
 of the Lamb shift of hydrogen consistently through order $\al^5\ln(1/\al)$.
 This showed that the level classification---$2S_{1/2}$ and $2P_{1/2}$ for
 this example---as dictated by rotational invariance was maintained in this example. 
 This calculation necessarily included one-loop electron
 kinetic mass renormalization which followed from coupling coherence.
 
 Neither example required the problematic
 diagrams mentioned in \cite{matrot} with an instantaneous fermion line 
 immediately preceding external self-energy shifts, thus  we could set the vertex mass
 to the kinetic mass (actually it was required to get correct results). However,
 since the transverse rotation generators are dynamic, more nonperturbative analysis
 (for the positronium example) was required than in an
 equal-time calculation. Analytic understanding of the bound-state
 problem in QED in a light-front hamiltonian approach has been advanced.


%
%

\appendix
\chapter{Summary of light-front conventions}\label{appendixlf}

In this appendix we  write our light-front
 conventions for the electron, proton,
and photon
system (including 
the antiparticles).
 The reader unaccustomed to light-front coordinates may want
to read Chapter~\ref{ch:2} before consulting this appendix. Our conventions are
\beaa
&\bullet&  V^{\pm}=V^0 \pm V^3~~\rm{where}~V^\mu ~{\rm is~ any~ four\!\!-\!\!vector,}\\
&\bullet& 
\gamma^+=\left[
\begin{array}{cc}
0&0\\
2 i&0
\end{array}
\right]~;~
\gamma^-=\left[
\begin{array}{cc}
0&-2 i\\
0&0
\end{array}
\right]\;,\\
&\bullet&
\alpha^i=\gamma^0 \gamma^i=\left[
\begin{array}{cc}
0&\sig^i\\
\sig^i&0
\end{array}
\right]~;~i=1,2~;~\sig^i~{\rm are}~SU(2)~{\rm Pauli}~{\rm matrices\;,}\\
&\bullet&\Lambda_+=\frac{1}{2}\gamma^0\gamma^+=
\left[
\begin{array}{cc}
1&0\\
0&0
\end{array}
\right]~;~
\Lambda_-=\frac{1}{2}\gamma^0\gamma^-=
\left[
\begin{array}{cc}
0&0\\
0&1
\end{array}
\right]\;,\\
&\bullet&\psi_\pm=\Lambda_\pm \psi~;~\psi=\psi_+ + \psi_-\;,\\
&\bullet&\psi_{e+}=\left[
\begin{array}{c}
\xi_e\\
0
\end{array}
\right]~;~
\psi_{p+}=\left[
\begin{array}{c}
\xi_p\\
0
\end{array}
\right]~;~ e~{\rm for~ electron}, ~p~{\rm for~ proton\;,}\\
&\bullet&\psi_{e-}=\left[
\begin{array}{c}
0\\
\frac{1}{i \partial^+}\left\{\left[
\sig^i\left(
i \partial^i+e A^i
\right)+i m
\right]\xi_e\right\}
\end{array}
\right]\;,\\
&\bullet&\psi_{p-}=\left[
\begin{array}{c}
0\\
\frac{1}{i \partial^+}\left\{\left[
\sig^i\left(
i \partial^i-e A^i
\right)+i m_p
\right]\xi_p\right\}
\end{array}
\right]\;,\\
&\bullet&
A^-= \frac{-2}{( \partial^+)^2} J^+ + 2 \frac{ \partial^i}{ \partial^+} A^i\;,\\
&\bullet& J^+=2 e \left(
\xi_p^\dagger \xi_p-\xi_e^\dagger \xi_e
\right)~;~ e > 0\;,\\
&\bullet&A^+=0\;.
\eeaa

In momentum space the dynamical
field
operators are expanded as (at $x^+ = 0$)
\beaa
A^i(x)&=&\sum_{s = \pm 1} \int_{q} (\eps^i_s a_s(q)
e^{-iq \cdot x}+h.c.)\;,\\
\xi_e(x)&=&\sum_{s = \pm 1}\chi_s \int_p \sqrt{p^+}(b_s(p)
e^{-ip \cdot x}+d_{{\overline s}}^{\dagger}(p)e^{+ip\cdot x})\;,\\
\xi_p(x)&=&\sum_{s = \pm 1}\chi_s \int_p \sqrt{p^+}(B_s(p)
e^{-ip \cdot x}+D_{{\overline s}}^{\dagger}(p)e^{+ip\cdot x})\;,\\
{\rm with}
&&\eps_1^i=\frac{-1}{\sqrt{2}}\left(\delta_{i,1} + i ~\delta_{i,2}\right)
\;,\;\eps_{-1}^i=\frac{1}{\sqrt{2}}\left(
\delta_{i,1} - i~ \delta_{i,2}\right)\;,\\
&&\chi_{_{1}}=
\left(\begin{array}{c}
1\\
0
\end{array}\right)\;,\;\chi_{_{\overline 1}}=
\left(\begin{array}{c}
0\\
1
\end{array}\right)
\;.
\eeaa
Note that ${\overline s} \equiv -s$. Also, we are using the shorthand
\beaa
\int_p f(p)&=&\int\frac{d^4p}{(2 \pi)^4}~2 \pi~ \del (p^2-m^2)~
\theta(p^0)~f(p)=
\int\frac{d^2p^\perp dp^+ \theta(p^+)}{16 \pi^3 p^+}f(p)
{\Biggr{|}}_{p^-=\frac{{p^\perp}^2+m^2}{p^+}}
~.
\eeaa
The fermion  helicity can only take
on the  values
$ \pm 1/2$, however we define 
$h_3=  s/2$; therefore, ``s" can only take
on the values
$\pm 1$. 
The non-vanishing
commutation (anti-commutation) relations and the free Fock states are given by
\beaa
&&[a_\lambda(q),a_{\lambda^\prime}^{\dagger}(q^\prime)]=16\pi^3q^+
\del^3(q-q^\prime)\del_{\lambda \lambda^\prime}\;,\;[\;\del^3(p)=
\del^2(p^\perp)\del(p^+)\;]\;,\\
&&\{b_s(p),b_{s^\prime}^{\dagger}(p^\prime)\}=
\{d_s(p),d_{s^\prime}^{\dagger}(p^\prime)\}=
16\pi^3p^+
\del^3(p-p^\prime)\del_{s s^\prime}\;,\\
&&\{B_s(p),B_{s^\prime}^{\dagger}(p^\prime)\}=
\{D_s(p),D_{s^\prime}^{\dagger}(p^\prime)\}=
16\pi^3p^+
\del^3(p-p^\prime)\del_{s s^\prime}\;,\\
&&\langle p_1 s_1 | p_2 s_2 \rangle = 16\pi^3p_1^+
\del^3(p_1-p_2)\del_{s_1 s_2}\;,\;|p_1 s_1\rangle=
b_{s_1}^{\dagger}(p_1)|0\rangle\;,\;{\rm etc}\;.
\eeaa

\chapter{Matrix elements of the canonical Hamiltonian used in
the dissertation}\label{brentrules}

In this appendix, the matrix elements
 of the 
canonical Hamiltonian $H$ in
the basis of the free Hamiltonian $h$ that were 
used to derive the results in this dissertation
are given. See 
Eqs.~(\ref{eq:can1})--(\ref{eq:instant}) for the operator form of $H$.
These matrix elements were worked out
by Allen  \cite{brent}.
 Note that these matrix elements correspond
 to normal-ordered interactions with 
 zero modes dropped.\footnote{As explained in the dissertation, coupling
 coherence fixes all counterterms, and it is our conjecture that 
 it is not necessary to start with
 ``self-inertias" or ``zero-modes." If they are required for the renormalized
 Hamiltonian $\hr$ to run coherently, they will arise through the process of
 renormalization.}
 
 First, some comments and notation for clarity
 and completeness of this appendix:
 $H=h+v$ where $h$ is the free field theoretic
  Hamiltonian.
  $h |i\rangle=\veps_i |i\rangle$ and the  matrix elements below are in
  this basis $\{|i\rangle\}$.
  ($e$, $p$, ${\overline e}$, $\gam$) labels represent
  (electron, proton, positron, photon) respectively. Transverse momenta are
  written as $p^i_1,\,p^i_2,\,\ldots$ ($i=1,\,2$ only as always) and 
  longitudinal momenta are written as 
  $p^+_1,\,p^+_2,\,\ldots$. Repeated transverse indices  are to be summed.
  The $1/2$ is factored out of the fermion helicities so all helicities
  below are written as $s_1,\,s_2,\,\ldots$, where $s_i=\pm 1$ only.
  Also note ${\overline s}_i=-s_i$.
  $m$ is the renormalized electron mass, $m_p$ is the renormalized proton mass
  and $-e$ [$e$] is the renormalized electric coupling of the electron 
  [proton] ($e>0$).
 The photon polarization vectors are
 \be
 \eps_1^j=\frac{-1}{\sqrt{2}}\left(\delta_{j,1} + i ~\delta_{j,2}\right)
\;,\;\eps_{-1}^j=\frac{1}{\sqrt{2}}\left(
\delta_{j,1} - i~ \delta_{j,2}\right)\;.
 \ee
 We also use
 \bea
p^j(s)&=&p^j+i~s~ \eps_{jk} ~ p^k
\;,
\eea
with $\eps_{12}=-\eps_{21}=1$ and 
$\eps_{11}=\eps_{22}=0$. A final convenient notation is
\be
\left\langle a\left|v\right
|b\right\rangle=16 \pi^3 \del^3(p_a-p_b) \left( a|v|b\right)
\;,
\ee
where $p_a$ and $p_b$ are total three-momenta of the respective state and
$\del^3(p)=\del^2(p^\perp)\del(p^+)$.

The eleven canonical matrix elements used in this dissertation are
\bea
\left( e_2\gam_3\left|v\right| e_1\right)&=&-e \sqrt{p_1^+p_2^+}\left[\del_{s_1 s_2}
\eps_{s_3}^{i\ast}\left(2\frac{p_3^i}{p_3^+}-\frac{p_1^i(s_1)}{p_1^+}-
\frac{p_2^{i\ast}(s_2)}{p_2^+}\right)\right.\nn\\
&&\hskip.85in +\left.\del_{{\overline s}_1s_2}\del_{s_1s_3}\frac{2i}{\sqrt{2}}{\overline s}_1 m
\left(\frac{1}{p_2^+}-\frac{1}{p_1^+}\right)\right]\;,\\
\left( p_2\gam_3\left|v\right| p_1\right)&=&e \sqrt{p_1^+p_2^+}\left[\del_{s_1 s_2}
\eps_{s_3}^{i\ast}\left(2\frac{p_3^i}{p_3^+}-\frac{p_1^i(s_1)}{p_1^+}-
\frac{p_2^{i\ast}(s_2)}{p_2^+}\right)\right.\nn\\
&&\hskip.85in +\left.\del_{{\overline s}_1s_2}\del_{s_1s_3}\frac{2i}{\sqrt{2}}{\overline s}_1 m_p
\left(\frac{1}{p_2^+}-\frac{1}{p_1^+}\right)\right]\;,\\
\left( {\overline e}_2\gam_3\left|v\right| {\overline e}_1\right)&=&
e \sqrt{p_1^+p_2^+}\left[\del_{s_1 s_2}
\eps_{s_3}^{i\ast}\left(2\frac{p_3^i}{p_3^+}-\frac{p_1^i(s_1)}{p_1^+}-
\frac{p_2^{i\ast}(s_2)}{p_2^+}\right)\right.\nn\\
&&\hskip.85in +\left.\del_{{\overline s}_1s_2}\del_{s_1s_3}\frac{2i}{\sqrt{2}}{\overline s}_1 m
\left(\frac{1}{p_2^+}-\frac{1}{p_1^+}\right)\right]
\;,\\
\left( e_3\left|v\right| e_1\gam_2\right)&=&-e \sqrt{p_1^+p_3^+}\left[\del_{s_1 s_3}
\eps_{s_2}^{i}\left(2\frac{p_2^i}{p_2^+}-\frac{p_1^i(s_1)}{p_1^+}-
\frac{p_3^{i\ast}(s_3)}{p_3^+}\right)\right.\nn\\
&&\hskip.85in +\left.\del_{{\overline s}_1s_3}\del_{{\overline s}_1s_2}\frac{2i}{\sqrt{2}} s_1 m
\left(\frac{1}{p_3^+}-\frac{1}{p_1^+}\right)\right]\;,\\
\left( p_3\left|v\right| p_1\gam_2\right)&=&e \sqrt{p_1^+p_3^+}\left[\del_{s_1 s_3}
\eps_{s_2}^{i}\left(2\frac{p_2^i}{p_2^+}-\frac{p_1^i(s_1)}{p_1^+}-
\frac{p_3^{i\ast}(s_3)}{p_3^+}\right)\right.\nn\\
&&\hskip.85in +\left.\del_{{\overline s}_1s_3}\del_{{\overline s}_1s_2}\frac{2i}{\sqrt{2}} s_1 m_p
\left(\frac{1}{p_3^+}-\frac{1}{p_1^+}\right)\right]\;,\\
\left( {\overline e}_3\left|v\right| {\overline e}_1\gam_2\right)&=&e \sqrt{p_1^+p_3^+}\left[\del_{s_1 s_3}
\eps_{s_2}^{i}\left(2\frac{p_2^i}{p_2^+}-\frac{p_1^i(s_1)}{p_1^+}-
\frac{p_3^{i\ast}(s_3)}{p_3^+}\right)\right.\nn\\
&&\hskip.85in +\left.\del_{{\overline s}_1s_3}\del_{{\overline s}_1s_2}\frac{2i}{\sqrt{2}} s_1 m
\left(\frac{1}{p_3^+}-\frac{1}{p_1^+}\right)\right]\;,\\
\left(\gam_3 \left|v\right| e_1{\overline e}_2\right)&=&-e \sqrt{p_1^+p_2^+}\left[\del_{s_1
{\overline s}_2}
\eps_{s_3}^{i\ast}\left(2\frac{p_3^i}{p_3^+}-\frac{p_1^i(s_1)}{p_1^+}-
\frac{p_2^{i}(s_2)}{p_2^+}\right)\right.\nn\\
&&\hskip.85in +\left.\del_{ s_1s_2}\del_{s_1s_3}\frac{2i}{\sqrt{2}} s_1 m
\left(\frac{1}{p_1^+}+\frac{1}{p_2^+}\right)\right]\;,\\
\left(e_2{\overline e}_3 \left|v\right| \gam_1\right)&=&-e \sqrt{p_2^+p_3^+}\left[\del_{
{\overline s}_2s_3}
\eps_{s_1}^{i}\left(2\frac{p_1^i}{p_1^+}-\frac{p_3^{i\ast}(s_3)}{p_3^+}-
\frac{p_2^{i\ast}(s_2)}{p_2^+}\right)\right.\nn\\
&&\hskip.85in +\left.\del_{ s_2s_3}\del_{s_1s_3}\frac{2i}{\sqrt{2}} {\overline s}_3 m
\left(\frac{1}{p_2^+}+\frac{1}{p_3^+}\right)\right]\;,\\
\left(e_3p_4 \left|v\right| e_1p_2\right)&=&-4e^2
\frac{\sqrt{p_1^+p_2^+p_3^+p_4^+}}{\left(p_1^+-p_3^+\right)^2}\del_{s_1s_3}\del_{s_2s_4}\;,\\
\left(e_3{\overline e}_4 \left|v\right| e_1{\overline e}_2\right)_{exchange}&=&-4e^2
\frac{\sqrt{p_1^+p_2^+p_3^+p_4^+}}{\left(p_1^+-p_3^+\right)^2}\del_{s_1s_3}\del_{s_2s_4}\;,\\
\left(e_3{\overline e}_4 \left|v\right| e_1{\overline e}_2\right)_{annihil}&=&4e^2
\frac{\sqrt{p_1^+p_2^+p_3^+p_4^+}}{\left(p_1^++p_2^+\right)^2}\del_{s_1{\overline s}_2}
\del_{s_3{\overline s}_4}
\;,
\eea
where these last two add as $\left(e_3{\overline e}_4 \left|v\right| e_1{\overline e}_2\right)=
\left(e_3{\overline e}_4 \left|v\right| e_1{\overline e}_2\right)_{exchange}+
\left(e_3{\overline e}_4 \left|v\right| e_1{\overline e}_2\right)_{annihil}$.

\chapter{4th order Bloch and 3rd order G-W}\label{app:bloch}

In this appendix we record the fourth order effective Bloch interactions
as promised in Eq.~(\ref{eq:blochtofourth}) of Chapter~\ref{ch3}. Also
we record the third order effective G-W interactions as promised at the end
of Section~\ref{G-W}. 

First, fourth order Bloch:
\bea
\langle a|v_\lam^{^{(4)}}|b\rangle&=&
\frac{1}{2}\sum_{i,j,k}\langle a | v_\Lam|i\rangle\langle i|v_\Lam|j\rangle\langle
 j|v_\Lam|k\rangle\langle k | v_\Lam|b\rangle\left[
 \frac{1}{\Del_{ai}\Del_{aj}\Del_{ak}}+\frac{1}{\Del_{bi}\Del_{bj}\Del_{bk}}
\right]
\nn\\
&&-\frac{1}{2}\sum_{i,c,j}\langle a | v_\Lam|i\rangle\langle i|v_\Lam|c\rangle\langle
 c|v_\Lam|j\rangle\langle j | v_\Lam|b\rangle\left[
 \frac{1}{\Del_{ai}\Del_{aj}\Del_{cj}}+\frac{1}{\Del_{bi}\Del_{bj}\Del_{ci}}
\right.
\nn\\
&&\left.\hskip0.5in+\frac{1}{4}\left\{
\frac{1}{\Del_{ai}\Del_{ci}\Del_{cj}}+\frac{1}{\Del_{ci}\Del_{cj}\Del_{bj}}
-\frac{1}{\Del_{ai}\Del_{ci}\Del_{bj}}-\frac{1}{\Del_{ai}\Del_{cj}\Del_{bj}}\right\}\right]
\nn\\
&&-\frac{1}{2}\sum_{i,j,c}\langle a | v_\Lam|i\rangle\langle i|v_\Lam|j\rangle\langle
 j|v_\Lam|c\rangle\langle c | v_\Lam|b\rangle\left[
 \frac{1}{\Del_{bi}\Del_{cj}}\left\{\frac{1}{\Del_{ci}}+\frac{1}{\Del_{bj}}\right\}
\right]
\nn\\
&&-\frac{1}{2}\sum_{c,i,j}\langle a | v_\Lam|c\rangle\langle c|v_\Lam|i\rangle\langle
 i|v_\Lam|j\rangle\langle j | v_\Lam|b\rangle\left[
 \frac{1}{\Del_{ci}\Del_{bj}}\left\{\frac{2}{\Del_{ai}}+\frac{2}{\Del_{bi}}\right\}
\right.\nn\\
&&\left.\hskip2.9in +\frac{1}{\Del_{ci}\Del_{aj}}
\left\{\frac{1}{\Del_{ai}}+\frac{1}{\Del_{cj}}\right\}\right]
\nn\\
&&+\frac{1}{2}\sum_{i,c,d}\frac{\langle a | v_\Lam|i\rangle\langle i|v_\Lam|c\rangle\langle
 c|v_\Lam|d\rangle\langle d | v_\Lam|b\rangle}{
 \Del_{bi}\Del_{ci}\Del_{di}}
\nn\\
&&+\frac{1}{2}\sum_{c,d,i}\frac{\langle a | v_\Lam|c\rangle\langle c|v_\Lam|d\rangle\langle
 d|v_\Lam|i\rangle\langle i | v_\Lam|b\rangle}{
 \Del_{ai}\Del_{ci}\Del_{di}}
\;.
\eea

And now we record third order G-W:
\bea
{\overline v}_{\lam ij}^{^{(3)}}&=&
\sum_{ab}\left[
{\overline v}_{\Lam ia}{\overline v}_{\Lam ab}{\overline v}_{\Lam bj}
\left\{
\frac{\xi_1}{\Del_{ja}\Del_{ab}}+\frac{\xi_2}{\Del_{ja}\Del_{jb}}\right.\right.\nn\\
&&\left.\left.~~~~~~~~~~+\frac{\xi_3}{\Del_{jb}\Del_{ia}}+\frac{\xi_4}{\Del_{jb}\Del_{ba}}
\right\}
+(i\longleftrightarrow j)^\dagger\right]
\;,
\eea
where
\beaa
&&\xi_1={\overline f}_{\lam ab} f_{\Lam ab} \Theta_{bai} \Theta_{abj} \Theta_{jab}+
{\overline f}_{\lam aj} f_{\Lam aj} \Theta_{jai} \Theta_{abj} f_{\Lam ab} \Theta_{baj}\;,\\
&&\xi_2={\overline f}_{\lam jb} f_{\Lam jb} \Theta_{jbia} \Theta_{jba} \Theta_{ajb}+
{\overline f}_{\lam aj} f_{\Lam aj} \Theta_{jai} \Theta_{jba} f_{\Lam jb} \Theta_{bja}\;,\\
&&\xi_3={\overline f}_{\lam bj} f_{\Lam bj} \Theta_{jbi} \Theta_{iabj} f_{\Lam ia}\Theta_{iab}\;,\\
&&\xi_4={\overline f}_{\lam bj} f_{\Lam bj} \Theta_{jbi} \Theta_{abj}f_{\Lam ab} \Theta_{bai}
\;.
\eeaa
This is for $f_{\lam ij}=\theta(\lam-|\Del_{ij}|)$. 
Recall $f_\lam+{\overline f}_\lam=1$ and $\Del_{ij}=\veps_i-\veps_j$.
Note that the labels `$i,~j,~ \ldots$' and `$a,~b, ~\ldots$' span the whole energy space
[unlike the notation used for Bloch;
see below Eq.~(\ref{eq:blochtofourth}) for the notation in this case].
$\Theta_{abcd}=\theta(|\Del_{ab}|-|\Del_{cd}|)$ 
and $\Theta_{abc}=\theta(|\Del_{ab}|-|\Del_{bc}|)$. 
Note that `$(i\longleftrightarrow j)^\dagger$'
 implies `${\overline v}_{\Lam ia}
 {\overline v}_{\Lam ab}{\overline v}_{\Lam bj}\longrightarrow
 {\overline v}_{\Lam ib}
 {\overline v}_{\Lam ba}{\overline v}_{\Lam aj}$ and $(i\longleftrightarrow j)$ 
 in the remaining pieces of the respective term'.
As advertised we see that the denominators are constrained to be between
$\lam$ and $\Lam$ ($\lam < \Lam$): there are no small energy denominators (as long as
 $\lam$ is 
not too small).

\chapter{Hyperspherical harmonics and Fock coordinate change}\label{app:hyper}

In this appendix we
 list some useful mathematical relations used in Chapter~\ref{ch:pos}.
 The conventions followed here are the same as in Ref.~\cite{judd1}. These
 hyperspherical 
 harmonics are given by
 \bea
 Y_\nu(\Omega)&\equiv&Y_{n,l,m}(\Omega)\equiv f_{n,l}(\omega)
 Y_{l,m}(\theta,\phi)
 \;,
 \eea
 where
 \bea
 0&\leq&|m|\leq l\leq n-1
 \;.
 \eea
 These function labels $\nu=(n,\,l,\,m)$ correspond with
 the standard nonrelativistic quantum numbers of hydrogen.
 These 
  spherical harmonics, $Y_{l,m}(\theta,\phi)$, are given by~\cite{jackson}
 \bea
 Y_{l,m}(\theta,\phi)&=&
 \sqrt{\frac{2 l+1}{4 \pi}\frac{(l-m)!}{(l+m)!}}
 \frac{(-1)^m}{2^l l!}
 (1-x^2)^{\frac{m}{2}}
 \frac{d^{l+m}}{dx^{l+m}}
 (x^2-1)^l
 e^{i m \phi}
 \;,
 \eea
 where $x=\cos\theta$.
 These other functions, $f_{n,l}(\omega)$, are given by
 \bea
 f_{n,l}(\omega)&=&
 (-i)^l \sqrt{\frac{2 n(n-l-1)!}
 {\pi (n+l)!}}
 \sin^l\omega\frac{d^l}{(d\cos\omega)^l}C_{n-1}(\cos\omega)
 \;,
 \eea
 where $C_{n-1}(\cos\omega)$ are Gegenbauer polynomials. The
 first few are~\cite{judd1prime}
 \bea
 C_0(y)&=&1\;,\;C_1(y)=2 y\;,\;C_2(y)=4 y^2-1\;,\;C_3(y)=8 y^3-4 y\nonumber~,\\
 C_4(y)&=&16 y^4-12 y^2+1 \;,\;C_5(y)=32 y^5-32 y^3+6 y
 \;.
 \eea
 The phase and orthonormality   relations  of these above functions are
 \be
Y_{n,l,m}= (-1)^{l+m} Y_{n,l,-m}^\ast\;,\;Y_{l,m}=(-1)^m
Y_{l,-m}^\ast\;,\;f_{n,l}=
(-1)^lf_{n,l}^\ast 
\ee
and
\bea
&&\int d\Omega^{^{(3)}} Y_{l,m}^\ast
 Y_{l^\pr,m^\pr}=
 \del_{l l^\pr} \del_{m m^\pr}
\;,\;\int
d\omega\sin^2\omega f_{n,l}^\ast
 f_{n^\pr,l}=\del_{n n^\pr}\nn\\
 &&\hskip.5in \Longrightarrow\int 
 d\Omega Y_\nu^\ast Y_{\nu^\pr}=\del_{\nu \nu^\pr}
\;,
\eea
where
$d\Omega\equiv d\Omega^{^{(4)}} \equiv  d\Omega^{^{(3)}} d\omega \sin^2\omega$,
 $0 \leq \omega\leq\pi$, $0\leq\theta\leq\pi$
 and $0\leq\phi\leq 2 \pi$.
Combining the spherical harmonics and respective derivative of the
Gegenbauer polynomials,    for
the first few hyperspherical harmonics gives
\bea
Y_{1,0,0}&=&\frac{1}{\sqrt{2 \pi^2}}~~,~~Y_{2,0,0}=\frac{\sqrt{2} \cos
\omega}{\pi}\nonumber\\
Y_{2,1,-1}&=&\frac{-i e^{-i\phi} \sin \omega \sin
\theta}{\pi}~~,~~Y_{2,1,0}=\frac{-i \sqrt{2}
 \sin \omega\cos \theta}{\pi}\nonumber\\
Y_{2,1,1}&=&\frac{i e^{i\phi} \sin \omega \sin \theta}{\pi}\;.
\eea
For further harmonics we
refer the interested
reader to Appendix~2 of Judd's text~\cite{judd2}, where this is done quite
nicely.

For the coordinate change in the Coulomb Schr{\"o}dinger equation 
of positronium
[see  Eq.~(\ref{eq:C}) and the discussion that follows it],
for ${\cal B}_{_{N}} < 0$, we define
\bea
 m {\cal B}_{_{N}}&\equiv&-{e_n}^2~,\\
u&\equiv&(u_0,{\bf u}) ~,\\
 u_0&\equiv&\cos(\omega)\equiv\frac{e_n^2-{\bf p}^2}{e_n^2+{\bf p}^2}~,\\
 {\bf u}&\equiv&\frac{{\bf p}}{\bp}\sin(\omega) \equiv \sin(\omega)\left[
\sin(\theta) \cos(\phi) ,sin(\theta) \sin(\phi) , \cos(\theta)\right]\nonumber\\
&\equiv& \frac{2 e_n {\bf p}}{e_n^2+{\bf p}^2}
 \;.\eea
 Note that $u_0^2+{\bf u}^2=1$. Conversely this coordinate change gives
 \bea
 {\bf p}&=&\frac{e_n}{1+u_0}{\bf u}~,\\
 e_n^2+{\bf p}^2&=&\frac{2 e_n^2}{1+u_0}
 \;.
 \eea
 We also have
 \bea
 &&d\Omega_p= \sin^2\omega d\omega d\Omega_p^{^{(3)}}=\left(
\frac{2 e_n}{e_n^2+{\bf p}^2} \right)^3 d^3p~,\\
&&\del^3(p-p^\pr)=\frac{(2 e_n)^3}{(e_n^2+{\bf p}^2)^3}
 \del(\Omega_p-\Omega_{p^\pr})=\frac{(1+u_o)^3}{e_n^3} \del(\p-\ppr)~.
 \eea

After this coordinate change, the nonrelativistic 
Schr{\"o}dinger equation of positronium 
is trivially
integrable as described below Eq.~(\ref{eq:C})
in Chapter~\ref{ch:pos}. Here we record the result for 
the bound wavefunctions and eigenvalues:
\be
\phi_\nu({\bf p})=\frac{4 (e_n)^{\frac{5}{2}}}{(e_n^2+{\bf p}^2)^2}Y_\nu(\Omega_p)
\ee
and
\be
{\cal B}_{_{n}}=-\frac{m \alpha^2}{4 n^2}\Longrightarrow e_n=\frac{m\,\al}{2\,n}
\ee
respectively.

 Finally, a most useful relation is
 \bea
 |{\bf p}-{\bf p}^\pr|^2&=&\frac{(e_n^2+{\bf p}^2)(e_n^2+{{\bf p}^\pr}^2)}{4
 e_n^2}|u-u^\pr|^2
 \;.
 \label{eq:1492}
 \eea
 This is useful because we can expand $|u-u^\pr|^2$ as follows
 \bea
 \frac{1}{|u-u^\pr|^2}&=&
 \sum_\nu \frac{2 \pi^2}{n} Y_\nu(\Omega_p) Y_\nu^\ast(\Omega_{p^\pr})
 \;.
 \label{eq:1900}
 \eea
 This completes the brief discussion on the hyperspherical harmonic 
 mathematical relations used in Chapter~\ref{ch:pos}; actually,
 as shown in Subsection~\ref{hard},
  the $\hard$ calculation requires some further formulas, which are given
 as they are needed.

\chapter{$M_{_{N}}^2$ -vs- $B_{_{N}}$ in positronium
as an illustration}\label{app:mvsb}

In this appendix we invert the equation
\bea
M_{_{N}}^2&\equiv&(2 m+B_{_{N}})^2
\;,
\eea
and obtain the $\alpha$-expansion for
 the binding energy, $-B_{_{N}}$. In Chapter~\ref{ch:pos}
  we set up a procedure to calculate the mass-squared
 $M_{_{N}}^2$ of positronium. This gave
 \bea
 M_{_{N}}^2&=&{\cal M}_N^2+b_4 m^2 \alpha^4+{\cal O}\left(\alpha^5\right)\;,
 \label{eq:eq}
 \eea
 with
 \bea 
 {\cal M}_N^2&\equiv&4 m^2+4 m {\cal B}_{_{N}}\;.
 \eea
 For the leading-order spectrum of $H_\lam$ we obtained
 \bea
 {\cal B}_{_{N}}&=& -\frac{1}{4} \frac{m \alpha^2}{n^2}\;.
 \eea
Taking a square root of $M_{_{N}}^2$ gives
 \bea
 B_{_{N}}&=& {\cal B}_{_{N}}
 +\frac{m \alpha^4}{2} \left( \frac{b_4}{2}-\frac{1}{32 n^4}     \right)
+{\cal O}\left(\alpha^5\right)
 \;.\label{eq:eee}
 \eea
 Recall that $\frac{m \alpha^4}{2}=\alpha^2 Ryd$. Now, in this work,
  Eqs.~(\ref{eq:1776})--(\ref{eq:1493}) 
 are the results of our calculation of the spin splittings of $M_{_{N}}^2$ in
 the ground state of
 positronium. These 
 were derived in the form of Eq.~(\ref{eq:eq}) with result
 \bea
 b_4(triplet)-b_4(singlet)&=&\frac{7}{3}
 \;.
 \eea
 Given Eq.~(\ref{eq:eee}) this implies
 \bea
 B_{_{triplet}}-B_{_{singlet}}&=&\frac{7}{6} \alpha^2 Ryd +{\cal
 O}\left(\alpha^5\right)
 \;.
 \eea
 This we recognize as the well known result for the positronium system.  A final
 note is
 that if the physical values of the fine structure constant and Rydberg energy 
 ($\frac{1}{137.0}$ and $13.60\,{\rm eV}$ respectively)
 are applied to this previous formula, the result agrees with experiment to
 one-half of a
  percent~\cite{IZ2}.

\chapter{Averaging over directions}\label{app:avg}

We start
 this appendix by calculating the following Coulomb matrix element
\bea
I_\perp&\equiv&\int d^3 p \int d^3 p^\pr\;\phi_\nu^{ \ast}({\bf p}^\pr) \frac
{\left({ p}^\perp-{ p}^{\pr \perp}\right)^2}{\pp}\phi_{\nu}({\bf p})
\;,
\eea
to verify the step taken from Eq~(\ref{eq:100}) to (\ref{eq:10122}).
It is useful to define another integral
\bea
I_z&\equiv&\int d^3 p \int d^3 p^\pr\;\phi_\nu^{ \ast}({\bf p}^\pr) \frac
{\left({ p}_z-{ p}^\pr_z\right)^2}{\pp}\phi_{\nu}({\bf p})
\;.
\eea
Now note that
\bea
I&\equiv&I_\perp+I_z=\int d^3 p \int d^3 p^\pr\;\phi_\nu^{ \ast}({\bf p}^\pr) 
\phi_{\nu}({\bf p})
=\frac{(2 \pi)^{3}} {\pi}\left(\frac{m \alpha}{n}\right)^3\del_{l,0}
\;,\label{eq:yyyy}
\eea
where in this last step we recalled Eq.~(\ref{eq:103}) and the fact that
the wave function at the origin is real.

For $l=0$, the wave function satisfies
\bea
&&\phi_{n,0,0}({\bf p})=\phi_{n,0,0}(|{\bf p}|)
\;.
\eea
Thus, by symmetry, for $l=0$,
\bea
&&I_z=\frac{1}{3} I=\frac{(2 \pi)^{3}} {3\pi}\left(\frac{m \alpha}{n}\right)^3
\;,
\eea
and
\bea
&&I_\perp=\frac{2}{3} I=\frac{2(2 \pi)^{3}} {3\pi}\left(\frac{m
\alpha}{n}\right)^3
\;.
\eea

For $l \neq 0$, first note that $I=0$. Thus, for $l \neq 0$,
\bea
I_\perp=-I_z
\;;\label{eq:131}
\eea
we will calculate $I_z$ below which then implies $I_\perp$.
Next
note that  in position space 
\bea
I&=&-2 \pi^2\int d^3 x~ \phi_\nu^\ast\left({\bf x}\right)\left(
{{\nabla}}^2\frac{1}{|{\bf x}|}
\right) \phi_\nu\left({\bf x}\right)
\;,
\eea
using ${\vec{\nabla}}^2\frac{1}{|{\bf x}|}=-4 \pi \del^3 \left({\bf x}\right)$.
 Thus, for $l \neq 0$,
 in position space
\bea
I_z&=&-2 \pi^2\int d^3 x ~\phi_\nu^\ast\left({\bf x}\right)\left(
{\nabla}_z^2\frac{1}{|{\bf x}|}
\right) \phi_\nu\left({\bf x}\right)
\;.
\eea
Note that there is no $|{\bf x}|\rightarrow 0$ ambiguity in this previous
equation because for $l \neq 0$, the wave function vanishes at the origin.
Carrying out the derivative gives
\bea
I_z&=&-2 \pi^2\int d^3 x ~\phi_\nu^\ast\left({\bf x}\right)\left(
\frac{-1+3z^2/|{\bf x}|^2}{|{\bf x}|^3}
\right) \phi_\nu\left({\bf x}\right)
\;.\label{eq:lll}
\eea
This matrix element was performed in the first appendix of
Bethe and Salpeter's textbook \cite{bethetext}. We use two of 
their formulas, (3.26) and (A.29).\footnote{Warning to the reader: In this text,
they use atomic units,
$\hbar=c=m=\alpha=1$, so $m$ and $\alpha$ have to be placed back into the
formulas.}
Eq.~(\ref{eq:lll}) integrated gives
\bea
I_z&=&-2 \pi^2 ~\overline{r^{-3}}~ c(l,m_l)
\;,
\eea
with
\bea
\overline{r^{-3}}&=&\frac{1}{l(l+1)(l+\frac{1}{2})} \left(\frac{m
\alpha}{n}\right)^3\;,
\\
c(l,m_l)&=&-1+3 \left(
\frac{2 l^2+2 l-1-2 m_l^2}
{(2 l+3)(2 l-1)}
\right)
\;.
\eea
Thus, recalling Eq.~(\ref{eq:131}), our result for $l\neq 0$ is
\bea
I_\perp&=&2 \pi^2 ~\overline{r^{-3}}~ c(l,m_l)
\;.\label{eq:nhy}
\eea

For $l=1$, $I_\perp$ is not zero, so what is going on?  The answer lies in the
fact 
that we really want to take matrix elements in 
the $|j,m_j\rangle$ basis not the
$|m_l,m_s\rangle$ 
basis,\footnote{Recall throughout this dissertation we have specified
electron helicity by $s_e/2=\pm 1/2$. Here we set $s_e/2\longrightarrow
m_s$. Also we use `$s$' to denote the electron spin, that
is $s=1/2$.} 
and based on rotational invariance
our results are assumed to be independent of $m_j$. 
To proceed, note that we assumed $m_s$ is constant in our calculation of the dominant
part of the Lamb
shift since it is a self-energy shift.
Also note that the result, Eq.~(\ref{eq:nhy}), is even under 
$m_l
\longrightarrow
-m_l$. Below we will use these facts to show that
 the Clebsch-Gordan coefficients (specifically we show it for the
 non-trivial case---the
$2 P_{\frac{1}{2}}$ states) imply
\bea
&&\langle j=1/2,m_j|V|j=1/2,m_j\rangle=\nn\\
&&\hskip1in \frac{1}{2l+1}\frac{1}{2s+1}
\sum_{m_l=-l}^{l}\sum_{m_s=-s}^{s}
\langle m_lm_s|V|m_lm_s\rangle
\;.\label{eq:wyatt2}
\eea
For now just take this result as given.
Now note that $I_\perp$ given by Eq.~(\ref{eq:nhy}) averaged over $m_l$
vanishes,
\bea
\frac{1}{2l+1}\sum_{m_l=-l}^{l}I_\perp=0
\;,\label{eq:GGG}
\eea
where  we used
\bea
\frac{1}{2l+1}\sum_{m_l=-l}^{l}m_l^2=\frac{1}{3}l(l+1)
\;,
\eea
an obvious result after the answer is known.
This result [Eq.~(\ref{eq:GGG})] was used in the step that led from
Eq.~(\ref{eq:100}) 
to Eq.~(\ref{eq:10122}) in the dissertation.

Now we show that given `$m_s={\rm constant}$', `the symmetry
under $m_l
\longrightarrow
-m_l$' and `independence of $m_j$' Eq.~(\ref{eq:wyatt2}) follows.
Using the standard Clebsch-Gordan coefficients \cite{expt},
the $2 P_{\frac{1}{2}}$ expectation values are 
\bea
\left\langle\frac{1}{2}\frac{1}{2}\right|V\left|\frac{1}{2}
\frac{1}{2}\right\rangle &=&
\frac{1}{3}\left\langle 0\frac{1}{2}\right|V\left|0\frac{1}{2}\right\rangle
+\frac{2}{3}\left\langle 1-\frac{1}{2}\right|V\left|1-\frac{1}{2}\right\rangle\nn\\
&&\hskip1in -\frac{2 \sqrt{2}}{3} \Re\left\langle
0\frac{1}{2}\right|V\left|1-\frac{1}{2}\right\rangle
\eea
and
\bea
\left\langle\frac{1}{2}-\frac{1}{2}\right|V\left|\frac{1}{2}-\frac{1}{2}\right\rangle &=&
\frac{1}{3}\left\langle 0-\frac{1}{2}\right|V\left|0-\frac{1}{2}\right\rangle
+\frac{2}{3}\left\langle -1\frac{1}{2}\right|V\left|-1\frac{1}{2}\right\rangle\nn\\
&&\hskip1in -\frac{2 \sqrt{2}}{3} \Re\left\langle
0-\frac{1}{2}\right|V\left|-1\frac{1}{2}\right\rangle
\;,
\eea
where on the left of the equal sign the states are in the 
$|jm_j\rangle$ basis,
and on the right of the equal sign the states are in the 
$|m_lm_s\rangle$ basis. $\Re$ implies ``the real part."
Given `$m_s={\rm constant}$' and the symmetry
under $m_l
\longrightarrow
-m_l$ these equations become
\bea
\left\langle\frac{1}{2}\frac{1}{2}\right|V\left|\frac{1}{2}\frac{1}{2}\right\rangle &=&
\frac{1}{3}\left\langle 0\frac{1}{2}\right|V\left|0\frac{1}{2}\right\rangle
+\frac{1}{3}\left\langle 1-\frac{1}{2}\right|V\left|1-\frac{1}{2}\right\rangle\nn\\
&&\hskip1in +\frac{1}{3}\left\langle -1-\frac{1}{2}\right|V\left|-1-\frac{1}{2}\right\rangle \;,\\
\left\langle\frac{1}{2}-\frac{1}{2}\right|V\left|\frac{1}{2}-\frac{1}{2}\right\rangle &=&
\frac{1}{3}\left\langle 0-\frac{1}{2}\right|V\left|0-\frac{1}{2}\right\rangle
+\frac{1}{3}\left\langle -1\frac{1}{2}\right|V\left|-1\frac{1}{2}\right\rangle\nn\\
&&\hskip1in +\frac{1}{3} \left\langle 1\frac{1}{2}\right|V\left|1\frac{1}{2}\right\rangle
\;.
\eea
Given $m_j$ independence we can take one-half of the sum of these
two terms giving
\bea
&&\frac{1}{2}\left[
\left\langle\frac{1}{2}\frac{1}{2}\right|V\left|\frac{1}{2}\frac{1}{2}\right\rangle
+
\left\langle\frac{1}{2}-\frac{1}{2}\right|V\left|\frac{1}{2}-\frac{1}{2}\right\rangle
\right]=\nn\\
&&\hskip.2in \frac{1}{2}\left[
\frac{1}{3}\left\langle 0\frac{1}{2}\right|V\left|0\frac{1}{2}\right\rangle
+\frac{1}{3}\left\langle 1-\frac{1}{2}\right|V\left|1-\frac{1}{2}\right\rangle
+\frac{1}{3}\left\langle -1-\frac{1}{2}\right|V\left|-1-\frac{1}{2}\right\rangle
\right.\nn\\
&&\left.\hskip.5in
+\frac{1}{3}\left\langle 0-\frac{1}{2}\right|V\left|0-\frac{1}{2}\right\rangle
+\frac{1}{3}\left\langle -1\frac{1}{2}\right|V\left|-1\frac{1}{2}\right\rangle
+\frac{1}{3} \left\langle 1\frac{1}{2}\right|V\left|1\frac{1}{2}\right\rangle
\right]
\eea
which is equivalent to Eq.~(\ref{eq:wyatt2}) 
for the specific case of the $2 P_{\frac{1}{2}}$ states,
as was to be shown.



\end{document}